\documentclass{article}

\usepackage{amsmath, amsthm, amscd, amsfonts, amssymb, graphicx, color}
\usepackage[bookmarksnumbered, colorlinks, plainpages]{hyperref}
\usepackage{pspicture,pst-all,multido}
\theoremstyle{definition}

\theoremstyle{remark}

\numberwithin{equation}{section}

\newcommand{\C}{\mathbb{C}}
\newcommand{\R}{\mathbb{R}}
\newcommand{\N}{\mathbb{N}}

\newcommand{\norm}[1]{\parallel\!\!#1\!\!\parallel}
\renewcommand{\d}{\operatorname{d}}
\newcommand{\e}{\operatorname{e}}
\renewcommand{\i}{\operatorname{i}}

\newcounter{envcount}%

{\vspace{\bigskipamount}\refstepcounter{envcount}\textbf{(\theenvcount)\enspace Asymptotic causality.}}%
  {\vspace{\bigskipamount}}

\newenvironment{Def}%
{\vspace{\bigskipamount}\refstepcounter{envcount}\textbf{(\theenvcount)\enspace Definition.}}%
  {\vspace{\bigskipamount}}

\newenvironment{Exa}%
{\vspace{\bigskipamount}\refstepcounter{envcount}\textbf{(\theenvcount)\enspace Example.}}%
  {\vspace{\bigskipamount}}

{\vspace{\bigskipamount}\refstepcounter{envcount}\textbf{(\theenvcount)\enspace Remarks.}}%
  {\vspace{\bigskipamount}}

\newenvironment{ANC}%
{\vspace{\bigskipamount}\refstepcounter{envcount}\textbf{(\theenvcount)\enspace Effectiveness  of NC}.}%
  {\vspace{\bigskipamount}}

\newenvironment{VLem}%
{\vspace{\bigskipamount}\refstepcounter{envcount}\textbf{(\theenvcount)\enspace Lemma.}}%
  {\vspace{\bigskipamount}}

{\vspace{\bigskipamount}\refstepcounter{envcount}\textbf{(\theenvcount)\enspace Note.}}%
  {\vspace{\bigskipamount}}

\newenvironment{The}%
{\vspace{\bigskipamount}\refstepcounter{envcount}\textbf{(\theenvcount)\enspace Theorem.}\itshape}%
  {\vspace{\bigskipamount}\upshape}
  
\newenvironment{Theo}%
{\vspace{\bigskipamount}\refstepcounter{envcount}\textbf{(\theenvcount)\enspace Theorem}\itshape}%
  {\vspace{\bigskipamount}\upshape}
  
\newenvironment{Pro}%
{\vspace{\bigskipamount}\refstepcounter{envcount}\textbf{(\theenvcount)\enspace Proposition.}\itshape}%
  {\vspace{\bigskipamount}\upshape}

\newenvironment{Cor}%
{\vspace{\bigskipamount}\refstepcounter{envcount}\textbf{(\theenvcount)\enspace Corollary.}\itshape}%
  {\vspace{\bigskipamount}\upshape}

\newenvironment{Lem}%
{\vspace{\bigskipamount}\refstepcounter{envcount}\textbf{(\theenvcount)\enspace Lemma.}\itshape}%
  {\vspace{\bigskipamount}\upshape}

\theoremstyle{definition}
\swapnumbers

\begin{document}
\setcounter{page}{1}
\pagenumbering{arabic}




\vspace{2mm}
\begin{center}
{\Large Causal Localizations of the Massive Scalar Boson}\\  

\vspace{1cm}
Domenico P.L. Castrigiano\\
Technischen Universit\"at M\"unchen, Fakult\"at f\"ur Mathematik, M\"unchen, Germany\\ 

\smallskip

{\it E-mail address}: {\tt
castrig\,\textrm{@}\,ma.tum.de}\\

\end{center}

\begin{abstract} The positive operator valued localizations (POL) of a massive scalar boson are constructed and a characterization and structural analyses of their kernels are obtained. 
In the focus of this article are the causal features of the POL. There is the well-known causal time evolution (CT). Recently a POL  by Terno and Moretti, which is a kinematical deformation of the Newton-Wigner localization (NWL) and belongs to the here fully analyzed class of finite POL, is shown by V.\,Moretti to comply  with CT. A further POL with CT treated here, which is in the same class,  is the only one being  the trace of a projection valued localization (like NWL) with CT. --- Causality  imposes a condition CC, which implies CT but  is more restrictive than CT. Extending Moretti's method   it is shown rigorously that  the POL of the class introduced by Petzold et al. satisfy CC. Their kernels are called causal kernels, of which  a  rather detailed description  is achieved. One the way there the  case of one spatial dimension is solved completely. This case is  instructive. In particular it directed Petzold et al.\,and subsequently Henning, Wolf to find their  basic one-parameter family $K_r$ of causal kernels. The causal kernels are, up to a fixed energy factor, normalized positive definite Lorentz invariant kernels. A full characterization of the latter is attained due to their close relation to the zonal spherical functions on the Lorentz group. Finally these considerations discharge into the main result that  $K_{3/2}$ is the absolute maximum, viz. $|K|\le K_{3/2}$ for all causal kernels $K$.
\end{abstract}

\section{Introduction}
As well-known relativistic massive particle position would be described in a fully satisfactory manner by the Newton-Wigner localization  (NWL), since above all NWL provides an abundance of  boundedly localized states, if there were not the requirement of causality.
The so-called Einstein causality in the most direct interpretation requires from 
localization that the probability of localization in a region of influence cannot be less than that in the region of actual localization. NWL violates blatantly this principle as it localizes frame-dependently. There is no state of the particle for which the particle is boundedly localized  by NWL in  different spacelike hyperplanes.\\
\hspace*{6mm}
So one is induced to determine the probability of localization in a region no longer by the expectation value of an orthogonal  projection  attributed to the region as occurs for NWL, but rather of a positive operator with spectrum in the unit interval  thus giving rise to a positive operator valued localization (POL).   By now PO-localization   is a settled concept. For a brief discussion and several references see for instance the passage in \cite[sec.\,III.F]{CL15}.\\
\hspace*{6mm}
In sec.\,\ref{CPOLMSB} we construct the POL for a massive scalar boson.  The result is the explicit formula in the theorem (\ref{APOL}), which  furnishes myriads of  different POL.  
In order to  recognize  their  features it is very effective to describe a POL  in a concise manner by its kernel.  A  useful characterization  and structural analyses of the POL  kernels are given by the theorem (\ref{CKPOLMB}).\\
\hspace*{6mm}
To every POL  a Euclidean covariant position operator  is attributed, which is  the  first moment of the POL.  The physical relevance of the POL is underlined by the fact that in case of a real smooth POL kernel the position operator coincides with the Newton-Wigner position operator. The proof is due to V. Moretti. Recently in   \cite{M23}  a
  POL $T^{TM}$ by Terno and Moretti 
 is studied in detail showing furthermore  that $T^{TM}$ is a kinematical deformation of  NWL with  almost localized states  for every region with nonempty  interior and  which obeys causal  time evolution (see CT in sec.\,\ref{POLCTE}). From general criteria for POL kernels we obtain  rather short proofs of these properties, see  sec.\,\ref{POLTM}, (\ref{SPOL}).  What is more,  (\ref{PLSS}) provides a simple explicit formula for a  sequence of localized states at  any point of the space. Actually  (\ref{PLSS}) is a byproduct of the proof of the easy criterion (\ref{SSL0})   in sec.\,\ref{LBR} on POL kernels for the existence of point localized sequences of states. 
 \\
\hspace*{6mm}
In sec.\,\ref{POLCTE} we treat also the POL with trace CT $T^{tct}$,  which has the noted  properties of $T^{TM}$ and which is the unique POL being the trace of a projection valued localization with CT.
 $T^{tct}$ is also a  kinematical deformation of NWL (\ref{FPOLKD}), all of which are given by the formula    (\ref{GKDNWL}).
\\
\hspace*{6mm}
However one cannot content oneself with POL obeying CT as the latter satisfies only partially the requirements   of  causality. If  $T$ is a POL, $\Delta$  a spacelike region, and $\sigma$ any spacelike hyperplane, then as expounded in sec.\,\ref{CKD} causality imposes the condition 
\begin{equation*}\label{POLCC}
T(\Delta)\le T(\Delta_\sigma) \tag{CC}
\end{equation*}
were $\Delta_\sigma$ denotes  the region of influence of $\Delta$ in $\sigma$, i.e., the set of all points  in $\sigma$, which can be reached  from some point  in $\Delta$ by a signal not moving faster than light. CT concerns just  the case that $\Delta$ and $\sigma$ are parallel. POL which satisfy CC are called causal. \\
\hspace*{6mm}
The causality principle formulated above let one think  of  the probability of localization as a conserved quantity reigned by an associated  density current. Petzold and his group \cite{GGP67} were apparently  the first to study the conserved covariant four-vector  currents with a positive definite zeroth component. The outcome on POL kernels is reported in sec.\,\ref{KKO}. See (\ref{CCKKK}) for a clear-cut result in this regard.
\\
\hspace*{6mm}
Petzold et al.\,\cite{GGPR68} argue that their  POL obey CT. Probably they were the first to introduce and to treat  the concept of CT. 
However they do not take under consideration  the full causality requirement CC.
Actually, extending Moretti's  method \cite{M23} we succeed   in (\ref{GCCPOL})  to prove rigorously 
that their POL  are causal.  Moreover, we mention particularly    a communication  by R.F. Werner, which  suggests  that most probably these are the only causal POL. Therefore we refer quite simply to their kernels (\ref{KKK}) as the causal kernels.\\
\hspace*{6mm}
A small digression (sec.\,\ref{CKOSD}) concerns  the case of one spatial dimension which we completely solve by the formula in (\ref{MROD}). This case is rather instructive since it directed Petzold et al.\,\cite{GGP67} and subsequently Henning, Wolf \cite{HW71} to find the particular causal kernels $K_r$ (\ref{KKKR}), which turn out to be fundamental.\\
\hspace*{6mm}
Multiplying a causal kernel by an element of $\mathcal{G}$ yields again a causal kernel.  Here $\mathcal{G}$ denotes the set of all continuous positive definite Lorentz invariant normalized kernels (\ref{LIPGK}). What is more, any
causal kernel  is  the product of some element of $\mathcal{G}$ and a fixed energy factor.
Therefore it is important to know $\mathcal{G}$. In  (\ref{MTGP}) a complete description of $\mathcal{G}$ is attained employing means of group representation theory.\\
\hspace*{6mm}
Each  $G\in\mathcal{G}$ determines by the Reproducing Kernel Hilbert Space (RKHS) construction a continuous unitary representation of $SL(2,\C)$ in separable Hilbert space (sec.\,\ref{PLKIV}). By this construction $G$ is closely related to the matrix element of  a $SU(2)$-invariant vector. Since $SL(2,\C)$ is locally compact tame  with countable basis, $SL(2,\C)$ admits an integral representation of the latter by the zonal  spherical functions on $SL(2,\C)$ (\ref{GLIPDK}).\\
\hspace*{6mm}
The question is to find out which $G\in\mathcal{G}$ determine a causal kernel. By (\ref{CKK}) these are only  $G$'s, which are derived from the principal series and are not extreme. We mention the inversion formula in (\ref{CIF}) concerning these elements.\\
\hspace*{6mm}
At this point we are ready for the main  result (\ref{MTKK}). Put $m>0$ mass, $k,p\in\R^3$ momentum, $\epsilon(k), \epsilon(p)$ energy.  The causal kernel 
\begin{center}
$K_{3/2}(k,p)=\frac{\epsilon(k)+\epsilon(p)}{2\sqrt{\epsilon(k)\epsilon(p)}}\,\,       \Big(\frac{2m^2}{m^2+\epsilon(k)\epsilon(p)-kp}\Big)^{3/2}$
\end{center}  
is maximal. Actually  $|K(k,p)| < K_{3/2}(k,p)$  holds for all $k\ne p$ and all causal kernels $K\ne K_{3/2}$ (\ref{IMTKK}).  The relation $|K|\le K_{3/2}\le  \textbf{1}$, where $ \textbf{1}$ is the kernel of the NWL distinguished by its  abundance of  boundedly localized states, suggests that the POL with kernel $K_{3/2}$ is the causal POL with the best localization features. Further studies should work on this distinguished POL, which satisfies several prerequisites. Above all it should be clarified whether there are  point localized sequences of states for this POL. \\
\hspace*{6mm}
Presumable, like  (\ref{GLEG})  in the case of one spatial dimension, not all $G\in\mathcal{G}$ with $G\le G_{3/2}$  (the second factor of $K_{3/2}$) determine a causal kernel. We rather guess that the causal kernels are exactly
\begin{center}
$K_r\,G$\; for\; $G\in\mathcal{G}$, $r\ge 3/2$
\end{center}
where it suffices to consider $r<2$, since $K_r/K_{3/2}\in\mathcal{G}$ for $r\ge 2$ by (\ref{PDGSOH}).\\

Finally here is a brief guide to the article.\\
\hspace*{6mm}
The sections \ref{SFREPG}-\ref{CPOLK} are concerned with the POL and their kernels. Their  content is summarized in (\ref{APOL}), (\ref{POLMB2}), and (\ref{CKPOLMB}).\\
\hspace*{6mm}
Section \ref{POLNWL} deals with  the class of finite POL. The concluding result is (\ref{GKDNWL}).  
In section \ref{POLCTE}   causal time evolution CT is introduced and two  finite POL   with CT are studied.\\
\hspace*{6mm}
Section \ref{CKD} deduces the causality condition CC for POL and introduces  the class of causal kernels (\ref{KKK}). There  in \ref{KKOR} is also an overview of the known results on causal kernels.
\\
\hspace*{6mm}
Section \ref{CKOSD} treats the case of one spatial dimension, which is completely solved by the formula in (\ref{MROD}). 
\\
\hspace*{6mm}
The following sections \ref{DGBKR}-\ref{TMCK} are concerned with a detailed description of the causal kernels. The main results in this regard are  (\ref{CKK}),\,(\ref{IFSFK}),\,(\ref{CIF}),  and (\ref{MTKK}),\,(\ref{IMTKK}).\\
\hspace*{6mm}
Section \ref{LBR} introduces the concept of a point localized sequence of states.  The criterion  (\ref{SSL0}) and its corollary (\ref{PLSS}) regard all POL kernels.\\
\hspace*{6mm}
The discussion in the  final   section \ref{D} incites  to find answers to two outstanding questions of great relevance.\\
\hspace*{6mm}
The appendices  \ref{EQOD} and \ref{AppA} provide space to  some cumbersome technical details, whereas  the appendices \ref{AppB} and \ref{AppC} house all which  regards the CT criterion (\ref{GCCT}) and the result on  CC  (\ref{GCCPOL}), respectively.

\section{Some facts on representations of the Euclidean and Poincar\'e group}\label{SFREPG} 
We recall some facts on the reps\footnote{In the following   rep  means a continuous unitary representation of a topological group.}
 of the universal covering groups  $\tilde{\mathcal{E}}=ISU(2)$  and  $\tilde{\mathcal{P}}=ISL(2,\C)$  
of the Euclidean group and the Poincar\'e group, respectively.  $\tilde{\mathcal{P}}$ acts on $\R^4$ as
\begin{equation}\label{PTUCH} 
g\cdot \mathfrak{x}:=\mathfrak{a}+\Lambda(A) \mathfrak{x}\quad  \text{ for } g=(\mathfrak{a},A)\in\tilde{\mathcal{P}}, \, \mathfrak{x}\in \R^4
\end{equation}
Here $\Lambda:SL(2,\C)\to O(1,3)_0$ is  the universal covering homomorphism onto the proper orthochronous Lorentz group. Identifying $\Lambda(SU(2))\equiv SO(3)$, $SU(2)$ acts on $\R^3$.
Representing Minkowski space by $\R^4$ the Minkowski product is given by 
\begin{equation}
\mathfrak{a}\cdot \mathfrak{a}':=a_0a'_0-a_1a'_1-a_2a'_2-a_3a'_3
\end{equation}

\subsection{Irreps of $\tilde{\mathcal{E}}$}  Up to unitary equivalence the irreducible mutually inequivalent reps of  $\tilde{\mathcal{E}}$ are $U^{0,j}$  on $\C^{2j+1}$ for $j\in \N_0/2$ and $U^{\rho,s}$ on $L^2(S_\rho^2)$ for $\rho>0$, $s\in \mathbb{Z}/2$ with the sphere $S_\rho^2:=\{p\in\R^3:|p|=\rho\}$ endowed with the rotational invariant measure normalized to $1$. Explicitly one has
\begin{itemize}
\item  $U^{0,j}(b,B):=D^{(j)}(B)$ \text{ and }  
\item $(U^{\rho,s}(b,B)f)(p):=\e^{-\i b p} \kappa(p,B)^{2s}f(B^{-1}\cdot p)$
\end{itemize}
where $\operatorname{diag}\big(\kappa(p,B),\overline{\kappa(p,B)} \,\big):=B(p)^{-1} B\,B(B^{-1}\cdot p)$. Here  $B(p)\in SU(2)$ denotes a cross section satisfying  $B(p)\cdot e_3=\frac{p}{|p|}$ for all $p\in\R^3\setminus\{0\}$, whence the Wigner rotation on the right hand side  leaves $e_3$ invariant and hence is diagonal.\footnote{The standard helicity cross section reads
\begin{equation*}\label{CCSEF}
B(p) =\left( \begin{array}{cc} a_+ & -\overline{b}\,a_-\\ b\,a_- & a_+ \end{array}\right), \quad a_\pm:=\sqrt{\frac{|p|\pm p_3}{2|p|}}, \quad b:=\frac{p_1+ip_2}{|p_1+ip_2|} \textrm{ for } (p_1,p_2)\ne 0
\end{equation*}
and $B(\alpha e_3)$  equals $I_2$ if $\alpha\ge 0$ and $-i\sigma_2$ if $\alpha <0$, see \cite[Sec.\,II, Helicity cross section]{CL15}.}

\subsection{Particular reps of $\tilde{\mathcal{E}}$}\label{PRE} We will be concerned with the reps $U^{(s)}$,  $s\in\mathbb{Z}/2$, of $\tilde{\mathcal{E}}$ on $L^2(\R^3)$ given by
\begin{itemize}
\item $(U^{(s)}(b,B)\varphi)(p):=\e^{-\i b p} \kappa(p,B)^{2s}\varphi(B^{-1}\cdot p)$
 \end{itemize}
for $p\ne 0$. Note that $\kappa(0,B)$ is not defined. The obvious unitary equivalence
\begin{equation}\label{RRIRG}
U^{(s)}\simeq  \int_0^\infty U^{\rho,s} 4\pi \rho^2\d \rho
 \end{equation}
yields the decomposition of $U^{(s)}$ into irreps.  It shows in particular that $U^{(s)}$ and $U^{(s')}$ for $s\ne s'$ are disjoint, i.e., that any two subreps of the latter are inequivalent.

\subsection{Induced reps of  $\tilde{\mathcal{E}}$ from $SU(2)$}\label{IRER}
A projection valued measure (PM) on $\R^3$, which is Euclidean covariant, is called a Wightman localization (WL). By Mackey's imprimitivity theorem  every WL is  Hilbert space isomorphic to a rep of $\tilde{\mathcal{E}}$ induced from the subgroup $SU(2)$ together with the related  system of imprimitivity.\\

 On  momentum space $L^2(\R^3,\C^{2j+1})$ 
the  rep $D^{(j)}_{\tilde{\mathcal{E}}}$ of $\tilde{\mathcal{E}}$ induced from the irrep $D^{(j)}$,  $j\in\N_0/2$, of $SU(2)$ and the system of imprimitivity $E$ read 
\begin{itemize}\label{PR}
\item  $(D^{(j)}_{\tilde{\mathcal{E}}}(b,B)\varphi)(p) =\e^{-\i bp} D^{(j)}(B)\,\varphi(B^{-1}\cdot p), \quad E=\mathcal{F}E^{can}\mathcal{F}^{-1}$
 \end{itemize}
The PM operator $E^{can}(\Delta)$ multiplies by the indicator function $1_\Delta$. $\mathcal{F}$ denotes the unitary Fourier integral transformation with kernel $(2\pi)^{-1/2}\e^{-\i xy}$ in $L^2$. The WL $(D^{(j)}_{\tilde{\mathcal{E}}},\mathcal{F}E^{can}\mathcal{F}^{-1})$, $j\in\N_0/2$, are irreducible, mutually inequivalent, and complete up to unitary equivalence.
The decomposition into  subreps is obtained by the unitary  transformation $X^{(j)}$ on  $L^2(\R^3,\C^{2j+1})$
\begin{itemize}
\item $(X^{(j)}\varphi)(p):=D^{(j)}(B(p)^{-1})\varphi(p)$
 \end{itemize}
for $p\ne 0$. Due to $D^{(j)}_{ss'}\big(B(p)^{-1} B\,B(B^{-1}\cdot p)\big) =\kappa(p,B)^{2s} \delta_{ss'}$ one yields  
\begin{equation}\label{DIRE}
X^{(j)}D^{(j)}_{\tilde{\mathcal{E}}}X^{(j)-1}= \oplus_{s}U^{(s)}, \quad s=-j,-j+1,\dots,j-1,j
 \end{equation}
which is completed to a WL  by the PM $X^{(j)}\mathcal{F}E^{can}\mathcal{F}^{-1}X^{(j)-1}$.\\

 One concludes  that every  sum $\oplus_{\iota}U^{(s_\iota)}$ is the subrep of a rep with an Euclidean covariant PM forming an WL. Hence generally this holds true for every rep, which is unitarily equivalent to such a  sum.

\subsection{Irreps of  $\tilde{\mathcal{P}}$ for positive masses}\label{IRPP}

The irreps are labeled by mass $m>0$, spin $j\in\N_0/2$,  sign $\eta=\pm$. The sign $\eta=-$ denotes  the antiparticle case.
Hence
up to unitary equivalence we consider the mutually inequivalent irreps of  $\tilde{\mathcal{P}}$ on 
\textbf{momentum space} $L^2(\R^3,\C^{2j+1})$ 
given by\footnote{Often one uses the antiunitarily equivalent $ \e^{-\i \mathfrak{a}\cdot \,\mathfrak{p}^{\eta}}$.}
\begin{itemize}
\item $\big(W^{m,j,\eta}(\mathfrak{a},A)\varphi\big)(p)=\sqrt{\epsilon(q^\eta)/\epsilon(p)}\, \e^{\i \mathfrak{a}\cdot \,\mathfrak{p}^{\eta}} \,D^{(j)}\big(R(\mathfrak{p}^\eta,A)\big)\,\varphi(q^\eta)$
\end{itemize}

for $p\ne 0$, where $\epsilon(p)=\sqrt{m^2+p^2}$,  $\mathfrak{p}^{\eta}=(\eta\epsilon(p),p)$, $\mathfrak{q}^\eta= (q_0^\eta,q^\eta):=A^{-1}\cdot \mathfrak{p}^\eta$, and $R$ the Wigner rotation with respect to the canonical cross section satisfying  $R( \mathfrak{p}^\eta,B)=B$ for $B\in SU(2)$, see e.g. \cite[sec.\,3.3 (1)]{C17}.

\subsection{Decomposition of  $W^{m,j,\eta}|_{\tilde{\mathcal{E}}}$}\label{DIRP} 
By the foregoing considerations it is  easy to verify
\begin{equation}\label{RERPG}
W^{m,j,\eta}|_{\tilde{\mathcal{E}}}\simeq \oplus_{s}U^{(s)}, \quad s=-j,-j+1,\dots,j-1,j
\end{equation}

\section{How to get  PO-Localizations}\label{HGPOL}

A positive operator valued measure $T$  on the Borel sets $\mathcal{B}(\R^3)$ (POM), which is Euclidean covariant, is called a PO-localization (POL). Recall that a POM  $E$ constituted by orthogonal projections is denoted by PM, and a Euclidean covariant PM is called a Wightman localization (WL).\\
\hspace*{6mm}
 Given a rep $W$ of  $\tilde{\mathcal{P}}$ and a POM $T$ such that $(W|_{\tilde{\mathcal{E}}},T)$ is a POL, then  by   \cite[(9) Theorem]{C17} there is a unique Poincar\'e covariant extension  $(W,T)$. This means that there is just one map on $\mathfrak{S}$, the set of all Lebesgue measurable subsets of spacelike hyperplanes of Minkowski space, still denoted by $T$, 
which satisfies  
\begin{equation}\label{PCEPOL}
W(g)T(\Delta)W(g)^{-1}=T(g\cdot \Delta) \quad \forall\; g\in\tilde{\mathcal{P}}, \;\Delta \in \mathfrak {S}
\end{equation}
with  $g\cdot \Delta=\{g\cdot \mathfrak{x}:\mathfrak{x}\in \Delta\}$, see (\ref{PTUCH}).
 The elements of  $ \mathfrak {S}$ are called regions.\\
\hspace*{6mm}
Henceforth  let  $W$ be a rep of $\tilde{\mathcal{P}}$ with positive mass spectrum and finite spinor dimension. This means equivalently that $W$ is a finite orthogonal sum of reps from 
sec.\,\ref{IRPP}.\\
\hspace*{6mm}
Regarding parts of  (\ref{CPCPOL}) see e.g. \cite[sec. II]{W86}  for the general theory. Let $(U,E)$ be a WL  acting on the Hilbert space $\mathcal{K}$. Let $\mathcal{H}$ be a $U$-invariant subspace of $\mathcal{K}$ and $j:\mathcal{H}\to \mathcal{K}$ the identical injection. Then the POL $j^*(U,E)j$ is called the trace (or compression) of $(U,E)$ on $\mathcal{H}$.

\begin{The}\label{CPCPOL}
Let $W$ be a rep of $\tilde{\mathcal{P}}$ on $\mathcal{H}$. 
\begin{itemize}
\item[\emph{(a)}] There is a Poincar\'e covariant POL $(W,T)$.
\item[\emph{(b)}]For every Poincar\'e covariant POL $(W,T)$,  the POL 
$(W|_{\tilde{\mathcal{E}}},T_{\mathcal{B}(\R^3)})$ is the trace on $\mathcal{H}$ of a WL on a separable Hilbert space.
\item[\emph{(c)}]
 Every  trace on $\mathcal{H}$ of a WL $(U',E)$, where $U'$ extends  $W|_{\tilde{\mathcal{E}}}$, determines uniquely a  Poincar\'e covariant POL $(W,T)$.
\end{itemize}
\end{The}
{\it Proof.} (a) Let $U:=W|_{\tilde{\mathcal{E}}}$. By sec.\,\ref{DIRP},  $U\simeq \oplus\, U^{(s)}$, which is a finite sum considering multiplicities. Therefore by sec.\,\ref{IRER} there is a WL
$(U',E)$ such that $U$ is a subrep of $U'$. So the trace of $(U',E)$ on $\mathcal{H}$ is a POL $(U,T)$, which by \cite[(9) Theorem]{C17} extends uniquely to a Poincar\'e covariant POL $(W,T)$. \\
(b) This is an application of the Principal Theorem on dilations in \cite[Appendix sec.\,6]{RN90} (see e.g.\,\cite{S77}, \cite{CH80}).\\
(c) See again \cite[(9) Theorem]{C17}.\qed

\section{POL for a Massive Scalar Boson}\label{CPOLMSB}

 We construct  all Poincar\'e covariant POL $(W^{m,0,\eta},T)$, $m>0$ following (\ref{CPCPOL}). For the result see (\ref{APOL}).
 \\
\hspace*{6mm} By (\ref{RERPG}), $W^{m,0,\eta}|_{\tilde{\mathcal{E}}}=U^{(0)}$. According to sec.\,\ref{IRER} the extensions $U'$ of $U^{(0)}$ to deal with are  Hilbert space isomorphic with $D_{\tilde{\mathcal{E}}}:=\oplus_j\, \nu_j D^{(j)}_{\tilde{\mathcal{E}}}$, which are countable sums considering  multiplicities $\nu_j\in\N_0\cup\{\infty\}$.\footnote{Let $\nu\in\{0,1,2,\dots\}\cup\{\infty\}$ and let $Z$ be a Hilbert space or a rep. The notation $\nu Z$ indicates the outer orthogonal sum of $\nu$ copies of $Z$. In particular,  $0Z$ is omitted and  $\infty Z$  is the outer orthogonal sum of countably infinite many copies of $Z$.} Due to (\ref{CPCPOL})(c) it is no restriction to assume at once $\nu_j=\infty$ for all $j$. Obviously half-integer spins do not contribute, whence  $j\in \N_0$ suffices. So $D_{\tilde{\mathcal{E}}}$ acts on $L^2(\R^3,\mathcal{S})$, where the outer orthogonal sum
 $$\mathcal{S}:=\oplus_{j\in\N_0}\infty\C^{2j+1}$$ 
 is a countably dimensional  Hilbert space called \textbf{spinor space},  
 by $$(D_{\tilde{\mathcal{E}}}(b,B)\varphi)(p)=\e^{-\i bp}D(B)\varphi(B^{-1}\cdot p) \text{ with } D:=\oplus_j\, \infty D^{(j)}$$ 
 Applying (\ref{DIRE}), $D_{\tilde{\mathcal{E}}}$ is diagonalized by $$X:=\oplus_j\infty X^{(j)}$$ yielding  
$XD_{\tilde{\mathcal{E}}}X^{-1}=\oplus_j\infty \oplus_{s=-j}^{j} U^{(s)}$. Hence by sec.\,\ref{PRE} its  component for $U^{(0)}$ is $\oplus_j\infty U^{(0)}$ with carrier space  $L^2(\R^3,\mathcal{S}_0)$, where $\mathcal{S}_0:= \oplus_j\infty \oplus_{s=0}\C \subset \mathcal{S}$.
The subspaces  of $L^2(\R^3,\mathcal{S})$, which are invariant under $XD_{\tilde{\mathcal{E}}}X^{-1}$ and reduce $XD_{\tilde{\mathcal{E}}}X^{-1}$ to a rep equivalent to $U^{(0)}$, are 
$$V_{\e}:=\{\varphi\in L^2(\R^3,\mathcal{S}): \varphi(p)=\phi(p)\e(|p|),\,\phi\in L^2(\R^3)\}$$
 for every measurable  
 $$\e:]0,\infty[\to \mathcal{S}_0 \text{ with } \norm{\e(\rho)}=1 \text{ for all } \rho$$ 
This is easy to verify because of  sec.\,\ref{PRE}.   Let  the map $\e$ be called  a \textbf{spinor choice}. \\
\hspace*{6mm}
Obviously the invariant subspaces $V_{\e}$ and  $V_{\e'}$ for the spinor choices $\e$ and $\e'$ are equal if and only if $\e$ and $\e'$ are equivalent, i.e.,
 $\e'(\rho)=s(\rho)\e(\rho) \text{ a.e with } |s(\rho)|=1$.\\
\hspace*{6mm}
Now recall $W^{m,0,\eta}|_{\tilde{\mathcal{E}}}=U^{(0)}$. We embed the carrier space of $U^{(0)}$ in $L^2(\R^3,\mathcal{S})$ mapping it onto $V_{\e}$ by the injection 
$$j_{\e}:L^2(\R^3)\to  L^2(\R^3,\mathcal{S}),\quad  j_{\e}\phi:=\phi\e(|\cdot|)$$
Recall  further that   $U^{(0)}$ is a subrep of $U'$. Denote by $P_0$ the projection onto the carrier space of $U^{(0)}$. Then  $U^{(0)}=P_0U'P_0^*$. Now let  $\iota$ be a Hilbert space isomorphism satisfying  $U'=\iota D_{\tilde{\mathcal{E}}}\iota^{-1}$ and put  $E':=\iota \mathcal{F}E^{can}\mathcal{F}^{-1}\iota^{-1}$. Then $(U',E')$ is a WL, and its trace 
$(U^{(0)}, T)$ for  $T:=P_0 E' P_0^*$ is the POL we are looking for.\\
\hspace*{6mm}
 Since $U^{(0)}=P_0U'P_0^*=(P_0\iota X^{-1}) \,(XD_{\tilde{\mathcal{E}}}X^{-1})\,(X\iota^{-1}P_0^*)$ it follows $X\iota^{-1}P_0^*=j_{\e}$ for some spinor choice $\e$. Indeed,  first it follows that $j:=X\iota^{-1}P_0^*$ is an isometry from $L^2(\R^3)$ into  $L^2(\R^3,\mathcal{S})$ with final space $V_{\e'}$ for some spinor choice $\e'$. Hence $j\phi=(S\phi) \e'(|\cdot|)$ for some unitary  $S$  on $L^2(\R^3)$. Therefore $(XD_{\tilde{\mathcal{E}}}(b,B)X^{-1})\,j\phi=U^{(0)}(b,B)(S\phi) \e'(|\cdot|)$ and further $j^*\,(XD_{\tilde{\mathcal{E}}}(b,B)X^{-1})\,j=S^*U^{(0)}(b,B)S$.
 So $S$ commutes with $U^{(0)}$. This implies $(S\phi)(p)=s(|p|)\phi(p)$ for some $s:]0,\infty[\to \C$ measurable, $|s(\rho)|=1$  for all $\rho$. One concludes $j=j_{\e}$ for $\e:=s(|\cdot|)\e'(|\cdot|)$. -- Hence $T$ coincides with $T_{\e}$ in (\ref{POLMB}).  We summarize.
 
  \begin{The}\label{APOL} The POL covariant with respect to $W^{m,0,\eta}|_{\tilde{\mathcal{E}}}$ are 
 \begin{equation}\label{POLMB}
 T_{\e}:=j_{\e}^*X\mathcal{F}E^{can}\mathcal{F}^{-1}X^{-1}j_{\e}
\end{equation}
where $\e$ runs through all spinor choices. 
Extending  $T_{\e}$ to $\mathfrak{S}$ by \emph{\cite[(9) Theorem]{C17}},
 the Poincar\' e covariant POL  are $(W^{m,0,\eta},T_{\e})$.
\end{The}

\begin{Def}\label{FSC}
The finite dimensional spinor spaces, considered as finite dimensional $D$--invariant subspaces of $\mathcal{S}$, are 
 $$\mathcal{S}^{fin}:=\oplus_{j=0}^J\oplus_j \nu_j\C^{2j+1}$$ for $J\in\N_0, \nu_j\in\N_0$ with
 $\dim (\mathcal{S}^{fin}) =\sum_{j=0}^J\nu_j (2j+1)$.\\
 \hspace*{6mm}
 A POL $T$ is called   \textbf{finite} if it is the trace of an WL with finite dimensional  spinor space. The   spinor choice $\e$ is called  \textbf{finite} if the range of $\e$ lies in a finite dimensional spinor space $\mathcal{S}^{fin}$.
\end{Def}\\
Obviously a POL $T$ is finite if and only if $T=T_{\e}$ for some finite spinor choice $\e$. We will be concerned with finite POL in sec.\,\ref{POLNWL} providing  an equivalent characterization of the latter as kinematic deformations of the NWL. This expression is due to Moretti \cite{M23} thus calling a POL proposed by Terno \cite{T14}. We give a rigorous definition of this term (\ref{TKD}). Besides the POL of Terno-Moretti  in  sec.\,\ref{POLCTE} a further  physically relevant finite POL is treated.

\hspace*{6mm}
We conclude this section by some remarks on the assignment $\e\to T_{\e}$.
If $\e$ and $\e'$ are equivalent, then $(W^{m,0,\eta}|_{\tilde{\mathcal{E}}}, T_{\e})$ and $(W^{m,0,\eta}|_{\tilde{\mathcal{E}}}, T_{\e'})$ are unitarily equivalent. Indeed, by assumption $\e'(\rho)=s(\rho)\e(\rho) \text{ a.e. with } |s(\rho)|=1$. Since $s(\rho)=\langle \e(p),\e'(\rho)\rangle$ a.e., $s$ is measurable. Therefore 
$(S\phi)(p):=s(|p|)\phi(p)$ defines a unitary operator on $L^2(\R^3)$, which commutes with $U^{(0)}=W^{m,0,\eta}|_{\tilde{\mathcal{E}}}$. Moreover, $j_{\e'}=j_{\e}S$. Hence  
$S^{-1}(W^{m,0,\eta}|_{\tilde{\mathcal{E}}}, T_{\e})S=(W^{m,0,\eta}|_{\tilde{\mathcal{E}}}, T_{\e'})$.\\
\hspace*{6mm}
Conversely, let $(W^{m,0,\eta}|_{\tilde{\mathcal{E}}}, T_{\e})$ and $(W^{m,0,\eta}|_{\tilde{\mathcal{E}}}, T_{\e'})$ be unitarily equivalent. Let the equivalence  be effected by the unitary operator $S$. Then $S$ and $U^{(0)}=W^{m,0,\eta}|_{\tilde{\mathcal{E}}}$ commute, whence $(S\phi)(p):=s(|p|)\phi(p)$ for some measurable $s$ with $|s(\rho)|=1$ for $\rho>0$. It follows 
$S^{-1}(W^{m,0,\eta}|_{\tilde{\mathcal{E}}}, T_{\e})S=(W^{m,0,\eta}|_{\tilde{\mathcal{E}}}, T_{\tilde{\e}})$, where $\tilde{\e}$ given by  $\tilde{\e}(\rho):=s(\rho)\e(\rho)$ is equivalent with $\e$. However this does not imply $\tilde{\e}=\e'$ a.e. In other words, $\e\mapsto T_{\e}$  is not injective. See the following general example and the concrete example in sec.\,\ref{TVFTCT}. A definite answer to this question is given in (\ref{URK}) and (\ref{CKPOLMB}).\\
\hspace*{6mm}
Let $C$ be a unitary operator on $L^2(\R^3,\mathcal{S})$, $(C\varphi)(p):=C_0\varphi(p)$, where $C_0$ is unitary on $\mathcal{S}$ such that  $C_0=\oplus_jC_{0,j}$ with $C_{0,j}$
acting on the multiplicity space  of $D^{(j)}_{\tilde{\mathcal{E}}}$.
Then $C$ and $X\mathcal{F}E^{can}\mathcal{F}^{-1}X^{-1}$ commute and $Cj_{\e}=j_{\e'}$ for $\e':=C_0\e$. Thus $T_{\e}=T_{\e'}$.\\

\section{WL  and NWL}\label{CWLMSB}  
  NWL is for Newton-Wigner localization. --- Let  $(W^{m,0,\eta},E)$ be a Poincar\'e covariant  WL. Recall  $W^{m,0,\eta}|_{\tilde{\mathcal{E}}}=D^{(0)}_{\tilde{\mathcal{E}}}$. According to the imprimitivity theorem reported in sec.\,\ref{IRER}, $(W^{m,0,\eta}|_{\tilde{\mathcal{E}}},E)=S( D^{(0)}_{\tilde{\mathcal{E}}},\mathcal{F}E^{can}\mathcal{F}^{-1})S^{-1}$ for some unitary $S$. This implies that $S$ and $D^{(0)}_{\tilde{\mathcal{E}}}=U^{(0)}$ commute, whence $(S\phi)(p)=s(|p|)\phi(p)$ with measurable $s:]0,\infty[\to \C$ , $|s(\rho)|=1$. So $E=S\mathcal{F}E^{can}\mathcal{F}^{-1}S^{-1}$. Hence referring to (\ref{POLMB}), $\mathcal{S}_0\equiv \C$ and 
$$E=j_{\e}^*\mathcal{F}E^{can}\mathcal{F}^{-1}j_{\e} \text{ for } \e=s$$
The NWL  $E^{NW}$ is given  by $\e=1$. About the uniqueness of $E^{NW}$ see  \cite[sec.\,2]{C17}.

\section{Kernels of the POL}\label{KPOL} Let $T$ be any $POL$ of a massive scalar boson. Then there is a spinor choice $\e$ with $T=T_{\e}$.
From  (\ref{POLMB}) it follows for $\phi\in L^2(\R^3)\cap L^1(\R^3)$
\begin{equation}\label{POLMB2}
\langle \phi,T(\Delta)\phi\rangle=(2\pi)^{-3}\int _\Delta\int\int \, \textsc{t}(k,p) 
\e^{\i(p-k)x}\overline{\phi(k)}\phi(p)   \,\d^3k\,  \d^3p\, \d^3x
\end{equation}
for every region $\Delta\subset\R^3$, where the kernel $\textsc{t}=\textsc{t}_{\e}$ of $T_{\e}$ reads
\begin{equation}\label{KPOLMB}
 \textsc{t}_{\e}(k,p):=\langle D(B(k))\e(|k|),D(B(p))\e(|p|)\rangle=\langle \e(|k|),D\big(B(k)^{-1}B(p)\big)\e(|p|)\rangle
\end{equation}

Obviously $\textsc{t}_{\e}$ is a measurable positive definite kernel\footnote{Let $X$ be a set. and let  $K:X\times X\to \C$ be a map. $K$ is called a positive definite kernel on $X$ if for any $n\in\N$, $(x_1,\dots,x_n)\in X^n$, and $(c_1,\dots,c_n) \in \C^n$ one has $\sum_{i,j=1}^n \overline{c_i}\,c_jK(x_i,x_j)\ge0$. If $K$ is  positive definite, then $K$ is Hermitian, i.e., $\overline{K(x,x')}=K(x',x)$. If $K$ is real-valued symmetric and if  the above condition holds for all $(c_1,\dots,c_n) \in \R^n$,  then $K$ is positive definite. }
on $\R^3\setminus\{0\}$ with $\textsc{t}_{\e}(p,p)=1$ for all $p$. It is also rotation invariant, i.e.,  $\textsc{t}_{\e}(Rk,Rp)= \textsc{t}_{\e}(k,p)$ for all rotations $R$, as we will see explicitly below. Clearly  $T$ and its kernel $\textsc{t}$ determine each other.\\

For later use we insert (\ref{FK}) and (\ref{FKD}). Recall the definition (\ref{FSC}). 

\begin{Cor}\label{FK} Let $(b^{(l)})_l$ be an orthonormal basis of  $\mathcal{S}$. 
Then 
\begin{equation*}
 \textsc{t}_{\e}(k,p)=\sum_l  \overline{f_l(k)}\,f_{l}(p) \tag{1}
 \end{equation*}
holds for all $k,p\ne 0$ and $f_l(p):=\langle\, b^{(l)},D\big(B(p)\big)\e(|p|)\,\rangle$. 
If $\e$ is a finite spinor choice then \emph{(1)} holds for an orthonormal basis  $(b^{(l)})_l$ of  $\mathcal{S}^{fin}$, and the sum is finite.
\end{Cor}

\begin{Def}\label{FKD} Let $T$ be a POL and $\textsc{t}$ its kernel. $\textsc{t}$ is called finite if
\begin{equation*}
 \textsc{t}(k,p)=\sum_l  \overline{f_l(k)}\,f_{l}(p) 
 \end{equation*}
holds for all $k,p\ne 0$ with finitely many measurable $f_l:\R^3\to \C$.
\end{Def}

For an explicit expression of $\textsc{t}_{\e}$ (\ref{KPOLMB}) choose an orthonormal basis  $(b^{(j,n)})_{j,n}$  of $\mathcal{S}_0$ according to the orthogonal sum  $\mathcal{S}_0= \oplus_j\infty \C$  
so that $\e(\rho)=\sum_{j=0}^\infty\sum_{n=1}^{\infty}e_{j,n}(\rho)\, b^{(j,n)}$ with $e_{j,n}(\rho):=
\langle b^{(j,n)},\e(\rho)\rangle$ holds for every spinor choice $\e$.  If $\e$ is a finite spinor choice then,  adapting the enumeration,  $e_{j,n}=0$ if $j>J$ or $n>\nu_j$ for some $j$.

\begin{Cor}\label{AKPOL}
Every set of measurable functions $0<\rho\mapsto e_{j,n}(\rho)\in\C$ with $\sum_j\sum_n|e_{j,n}(\rho)|^2=1$ determines a kernel of a POL for the massive scalar boson  
represented by  $W^{m,0,\eta}$ and vice versa by
\begin{equation}\label{POLMB3}
\textsc{t}_{\e}(k,p)=\sum_{j=0}^\infty\Big(\sum_{n=1}^{\infty}\overline{e_{j,n}(|k|)}\,e_{j,n}(|p|)\Big)D^{(j)}(B(k)^{-1}B(p))_{00}
\end{equation}
 for $p\ne0,k\ne0$, where
\begin{equation}\label{POLMB4}  
D^{(j)}(B(k)^{-1}B(p))_{00}=2^{-j}\sum_{i=0}^{j} {j\choose i}^2 \Big(1+\frac{kp}{|k||p|}\Big)^{j-i}\Big(-1+\frac{kp}{|k||p|}\Big)^i= P_j\Big(\frac{kp}{|k||p|}\Big)
 \end{equation}
 Here Szeg\"o's notation  for the Legendre polynomials  $P_n=P^{(0,0)}_n$ \emph{\cite[10.10]{E53}} is used.
\end{Cor}\\
 {\it Proof.} Recall the  defining relation $\e(\rho)=\sum_{j=0}^\infty\sum_{n=1}^{\infty}e_{j,n}(\rho)\, b^{(j,n)}$ for the first part.
 It remains to show (\ref{POLMB4}).  Note $D^{(j)}(M)_{00}=\sum_{i=0}^{j} {j\choose i}^2(M_{11}M_{22})^{j-i}(M_{12}M_{21})^i $ for $j \in \N_0$ and $M\in M_2(\C)$, whence $D^{(j)}(M)_{00}=\sum_{i=0}^{j} {j\choose i}^2|M_{11}|^{2(j-i)}\big(|M_{11}|^{2}-1\big)^i$ for $M\in SU(2)$. So using the explicit expression for $B(p)$ 
 (see sec.\,\ref{IRER}), one finds $|M_{11}|^2=\frac{1}{2}\big(1+\frac{kp}{|k||p|}\big)$ for $M:=B(k)^{-1}B(p)$.\qed\\

\begin{Cor}\label{KWL} In case of a WL $E$  (see \emph{sec.\,\ref{CWLMSB}}) one has $\textsc{e}_{\e}(k,p)=\overline{s({|k|})}s(|p|)$ with $|s|=1$, and the kernel of the NWL $E^{NW}$ is constant $1$.
\end{Cor}

Recall $|P_n(x)|\le 1$ for $|x|\le 1$. The  series  $\big(\sqrt{n+\frac{1}{2}}P_n\big)_n$ is  an ONB of $L^2(-1,1)$. The first three Legendre  polynomials read
 
 $$P_0(x)=1,\quad P_1(x)=x,\quad P_2(x)=\frac{3}{2}x^2-\frac{1}{2}$$\\

 For $j\in\N_0$, $\sigma,\rho>0$ put 
 $$k_j(\sigma,\rho):=\sum_{n=1}^{\infty}\overline{e_{j,n}(\sigma)}\,e_{j,n}(\rho)$$
  Clearly $k_j$ is a measurable positive definite kernel on $]0,\infty[$. By Cauchy-Schwarz inequality $|k_j(\sigma,\rho)|\le 
 \sqrt{k_j(\sigma,\sigma)}\sqrt{k_j(\rho,\rho)}$, whence again by Cauchy-Schwarz inequality $\sum_j |k_j(\sigma,\rho)| \le \sum_j k_j(\sigma,\sigma) \sum_j k_j(\rho,\rho)=1$ as 
 $\sum_{j,n}|e_{j,n}(\cdot)|^2=1$. So for all  $\sigma,\rho>0$
 \begin{equation}\label{PDKF}
 \sum_j k_j(\rho,\rho)=1,\quad \sum_j |k_j(\sigma,\rho)|\le 1
 \end{equation}

 In view of the comment on (\ref{APOL}) we note
 
  \begin{Cor}\label{URK} Let $\e$ and $\e'$ be two spinor choices.  Suppose $\textsc{t}_{\e}=\textsc{t}_{\e'}$. Then for every $j\in\N_0$ and  all $(\sigma,\rho)$ 
 $$\sum_{n=1}^{\infty}\overline{e_{j,n}(\sigma)}\,e_{j,n}(\rho)=\sum_{n=1}^{\infty}\overline{e'_{j,n}(\sigma)}\,e'_{j,n}(\rho)$$ 
 \end{Cor}\\
{\it Proof.} For fixed $\sigma>0$, $\rho>0$, consider $f:[-1,1]\to\C$, $f:=\sum_j \lambda_j (j+\frac{1}{2})^{1/2}P_j$ with $\lambda_j:=(j+\frac{1}{2})^{-1/2}k_j(\sigma,\rho)$ and similarly $f'\:=\sum_j \lambda'_j P_j$. Check $f(x)=\textsc{t}_{\e}(k,p)$ and $f'(x)=\textsc{t}_{\e'}(k,p)$ for $k:=\sigma\,(1,0,0)$ and $p:=\rho\,(x,\sqrt{1-x^2},0)$ by (\ref{POLMB3}),\,(\ref{POLMB4}). Hence by assumption $f=f'$. This implies $\lambda_j=\lambda'_j$ being the coefficients of the expansion in orthonormal Legendre polynomials in $L^2(-1,1)$.\qed

\section{Characterization of the POL kernels} \label{CPOLK}

For a  map
\begin{equation}\label{GPOLK}
K:\R^3\setminus\{0\} \times \R^3\setminus\{0\} \to \C
\end{equation} 
consider the properties    (i) to be  measurable,  (ii) to be normalized, i.e.,  $K(p,p)=1$ for all $p$,  (iii)  to be a positive definite kernel, (iv)  to be a positive definite separable kernel, i.e. 
the RKHS associated to $K$ (see e.g.\,\cite{F08}) is separable,
and (v)  to be  rotation invariant, i.e., $K(Rk,R p)=Kk,p)$ for all $k,p$ and rotations $R$.\footnote{The kernel $K(k,p):=\delta_{kp}$ satisfies (i)-(v) except (iv), i.e., it is not separable.} 
Recall that (iii) implies 
\begin{equation}\label{CSK}
|K(k,p)|^2\le K(k,k)K(p,p)
\end{equation}
 whence, if (ii) holds, $|K|\le 1$.

\begin{Pro}\label{TCK} Let $K$ satisfy \emph{(i)-(iv)}. Then
there is a measurable feature map $J$ on $\R^3\setminus\{0\}$ in a  separable  Hilbert space 
$\mathcal{H}_J$ for $K$, i.e. $K(k,p)=\langle J_k,J_p\rangle$, such that 
\begin{equation}\label{AKPOLMB}
T=j^*\mathcal{F}E^{can}\mathcal{F}^{-1}j
\end{equation}
with $j: L^2(\R^3)\to L^2(\R^3,\mathcal{H}_J)$, $(j\phi)(p):=\phi(p)J_p$, is  a POM $T$ on $L^2(\R^3)$ with  kernel  $K$. It is translation covariant with respect to the rep $(U^{(0)}(b)\phi)(p)=\e^{-\i bp}\phi(p)$ of translations $b$, i.e.,  $U^{(0)}(b)T(\Delta)U^{(0)}(b)^{-1}=T(b+\Delta)$.
\end{Pro}

{\it Proof.} Let $\mathcal{H}_J$ denote the  RKHS associated to the kernel $K$  and put $J_p:=K(\cdot, p)$. Since $\norm{J_p}^2=K(p,p)=1$, $j$ is an injection. Note that $j$ is well-defined just because $\mathcal{H}_J$ is separable. Hence $T$ is a POM. Its kernel is $K$, since 
$\langle \phi,T(\Delta)\phi\rangle= \langle j \phi,\mathcal{F}E^{can}(\Delta)\mathcal{F}^{-1} j\phi\rangle=
(2\pi)^{-3}\int_\Delta\int\int \langle J_k,J_p\rangle\,\e^{\i(p-k)x}\overline{\phi(k)}\phi(p)\d^3k\,\d^3p\, \d^3x $ for $\phi \in L^{2}\cap L^1$, $\Delta\subset\R^3$ measurable. Hence translation covariance of $T$ easily follows.\qed\\

 Obviously every POL kernel $\textsc{t}_{\e}$ satisfies (i)-(v). In particular see (\ref{KPOLMB}). The converse holds true, too.

\begin{Cor}\label{EPDKKPOL} Let $K$ satisfy  \emph{(i)-(v)}. Then \emph{(\ref{AKPOLMB})} is a Euclidean covariant POM with respect to $U^{(0)}$. Hence  $(W^{m,0,\eta}|_{\tilde{\mathcal{E}}},T)$ 
 is a POL for the massive scalar boson and $K$ is its kernel.
\end{Cor}

{\it Proof.}  Recall $\big(U^{(0)}(B)\phi\big)(p)=\phi(B^{-1}\cdot p)$. It suffices to check rotation covariance $U^{(0)}(B)T(\Delta)U^{(0)}(B)^{-1}=T(B\cdot \Delta)$. This is easy using the formula on $\langle \phi,T(\Delta)\phi\rangle$ in the proof of (\ref{TCK}).\qed\\

Let $K$ be rotation invariant. Introduce the map
\begin{equation}\label{SRIK}
\textsc{k}:\;]0,\infty[\times  ]0,\infty[ \times [-1,1]\to \C,\quad \textsc{k}(\sigma,\rho,x):=K\big(\sigma(0,0,1),\rho(0,\sqrt{1-x^2},x)\big)
\end{equation}
Due to rotation invariance  one has $K(k,p)=\textsc{k}\big(|k|,|p|,\frac{kp}{|k||p|}\big)$  for all $k,p\ne 0$. 

\begin{The}\label{CKPOLMB}  The kernels of the POL of the massive scalar boson are exactly the measurable normalized rotational invariant positive definite  separable kernels $K$ on $\R^3\setminus\{0\}$. They  are given by 
\begin{equation}\label{PDEC}
K(k,p)=\sum_{j=0}^\infty k_j(|k|,|p|)P_j\left(\frac{kp}{|k||p|}  \right)
\end{equation}

for all $k,p\ne 0$ with $(k_j)_{j=0,1,2,\dots}$ any sequence  of measurable positive definite kernels $k_j$  on $]0,\infty[$  (not excluding $k_j=0$ for some $j$) satisfying $\sum_jk_j(\rho,\rho)=1$ for all $\rho$. 
  $K$ and $(k_j)_j$ determine each other uniquely.
  The sum converges everywhere.  Recall \emph{(\ref{SRIK})}.
 One has $$k_j(\sigma,\rho)=\int_{-1}^1\textsc{k}(\sigma,\rho,x) \,(j+\frac{1}{2})P_j(x)\d x,\; j\in\N_0$$
\end{The} \\
{\it Proof.} In view of (\ref{EPDKKPOL}) and (\ref{AKPOL}) it remains to add the following consideration regarding the POL kernels.  Choose a measurable feature map $\Phi_j$ for $k_j$ (see e.g.\,\cite{F08})  and let $e_{j,n}(\rho)$ be the coefficients of $\Phi_j(\rho)$ with respect to some ONB. Then $k_j(\sigma,\rho)=\langle \Phi_j(\sigma),
 \Phi_j(\rho)\rangle =\sum_{n=1}^{\infty}\overline{e_{j,n}(\sigma)}\, e_{j,n}(\rho)  $ for all $\sigma,\rho, j$. Further by the assumption on $(k_j)$ one has $\sum_{j,n}|e_{j,n}(\rho)|^2=1$  for all $\rho$. Let $\textsc{t}_{\e}$ denote the kernel corresponding to $(e_{j,n})$ according to (\ref{AKPOL}). Note  $\textsc{k}(\sigma,\rho,\cdot)=\sum_jk_j(\sigma,\rho)P_j$ in $L^2(-1,1)$, where  the sum converges everywhere due to (\ref{POLMB3}). So $K$ is the kernel of $T_{\e}$ and hence equals $\textsc{t}_{\e}$.\qed

 \begin{Cor}\label{CRIK} Fix $\sigma,\rho>0$.  Then  $\big(k_j(\sigma,\rho)\big)_j\in \ell^1$,  $\textsc{k}(\sigma,\rho,x)=\sum_jk_j(\sigma,\rho)P_j(x)$ for $|x|\le 1$, and $\textsc{k}(\sigma,\rho,\cdot)$ is continuous on $[-1,1]$. 
 \end{Cor}
 
 {\it Proof.} Recall (\ref{PDKF}) and $|P_j(x)|\le1$. Then continuity of $\textsc{k}(\sigma,\rho,\cdot)$ holds by dominated convergence.\qed
 
Clearly   the sequence of the coefficients of the expansion of $\textsc{k}(\sigma,\rho,\cdot)$ in orthonormalized Legendre polynomials is $\big((j+\frac{1}{2})^{-1/2}k_j(\sigma,\rho)\big)_j$ with $\big(k_j(\sigma,\rho)\big)_j\in \ell^1$.

 \section{Change to shell  rep}\label{CTSR}
  In place of the momentum reps $W^{m,j,\eta}$ of $\tilde{\mathcal{P}}$  from sec.\,\ref{IRPP}  frequently  it is convenient, because of the simpler transformation formulae,  to use the unitarily equivalent shell reps on $L^2(\mathcal{O}^{m,\eta},\C^{2j+1})$ of functions on the mass shell $\mathcal{O}^{m,\eta}:=\{\mathfrak{p}\in\R^4:p_0=\eta\epsilon(p)\}$ equipped with the Lorentz invariant measure 
\begin{itemize}
\item $\big(W^{shell, m,j,\eta}(\mathfrak{a},A)\Phi\big)(\mathfrak{p})=\e^{\i\mathfrak{a}\cdot\mathfrak{p}}D^{(j)}\big(R(\mathfrak{p},A)\big)\,\Phi(A^{-1}\cdot\mathfrak{p})$
\end{itemize}
The Hilbert space isomorphism from $L^2(\mathcal{O}^{m,\eta},\C^{2j+1})$ onto $L^2(\R^3,\C^{2j+1})$
\begin{itemize}
\item  $(X^{m,\eta}\Phi)(p)=\epsilon(p)^{-1/2}\Phi(\mathfrak{p}^\eta)$
\end{itemize}
satisfies $W^{m,j,\eta}=X^{m,\eta}W^{shell, m,j,\eta}(X^{m,\eta})^{-1}$. Note that for this $X^{m,\eta}$ is unique up to a constant factor of modulus $1$. \\
\hspace*{6mm}
Hence in the case of the massive scalar Boson the Poincar\'e covariant POM $(W^{shell, m,0,\eta},T^{shell}_{\e})$ are given by (cf.\,(\ref{POLMB}))
 \begin{equation}\label{PCPOLS}
T^{shell}_{\e}:=(X^{m,\eta})^{-1}T_{\e} X^{m,\eta}=    (X^{m,\eta})^{-1}  j_{\e}^*X\mathcal{F}E^{can}\mathcal{F}^{-1}X^{-1}j_{\e}X^{m,\eta}, \quad \e \text{ spinor choice}
\end{equation}
 Then for  $\Phi, \Phi'$ square integrable and $\sqrt{\epsilon}\, \Phi$,   $\sqrt{\epsilon}\, \Phi'$ integrable 
 \begin{equation}\label{POLSR}
\langle \Phi,T^{shell}_{\e}(\Delta)\,\Phi\rangle=(2\pi)^{-3}\int_\Delta \int\int       \textsc{t}^{shell}_{\e}(\mathfrak{k}^{\,\eta},\mathfrak{p}^\eta)\,       \e^{\i(p-k)x}\overline{\Phi'(\mathfrak{k}^{\,\eta})}\Phi(\mathfrak{p}^\eta) \frac{\d^3k}{\epsilon(k)}\, \frac{\d^3p}{\epsilon(p)}\,   \d^3x 
\end{equation}
 determines the kernel $\textsc{t}^{shell}_{\e}$. Due to $\langle \Phi',T^{shell}_{\e}[g]\,\Phi\rangle=\langle X^{m,\eta}\Phi',T_{\e}X^{m,\eta}\Phi\rangle$ one gets immediately
\begin{equation}\label{KPOLS}
\textsc{t}^{shell}_{\e}(\mathfrak{k},\mathfrak{p})=\sqrt{\epsilon(k)\epsilon(p)}\,\,\textsc{t}_{\e}(k,p)
\end{equation}

In particular  with respect to the shell representation the kernel of the NWL $E^{NW}$ reads $\sqrt{\epsilon(k)\epsilon(p)}$.

\section{How to get a POL from  Newton-Wigner Localization}\label{POLNWL}

Moretti \cite{M23} proves a formula  for the POL  proposed by Terno \cite{T14}  (sec.\,\ref{POLTM}) calling it a kinematic deformation  of the NWL. Generalizing we like to use Moretti's expression   for a certain kind of POL (\ref{TKD}). It turns out that just the finite POL  (\ref{FSC})  are of this kind (\ref{FPOLKD}).\\

For the momentum operator $P=(P_1,P_2,P_3)$  and a  bounded measurable function $f:\R^3\to \C$ let  $f(P)$ be the  related operator defined by the functional calculus. On momentum state space $L^2(\R^3,\C)$,  $f(P)$ is the multiplication operator by $f$ given by $(f(P)\phi)(p):=f(p)\phi(p)$.
Obviously  $f(P)^*E^{NW}f(P)$, i.e. $\Delta\to f(P)^*E^{NW}(\Delta)f(P)$,  is a POM with value $||f||_\infty^2 I$ at $\R^3$.

\begin{Lem} \label{KKDNWL}
The kernel 
 of  $f(P)^*E^{NW}f(P)$
reads 
$(k,p)\mapsto \overline{f(k)}f(p)$.
\end{Lem}\\
{\it Proof.} As $\langle \phi,  f(P)^*E^{NW}(\Delta)f(P)\phi\rangle=\langle f(P)\phi,  E^{NW}(\Delta)f(P)\phi\rangle=\langle f\phi,  E^{NW}(\Delta)(f\phi)\rangle$ the result follows from (\ref{POLMB2}) and (\ref{KWL}).\qed

\begin{Def}\label{TKD} A POL $T$  is called a  kinematic deformation  of the NWL if
\begin{equation}\label{KDNWL}
 T=\sum_l \,f_l(P)^*\,E^{NW}\,f_l(P)
 \end{equation} 
 holds for some finitely many measurable bounded $f_l:\R^3\to\C$.
\end{Def}

\begin{The}\label{FPOLKD} Let $T$ be a POL and $\textsc{t}$ its kernel. Then the statements \emph{(a)}, \emph{(b}), and \emph{(c)} are equivalent.

\emph{(a)} $T$ is finite \emph{(\ref{FSC})}.

\emph{(b)} $\textsc{t}$ is finite  \emph{(\ref{FKD})}.

\emph{(c)}  $T$  is  a  kinematic deformation   of the NWL  \emph{(\ref{TKD})}.
  \end{The}

{\it Proof.} Assume (a).  Then by (\ref{FK}) $\textsc{t}(k,p)=\sum_l \,\overline{f_l(k)}\,f_l(p)$ for all $p,k\ne 0$ with some finitely many measurable bounded $f_l:\R^3\to\C$ thus showing (b). (b) implies (c) since (\ref{KDNWL}) follows by (\ref{KKDNWL}).\\
\hspace*{6mm}
Now assume (c). Then (\ref{KDNWL}) and (\ref{KKDNWL}) imply $\textsc{t}(k,p)=\sum_{l=1}^L \,\overline{f_l(k)}\,f_l(p)$ for all $k,p\ne 0$, where $\textsc{t}$ is the kernel of $T$. ---
We enter into the proof of (\ref{TCK}) for $K=\textsc{t}$. The feature map $J$ reads $J_p=\textsc{t}(\cdot,p)=(f_1(p),\dots,f_L(p))\in\C^L$, whence $\mathcal{H}_J$ is a subspace of $\C^L$.  For $B\in SU(2)$,  due to  the rotational invariance of $\textsc{t}$, $||\sum_i\lambda_iJ_{B\cdot p_i}||^2=
\sum_{i,j}\overline{\lambda_i}\lambda_j\textsc{t}(B\cdot p_i,B\cdot p_j)=\sum_{i,j}\overline{\lambda_i}\lambda_j\textsc{t}( p_i,p_j)=||\sum_i\lambda_iJ_{p_i}||^2$ holds 
for every linear combination of the $J_p$. Therefore $D'(B)J_p:=J_{B\cdot p}$ determines a unitary operator $D'(B)$  and $B\mapsto D'(B) $ a rep of $SU(2)$ on 
$\mathcal{H}_J$. The latter induces the rep  $(U'(b,B)\varphi)(p):=\e^{-\i bp} D'(B)\varphi(B^{-1}\cdot p)$ of $\tilde{\mathcal{E}}$ on $L^2(\R^3, \mathcal{H}_J)$. Hence $T$ is the POL
(\ref{AKPOLMB}) with kernel $\textsc{t}$. Since $\mathcal{H}_J$ is finite dimensional, (a) follows.\qed

In conclusion we specify how to get the kinematical deformations $T$  of the NWL. (\ref{FKDNWL}) is an application of (\ref{FPOLKD}) and (\ref{CKPOLMB}).

\begin{Cor}\label{FKDNWL}
Let $(g_{j,n})_{j,n}$, $j=0,1,2,\dots,J$, $n=1,2,3,\dots,\nu$ be finitely many measurable complex-valued functions on $]0,\infty[$ satisfying $\sum_{j,n}|g_{j,n}(\rho)|^2=1$.
Then
\begin{equation}
\textsc{t}(k,p):=\sum_j\sum_n \overline{g_{j,n}(|k|)}g_{j,n}(|p|)\,P_j\left(\frac{kp}{|k||p|}\right)
\end{equation}
is a finite  POL kernel. Every finite POL kernel is of this kind. Write $P_j\left(\frac{kp}{|k||p|}\right)=\sum_{0\le s_1+s_2+s_3\le j} c^j_{s_1s_2s_3}
\prod_{i=1}^3\big(\frac{k_i}{|k|}\big)^{s_i}\prod_{i=1}^3\big(\frac{p_i}{|p|}\big)^{s_i}$. Then the kinematical deformations of the NWL are
\begin{equation}\label{GKDNWL}
T=\sum_j\sum_n \sum_{0\le s_1+s_2+s_3\le j} c^j_{s_1s_2s_3}\,\,g_{j,n}(|P|)^*\prod_{i=1}^3\Big(\frac{P_i}{|P|}\Big)^{s_i} \,E^{NW}\, \prod_{i=1}^3\Big(\frac{P_i}{|P|}\Big)^{s_i}g_{j,n}(|P|)
\end{equation}
\end{Cor}

The first coefficients $c^j_{rst}$ read
\begin{itemize}
\item $c^0_{000}=1$
\item $c^1_{000}=0$, $c^1_{100}=c^1_{010}=c^1_{001}=1$
\item $c^2_{000}=-\frac{1}{2}$, $c^2_{110}=c^2_{101}=c^2_{011}=3$, $c^2_{200}=c^2_{020}=c^2_{002}=\frac{3}{2}$
\end{itemize}

The following section \ref{POLCTE} is concerned with two important examples of NWL deformations.

\section{POL with causal time evolution}\label{POLCTE}

Let $(U,T)$ be a POL and $(V,U)$ a rep of the little kinematical group $\R\otimes \tilde{\mathcal{E}}$. Then $(V,U,T)$ is said to be a POL with causal time evolution  if 
\begin{equation*}\label{CTEPOL}
V(t)T(\Delta)V(t)^{-1}\le T(\Delta_t)\tag{CT}
\end{equation*}
holds for all regions $\Delta\subset R^3$ and all times $t\in\R$ Here $\Delta_t:=\{y\in\R^3:\,\exists \,x\in\Delta  \text{ with } |y-x|\le |t|\}$ is the region of influence of $\Delta$.  CT simply means that  after time $t$, respectively before time $t$,  the  probability of localization in $\Delta_t$ is not less than originally in $\Delta$.
 In case of a POL for a massive scalar boson  $V(t)$ is given by $W^{m,0,\eta}(t)$.

Probably the first to introduce and to treat  this concept of CT is the Petzold group \cite[sec.\;3]{GGPR68}. There CT follows from the existence of a nonspacelike conserved  current associated to the probability density of localization. This relation is adapted by Moretti \cite{M23}  in an advanced manner thus showing rigorously CT for the POL  introduced  by Terno \cite{T14}. In case of WL, CT is examined in \cite{Sch71} and  thoroughly studied in \cite{C84} and \cite{CL15}, \cite{C17}.

The POL with CT for a massive scalar boson known so far in the literature are   the  (not finite) POL  with conserved covariant  kernels by Petzold et al. \cite{GGP67} and Henning,\,Wolf \cite{HW71}, which we study in the following sections,
and the POL by Terno-Moretti.  We add here the finite POL with trace CT.

Henceforth we will consider the particle case $\eta=+$ only, as the antiparticle case $\eta=-$ is quite analogous.
Let $H$ denote the energy operator, which in momentum rep equals the multiplication operator by $\epsilon(p)=\big(m^2+|p|^2\big)^{1/2}$.

\subsection{POL by Terno-Moretti}\label{POLTM}

In Moretti \cite{M23} it is shown that $T^{TM}$  (\ref{TMSPOL}) is a POL with CT. Below we give a simplified proof of this based on the general criteria (\ref{CKPOLMB}), (\ref{FKDNWL}),
and (\ref{GCCT}).

\begin{Pro}
\begin{equation}\label{KTM}
\textsc{t}^{TM}(k,p):=\frac{1}{2}\left(1+\frac{m^2+kp}{\epsilon(k)\epsilon(p)}\,\right)
\end{equation}
is the kernel of the finite POL  with CT
\begin{equation}\label{TMSPOL}
T^{TM}= \frac{1}{2}E^{NW}+\frac{m}{\sqrt{2}H}E^{NW}\frac{m}{\sqrt{2}H}+\sum_{i=1}^3\frac{P_i}{\sqrt{2}H} E^{NW} \frac{P_i}{\sqrt{2}H}
\end{equation}
\end{Pro}

 {\it Proof.} Obviously $\textsc{t}^{TM}$ is a  finite continuous rotational invariant positive definite normalized  kernel on $\R^3\setminus\{0\}$ and hence a finite POL kernel (\ref{CKPOLMB}). Recall (\ref{FKDNWL}). It remains to show CT. For this we apply (\ref{GCCT}). Indeed, induced by \cite{M23} we show that $\textsc{t}^{TM}$ is conserved  timelike definite 
 due to $\textsc{j}=( \textsc{j}_1, \textsc{j}_2, \textsc{j}_3)$ given by
 \begin{equation}
 \textsc{j}(k,p)):=\frac{\epsilon(p)k+\epsilon(k)p}{2\epsilon(k)\epsilon(p)}
 \end{equation}
 
Show  condition (\ref{DCPOLK})(i): $2\,\epsilon(k)\epsilon(p)\, (k-p)\textsc{j}(k,p))
=(\epsilon(k)-\epsilon(p))(m^2+kp+\epsilon(k)\epsilon(p))$. 

Show  condition (\ref{DCPOLK})(ii): Since $\textsc{j}$ is real symmetric, it suffices to consider real coefficients $c_a$. Put $C:=\sum_ac_a$, $D_i:=\sum_ac_ap_{a.i}/\epsilon(p_a)$,  and $M:=\frac{m^2}{2}\sum_ac_a/\epsilon(p_a)$. Then $ \sum_i\big(\sum_{a,b}c_ac_b\textsc{j}_i(p_a,p_b)\big)^2=C^2\sum_iD_i^2$ and $\big(\sum_{a,b}c_ac_b\textsc{t}(p_a,p_b)\big)^2=\big(\frac{1}{2}C^2+\sum_iD_i^2+M^2\big)^2$, whence the claim.\qed
\subsubsection{Spinor choice for $T^{TM}$}

From (\ref{AKPOL}) one obtains  $T^{TM}=T_{\e}$ for

\begin{equation}\label{ETM}
\e(\rho):=\frac{1}{\sqrt{2}}\,b^{(0,1)}+\frac{m}{\sqrt{2}\epsilon(\rho)}b^{(0,2)}+\frac{\rho}{\sqrt{2}\epsilon(\rho)}b^{(1,1)}
\end{equation}
Hence a dilation of $(W^{m,0,+}|_{\tilde{\mathcal{E}}}, T^{TM})$ corresponding to (\ref{ETM}) is  the WL $(W'|_{\tilde{\mathcal{E}}},E)$ for $$W':=W^{m,0,+}\oplus W^{m,0,+}\oplus W^{m,1,+}
\text{ and \;} E:=\mathcal{F}E^{can}\mathcal{F}^{-1}$$

\subsection{POL with trace CT}\label{TCT}

\begin{Pro}
 \begin{equation}\label{KPOLCT}
\textsc{t}^{tct}(k,p):=\frac{1}{2}\Big(\epsilon(k)(m+\epsilon(k))\epsilon(p)(m+\epsilon(p))\Big)^{-1/2}\Big((m+\epsilon(k))(m+\epsilon(p)) +kp \Big)
\end{equation} 
is the kernel of the finite  POL    with CT
\begin{equation}\label{POLTCT}
T^{tct}= \left(\frac{m+H}{2H}\right)^{1/2}E^{NW}\left(\frac{m+H}{2H}\right)^{1/2}+\sum_{i=1}^3\frac{P_i}{\big(2H(m+H)\big)^{1/2}} E^{NW}\frac{P_i}{\big(2H(m+H)\big)^{1/2}}
\end{equation}
 Actually,  $(W^{m,0,+}|_{\R\otimes \tilde{\mathcal{E}}},T^{tct})$
is the only  one which is the trace of a relativistic quantum system of finite spinor dimension with CT by a WL. 
\end{Pro}

The proof is postponed to app.\,\ref{AppA}.

\subsubsection{Two spinor choices for $T^{tct}$}\label{TVFTCT}

\begin{Pro}
 The spinor choice $\e$  for which $T^{tct}=T_{\e}$  according  to the construction in \emph{sec.\,\ref{CPOLMSB}} is 
 \begin{equation}\label{ETCT}
 \e(\rho):=\big(4\epsilon(\rho)(m+\epsilon(\rho)) \big)^{-1/2}    \Big[\big(m+\epsilon(\rho)\big)b^{(0,1)}+\big(m+\epsilon(\rho)\big)b^{(0,2)}+\rho\, b^{(1,1)}-\rho\, b^{(1,2)}\Big]
\end{equation}
where $\epsilon(\rho)=\sqrt{m^2+\rho^2}$.
\end{Pro}

The proof is postponed to app.\,\ref{AppA}.

On the other hand, following (\ref{AKPOL}) one obtains immediately $T^{tct}=T_{\e'}$ for

\begin{equation}\label{ETCT'}
\e'(\rho):=\Big(\frac{m+\epsilon(\rho)}{2\epsilon(\rho)}\Big)^{1/2}\,b^{(0,1)}+\rho\,\big(2\epsilon(\rho)(m+\epsilon(\rho))\big)^{-1/2}b^{(1,1)}
\end{equation}
Hence a dilation of $(W^{m,0,+}|_{\tilde{\mathcal{E}}}, T^{tct})$ corresponding to (\ref{ETCT'}) is  the WL $(W'|_{\tilde{\mathcal{E}}},E')$ for $$W':=W^{m,0,+}\oplus W^{m,1,+}
\text{ and \;} E':=\mathcal{F}E^{can}\mathcal{F}^{-1}$$

 This is rather interesting, since here $(W',E')$ is the NWL and therefore the time evolution of $(W',E')$ is not causal \cite[(2) Theorem]{C17}.

\section{Causal POL} \label{CKD}

CT for a POL  is a requirement of the fact that there is no propagation faster than light. Clearly limited propagation affects  all kinematical transformations. For instance, according to NWL, if   a massive particle localized in a bounded region  is subjected to a boost of however small rapidity then  there is a non-vanishing probability to observe the particle arbitrarily far away. This  frame-dependence of NWL (already described by Weidlich, Mitra 1964 \cite{WM63}  and still currently discussed, see e.g. \cite[3]{C17})  is an acausal behavior. In fact,  causality  requires from 
 localization that the probability of localization in a region of influence cannot be less than that in the region of actual localization. \\
\hspace*{6mm}
More precisely, let $\Delta$ be a region, i.e., a measurable subset of a spacelike hyperplane of Minkowski space, and let $\sigma$ be any spacelike hyperplane, then 
$$\Delta_\sigma:=\{\mathfrak{x}\in\sigma: \exists \,\mathfrak{z}\in\Delta \text{ with } (\mathfrak{x}-\mathfrak{z}\}\cdot (\mathfrak{x}-\mathfrak{z}\}\ge 0\}$$
is the region of influence of $\Delta$ in $\sigma$.  It is the set of all points  $\mathfrak{x}$ in $\sigma$, which can be reached  from some point $\mathfrak{z}$ in $\Delta$ by a signal not moving faster than light.  Therefore  $(W,T)$ (sec.\,\ref{HGPOL}) is called a \textbf{causal POL} if
\begin{equation*}\label{POLCC}
T(\Delta)\le T(\Delta_\sigma) \tag{CC}
\end{equation*}
holds  for all $\Delta\in\mathfrak{S}$, $\sigma$ spacelike hyperplane. The causality condition CC is thoroughly studied in \cite{C17}. There causal POL for the Dirac electron and the Weyl fermions are revealed.  Apparently the causality condition  has  not  been analyzed  elsewhere.  Recently, along a tentative \textquotedblleft new and different  operational interpretation of the  notion  of spatial position\textquotedblright\,   in \cite{M23} a causality condition corresponding to CC is shown to hold.
\\
\hspace*{6mm}
Let us verify explicitly that CT    is a special case CC. Indeed,  for $\Delta\subset \{0\}\times \R^3$, $t\in\R$, and $\sigma:=\{-t\}\times \R^3$ check $\Delta_\sigma=(-t,0,I_2)\cdot \Delta_t$, whence  CT by covariance of $T$.

\subsection{Causal kernels}\label{KKO}

 As argued above CC, which implies CT, is an  indispensable property for a localization. For this  one is highly  interested in the  kernels of causal POL.  We study  the promising class  $\mathcal{K}$ of  conserved covariant POL kernels (\ref{CKKK})  introduced   by Petzold et al.\,\cite{GGP67}. Indeed,   (\ref{GCCPOL}) shows that a POL with kernel $K\in\mathcal{K} $ is causal. Moreover, most probably $\mathcal{K}$ comprises even all kernels of causal POL. 
 Therefore we refer quite simply to their kernels (\ref{KKK}) as the \textbf{causal kernels}.\\
 \hspace*{6mm} 
  One deals with  the kernels on $\R^3$ of the kind
\begin{equation}\label{KKK}
K(k,p):=\frac{\epsilon(k)+\epsilon(p)}{2\sqrt{\epsilon(k)\epsilon(p)}}\,\,g\big(\epsilon(k)\epsilon(p)-kp\big)
\end{equation}
 with $g: [m^2,\infty[ \to \R$  continuous and normalized by $g(m^2)=1$, which are positive definite. Occasionally $g$ is said to determine a causal kernel.\\
\hspace*{6mm} 
 Recall  $\epsilon(p)=\big(m^2+|p|^2\big)^{1/2}$. Put $\breve{g}(k,p):=g\big(\epsilon(k)\epsilon(p)-kp\big)$ for $k,p\in\R^3$.   Note  $\breve{g}(p,p)=1$ for all $p\in\R^3$.

 \subsection{Known results}\label{KKOR}
 The problem  is  to find the functions $g$ so that $K$ is positive definite.
 Let us report the results, which in the literature are  achieved so far including the personal communications of R.F. Werner with proofs. Put
 \begin{equation}\label{KKKE}
 g_r(t):=(2m^2)^r(m^2+t)^{-r}, \;t\ge m^2, \;r>0
 \end{equation}
 
\begin{itemize}
\item[(a)] 
The known solutions for $K$ in  (\ref{KKK}) are deduced in  \cite{GGP67} and  \cite{HW71}. They read

\begin{equation}\label{KKKR}
K_r=K  \text{ where } g=g_r \,\text{ for } r\ge 3/2 
\end{equation}
Further solutions are the pointwise limits of their convex  combinations as  e.g. $K$ determined by $g(t):=\big(\frac{(1+c)m^2}{cm^2+t}\big)^n$ for $-1<c\le1$, $n=2,3,\dots$. Note  that $\frac{(1+c)m^2}{cm^2+t}\le g_1(t)$.
 \item[(b)] 
As observed by R.F. Werner, obviously
 \begin{equation}\label{GMWRTK}
 K\breve{h}\in\mathcal{K}
 \end{equation} 
 for every   $K\in\mathcal{K}$ and continuous  $h:[m^2,\infty[\to \R$ with $h(m^2)=1$, $\breve{h}$ positive definite.  $\breve{h}$ is called a noise factor.
\item[(c)] 
As pointed out by R.F. Werner, one has the important implication
\begin{equation}\label{KIPDG}
 K\in\mathcal{K} \Rightarrow \;\breve{g} \text{ \,positive definite}
 \end{equation}
Indeed,
$\breve{g}(k,p)
=\frac{2\,\sqrt{\epsilon(k)\epsilon(p)}}{\epsilon(k)+\epsilon(p)}K(k,p)$ and by  the formula by R.F. Werner 
\begin{equation}\label{PDIFF}
\frac{2\sqrt{xy}}{x+y}=\int_0^\infty\big(2\sqrt{x}\e^{-\lambda x}\big)\big(2\sqrt{y}\e^{-\lambda y}\big)\d \lambda
\end{equation}
 the first factor is positive definite. 
 \item[(d)] 
 For $K$ in (\ref{KKK}), one has $|K|\le 1$  if and only if  $|g|\,\le \,g_{1/2}$.  This is  a communication by R.F. Werner and follows from  (\ref{PDDOO})(b). Hence
 \begin{equation}\label{NCKC}
 K\in\mathcal{K} \Rightarrow |g|\,\le \,g_{1/2}
 \end{equation}
\item [(e)]  In \cite{GGPR68} it is argued that a POL with kernel $K\in\mathcal{K} $ obeys CT.
\end{itemize} 

Henceforth let  $m=1$ without restriction.

\section{Causal kernels for one spatial dimension}\label{CKOSD}
As recommended by \cite[footnote to (3.2)]{GGP67} and since there is still some interest in this topic  \cite{KHB18}, we look first at the instructive case of one spatial dimension.
 Here, indeed,   $g_{1/2}$ arises naturally and turns out to furnish  the maximal solution $K^{1}_{1/2}$ (\ref{PDDOO})(c). A complete description of the  causal kernels is given in  (\ref{MROD}). Accordingly,  the product of $K^{1}_{1/2}$ with a normalized positive definite kernel is a causal kernel and every causal kernel is of this kind. --- Call
\begin{equation}\label{KKOSD}
K^{1}:\R\times\R\to \R, \quad K^{1}(k,p)=
\frac{\epsilon(k)+\epsilon(p)}{2\sqrt{\epsilon(k)\epsilon(p)}}\,\,g\big(\epsilon(k)\epsilon(p)-kp\big)
\end{equation}
 with $\epsilon(p):=\sqrt{1+p^2}$ and $g:[1,\infty[\to \R$ measurable normalized by $g(1)=1$,
  a \textbf{causal kernel } if it is positive definite. Put $\tilde{g}(k,p):=g\big(\epsilon(k)\epsilon(p)-kp\big)$.\\
 \hspace*{6mm}
The question is about the functions $g$ of causal kernels on $\R$.

\begin{Lem}\label{AMKK} One has 

\begin{itemize}
\item[\emph{(i)}] $\frac{\epsilon(k)+\epsilon(p)}{2\sqrt{\epsilon(k)\epsilon(p)}}=\left(\frac{1+\epsilon(k)\epsilon(p)+kp}{2\epsilon(k)\epsilon(p)}\right)^{1/2}\frac{1}{\tilde{g}_{1/2}(k,p)}$
\end{itemize}
With respect to the variables  $x:=\sinh^{-1}(k)$, $y:=\sinh^{-1}(p)$  \emph{(i)} reads
\begin{itemize}
\item[\emph{(ii)}] $\frac{\epsilon(\sinh x)+\epsilon(\sinh y)}{2\sqrt{\epsilon(\sinh x)\epsilon(\sinh y)}}=\,h(x,y)\,\frac{1}{g_{1/2}\big(\cosh(x-y)\big)}$
\end{itemize}
where the kernel $h$
\begin{itemize}
\item[\emph{(iii)}] $h(x,y):=\frac{\cosh x/2}{\sqrt{\cosh x}} \frac{\cosh y/2}{\sqrt{\cosh y}} \,\big(1+\tanh x/2\;\tanh y/2\big)$
\end{itemize}
 is positive definite.  
\begin{itemize}
 \item[\emph{(iv)}] $K^1(\sinh x,\sinh y)\;= h(x,y)
 \,\frac{g\big(\cosh(x-y)\big)}{g_{1/2}\big(\cosh(x-y)\big)}$
\end{itemize}
\end{Lem}
{\it Proof.} Verify the claim by elementary computations. Positive definiteness of $h$ is obvious.\qed

 \begin{Lem}\label{PDDOO} Consider $K^{1}$ in \emph{(\ref{KKOSD})}.
\begin{itemize}
\item[\emph{(a)}] $(k,p)\to  \frac{2\sqrt{\epsilon(k)\epsilon(p)}}{\epsilon(k)+\epsilon(p)}$ is a positive definite kernel on $\R$. If $K^{1}$ is positive definite  then so is 
 $\tilde{g}$.
\item[\emph{(b)}] If $|K^{1}|\le 1$, then $|g|\le g_{1/2}$.
\item[\emph{(c)}] $K^{1}_{1/2}:=K^{1}$ for $g=g_{1/2}$ is continuous positive definite. $\tilde{g}_{1/2}$ is positive definite.  Explicitly one has $K^{1}_{1/2}(k,p)=\left(\frac{1+\epsilon(k)\epsilon(p)+kp}{2\epsilon(k)\epsilon(p)}\right)^{1/2}$.
\end{itemize}
\end{Lem} 
{\it Proof.}
 (a) follows by  the formula (\ref{PDIFF}). --- (b) Assume $|K^{1}|\le 1$. By (\ref{AMKK})(iii) the claim is $\sup_{x,y} h(x,y)=1$ under the constraint that $\cosh(x-y)$ is constant. Since $h$ is symmetric it is no restriction assuming $x\ge y$. Now, the constraint means $y=x+c$ for  some constant $c$, and obviously $\lim_{x\to\infty}h(x,x+c)=1$. ---
 (c) Apply  (\ref{AMKK})(iv),(iii), part (a), and  (\ref{AMKK})(i).\qed

\begin{The}\label{MROD} $K^{1}$ in \emph{(\ref{KKOSD})} is causal if and only if
$$K^{1}=K^{1}_{1/2}\; \tilde{g}$$
where $ g:[1,\infty[\to \R$  is measurable with $g(1)=1$ and $\tilde{g}$ positive definite. In particular $K^1\le K^1_{1/2}$ for all causal $K^1$.
\end{The}\\
{\it Proof.} Clearly the condition is sufficient. Now assume that $K^{1}$ is positive definite. Put $k:=\frac{g\,\circ\, \cosh }{g_{1/2\,}\circ\, \cosh}$. 
 By  (\ref{AMKK})(iv),(iii) it suffices to show that $(x,y)\mapsto k(x-y)$ is positive definite on $\R$.\\
 \hspace*{6mm} 
 (a) First we note that a real kernel $G(x,y)$ on $\R$ is positive definite if and only if the kernel $G_1(x,y):= h_1(x)h_1(y)G(x,y)$ for $h_1(x):=\frac{\cosh x/2}{\sqrt{\cosh x}}$  is positive definite. Indeed, as to the less trivial implication let  $G_1$ be positive definite. Let $c_1,\dots, c_n \in\R$ and $x_1,\dots, x_n\in\R$. Put $c_i':=c_i/h_1(x_i)\in\R$,  $i=1,\dots, n$. Then by assumption $0\le \sum_{i,j}c_i'c_j'G_1(x_i,x_j)=\sum_{i,j}c_ic_jG(x_i,x_j)$. Hence $G$ is positive definite.\\
 \hspace*{6mm}
(b)  Recall  (\ref{AMKK})(iii),(iv) and apply the result (a) to $G(x,y):=\big(1+h_2(x)h_2(y)\big)k(x-y)$ for $h_2(x):=\tanh x/2$.  Hence $G$ is positive definite. Now assume that $(x,y)\mapsto k(x-y)$ is not positive definite. Then there are $c_1,\dots, c_n \in\R$ and $x_1,\dots, x_n\in\R$ such that $\sum_{i,j}c_ic_jk(x_i-x_j)=:\gamma<0$. Let $b\in\R$. It follows 
$0\le \sum_{i,j}c_ic_jG(x_i+b,x_j+b)= \sum_{i,j}c_ic_j\big(1+h_2(x_i+b)h_2(x_j+b)\big)k(x_i-x_j)\to \sum_{i,j}c_ic_j(1+1)\,k(x_i-x_j)=2\gamma<0$ for $b\to \infty$, which is a contradiction.\qed\\

 By (\ref{AMKK})(iv) a measurable normalized function $g:[1,\infty[\to \R$, $g(1)=1$ yields in (\ref{KKOSD}) a causal kernel $K^{1}$ if and only if $h\,\circ\, \cosh$ for $h:=g/g_{1/2}$ is normalized of positive type. The functions $g=g_r$, $r\ge 1/2$ (see (\ref{KKKE})) are of this kind as follows from  (\ref{IDGOH}). Other explicit examples can be  obtained applying the  corollary (\ref{PDDOO3})(a),(b)   to Bochner's theorem.

\begin{Lem}\label{IDGOH} The Fourier transform of
 $g_{1/2}\circ \cosh=  [\cosh \frac{1}{2}(\cdot)]^{-1}$ is infinitely divisible. 
$\tilde{g}_r $ is positive definite for   $r>0$.
\end{Lem}

{\it Proof.}   $\ln\, [\cosh \,x/2]^{-1}=-\frac{1}{2}\int_0^x\tanh\frac{t}{2}\d t
=-\int_0^x\int_0^\infty\frac{\sin ty}{\sinh\pi y}\d y\d t$  by \cite[17.33(25)]{GR07}. By Fubini  this becomes  $=-\frac{1}{2}\int_{-\infty}^\infty\int_0^x\frac{\sin ty}{\sinh\pi y}\d t\d y=
\int_{-\infty}^\infty\frac{\cos xy \,-1}{y}\frac{1}{2\sinh \pi y}\d y=\int \frac{\cos xy\, -1}{y^2} (1+y^2) \d \mu(y)$ for the finite measure $\d \mu(y):= \frac{y}{2(1+y^2)\sinh \pi y}\d y$, whence the first part of the claim by the L\'evy-Khinchin formula implying the second part.\qed
 
\begin{Exa}\label{GLEG}  $K^1$ in  (\ref{KKOSD}) need  not be positive definite despite $0<g\le g_{1/2}$ and $\tilde{g}$ positive definite. The example is furnished by the  Gaussian function with standard deviation $0<\varsigma\le 2$. We apply (\ref{PDDOO3}). \\ 
\hspace*{6mm}
Indeed, let $g:=k\circ \cosh^{-1}$, where
  $k(x):=\exp(-\frac{x^2}{2\varsigma^2})$, $\varsigma>0$  is of positive type (since $\hat{k}_\varsigma=\varsigma\, k_{1/\varsigma}>0$) and $k(0)=1$.
\\
\hspace*{6mm}
Put $f(x):=\frac{k(x)}{[\cosh\frac{x}{2}]^{-1}}=\frac{1}{2}\big(\exp(\frac{x}{2}-\frac{x^2}{2\varsigma^2})+\exp(-\frac{x}{2}-\frac{x^2}{2\varsigma^2})\big)$. Then $\hat{f}(y)=\varsigma \e^{\varsigma^2/8}\cos\big(\frac{\varsigma^2}{2}y\big)\exp(-\frac{\varsigma^2}{2}y^2)$. Therefore, by (\ref{PDDOO3})(c), $f$ is not of positive type.\\
\hspace*{6mm}
It remains to check $f\le1$. Note $f(0)=1$.  Verify $f'(x)=0$ $\Leftrightarrow$ $\tanh\frac{x}{2}=\frac{2}{\varsigma^2}x$. This equation has the only solution $x=0$ if $0<\varsigma\le 2$, whence $f(0)=1$ is the maximum. (If $\varsigma>2$, there are exactly three solutions, whence $f(0)=1$ is a local minimum.)
\end{Exa}

There is a consequence  of    (\ref{MROD})   for  $K$ in (\ref{KKK}). Choose any $p_0\in\R^3$ with $|p_0|=1$. One observes  that $K(ap_0,bp_0)$ equals $K^{1}(a,b)$ for $a,b\in \R$. Therefore $K^{1}$ is positive definite if
$K$ is positive definite. By  (\ref{MROD})  this implies the following necessary condition for $K$ being causal.

\begin{Cor}\label{NCCK}
 Let $K$ be  causal. Then  $g=g_{1/2}\,h$,
 where $ h:[1,\infty[\to\R$  is continuous with $h(1)=1$ and $\tilde{h}$ positive definite, i.e., $h\,\circ \,\cosh$  of positive type.
\end{Cor}

 \section{Definiteness of  $\breve{g}_r$ and $K_r$}\label{DGBKR}
 
 We return to causal kernels on $\R^3$.
 Recall $g_r(t)=2^r(1+t)^{-r}$ , $r>0$  and let $K_r$ denote the map $K$ in (\ref{KKK})  for $g=g_r$. Recall that $K_r$ is positive definite for $r\ge 3/2$ (\ref{KKKR}).\\
 \hspace*{6mm}
 The literature  is not clear what about the values $0<r<3/2$. By \cite[Eq.\,(3.10)]{GGP67}  $K_1$ is not positive definite. 
   In (\ref{NPDKR}) we show that $K_r$  is not positive definite for $0<r<3/2$.
 \\
\hspace*{6mm} 
  Similarly there is the question about the definiteness of  $\breve{g}_r$. Clearly it is positive definite for $r\ge 3/2$. Actually it holds true  exactly for $r\ge 1/2$ as we are going to show simply adapting the proof of \cite[Theorem]{HW71}.\\
\hspace*{6mm} 
 Put $$h:=\frac{\sigma\rho}{(1+\epsilon(\sigma))(1+\epsilon(\rho))}\text{  for } \sigma,\rho> 0$$ with $\epsilon(\rho)=\sqrt{1+\rho^2}$.
Verify $1+\epsilon(\sigma)\epsilon(\rho) -\sigma\rho \,x=\frac{\sigma\rho}{2h}+\frac{h\sigma\rho}{2} -\sigma\rho \,x$,  whence for $r\in\R$, $-1\le x\le 1$ \cite[(7),\,(10)]{HW71}
\begin{equation}\label{EF}
2^r\big(1+\epsilon(\sigma)\epsilon(\rho) -\sigma\rho \,x\big)^{-r}=\big(4h/\sigma\rho\big)^r\big(1-2hx+h^2\big)^{-r}=\big(4h/\sigma\rho\big)^r \,\sum_{n=0}^\infty C_n^r(x)h^n
\end{equation}
by the definition of the Gegenbauer polynomials $C_n^r$ \cite[3.15.1]{E53}.

\begin{Lem}\label{PDGSOH} $\breve{g}_r$ is positive definite for $r\ge 1/2$, it is not positive definite for $0<r<1/2$.   The expansion \emph{(\ref{PDEC})} of $\breve{g}_{1/2}$ is

$$\breve{g}_{1/2}(k,p)=\sum_{j=0}^\infty      \frac{\sqrt{2}\,|k|^j}{   \big(1+\epsilon(k)\big)^{j+1/2}  }    \frac{\sqrt{2}\,|p|^j}{   \big(1+\epsilon(p)\big)^{j+1/2}  }   P_j\left(\frac{kp}{|k||p|}  \right)$$
\end{Lem}

{\it Proof.} (a) Let $r=1/2$. Note $C_j^{1/2}=P_j$ \cite[10.10 (3)]{E53}. Hence (\ref{EF}) yields the claimed expansion  (\ref{PDEC}) of $\breve{g}_{1/2}$ by (\ref{CKPOLMB}), whence $\breve{g}_{1/2}$ is positive definite. This implies that $\breve{g}_1$ is positive definite, too. \\
\hspace*{6mm} 
 (b) So let $r>1/2$ and $r \ne 1$. Polar coordinates $k=\sigma(\sin \theta\cos \phi,\sin \theta \sin \phi, \cos \theta)$, $p=\rho(\sin \theta'\cos \phi',  \sin \theta' \sin \phi', \cos \theta')$ yield 
 $x=\frac{kp}{|k||p|}=\cos \theta \cos \theta'+\sin\theta\sin\theta'\cos (\phi-\phi')$. This expression substitutes $x$ in (\ref{EF}). One obtains the separation of the primed and unprimed variables applying the addition theorem   \cite[10.9(34)]{E53} twice: 
 \begin{itemize}
 \item $C_n^r(\cos \theta \cos \theta'+\sin\theta\sin\theta'\cos \varphi)=\\ \sum_{m=0}^n a(r;n,m) (\sin \theta)^mC_{n-m}^{r+m}(\theta)(\sin \theta')^mC_{n-m}^{r+m}(\theta')C_m^{r-1/2}(\cos \varphi)$
 \item  $C_m^{r-1/2}(\cos \varphi)=\\ \sum_{l=0}^m a(r-1/2;m,l) (\sin \phi)^l C_{m-l}^{r-1/2 +l}(\phi) (\sin \phi')^l C_{m-l}^{r-1/2 +l}(\phi') C_l^{r-1}(1)$  for
 
 $\cos\varphi=\cos(\phi-\phi')=\cos \phi \cos\phi'+\sin\phi\sin\phi'$. 
\end{itemize}  
Finish the proof checking that the coefficients $a(r;n,m) a(r-1/2;m,l) C_l^{r-1}(1)$ are positive: $a(r;n,m)=2^m(n-m)!((r)_m)^2\frac{2r+2m-1}{(2r-1)_{n+m+1}} >0$, and $a(r-1/2;m,l) C_l^{r-1}=2^l(m-l)!((r)_l)^2\frac{1}{l!}\,\frac{(2r+2l-2)(2r-2)_l}{(2r-2)_{m+l+1}}$ is positive, too, since for $l=0$ the fraction becomes $\frac{(2r-2)\cdot 1}{(2r-2)\dots (2r-2+m)}$.\\
\hspace*{6mm}
(c) Let $0<r<1/2$. Put $a:=1+\epsilon(\sigma)\epsilon(\rho)$, $b:=\sigma\rho$. Referring to the expansion  (\ref{PDEC}) of $\breve{g}_r$ it suffices to show the non-positive definiteness of the $0$th coefficient $\gamma^{(r)}_0(\sigma,\rho):=\frac{1}{2}\int_{-1}^12^r(a-bx)^{-r}\d x=\frac{2^{r-1}}{1-r}b^{-1}\big((a+b)^{1-r}-(a-b)^{1-r}\big)$. The expansion (\ref{EF}) in Gegenbauer polynomials for $1-r$ in place of $r$ and for $x=1,-1$ yields 
\begin{equation}\label{OCEGR}
\gamma^{(r)}_0(\sigma,\rho)=\left(\frac{4h}{\sigma\rho}\right)^r\Big(1-\sum_{l=1}^\infty a(r;l)h^{2l}   \Big), \quad a(r;l):=(1-2r)\frac{(2r)_{2l-1}}{(2l+1)!}
\end{equation}
Note $a(r;l)>0$ and $h=h_1(\sigma)h_1(\rho)$ for $h_1(\rho):=\rho/(1+\epsilon(\rho))$, which is injective. Hence for $\rho_1\ne\rho_2$  and  $(c_1,c_2):=(1,-1)$ one easily checks 
 $\sum_{i,j=1}^2c_ic_j\gamma^{(r)}_0(\rho_i,\rho_j)<0$,  thus ending the proof.\qed

 Exploiting a further formula  \cite[(8)]{HW71}
 \begin{equation}\label{FE2}
 \epsilon(\sigma)+\epsilon(\rho)=\frac{\sigma\rho}{2h}\,\big(1-h^2\big)
  \end{equation}
 (one verifies it using $\big(\epsilon(\rho)+1\big)^{-1}=\rho^{-2}(\epsilon(\rho)-1)$\,) we complete the result by

 \begin{Lem}\label{NPDKR}
 $K_r$ is not positive definite for  $0<r<3/2$.
 \end{Lem}\\
{\it Proof.} Since $\breve{g}_r$ is not positive definite for $0<r<1/2$, so is $K_r$. Hence we may assume $1/2\le r<3/2$.\\
\hspace*{6mm}
We use again  the expansion  (\ref{PDEC}) and show that the $0$th coefficient of $K_r$ is not positive definite. By  (\ref{FE2}) the latter equals 
$k_0^{(r)}(\sigma,\rho)=\frac{1}{2}\big(\epsilon(\sigma)\epsilon(\rho)\big)^{-1/2}\frac{\sigma\rho}{2h}(1-h^2)\gamma^{(r)}_0(\sigma,\rho)$. Then (\ref{OCEGR}) yields 
\begin{equation}
k_0^{(r)}(\sigma,\rho)=\big(\epsilon(\sigma)\epsilon(\rho)\big)^{-1/2}\left(\frac{4h}{\sigma\rho}\right)^{r-1}\Big(1-\sum_{l=1}^\infty b(r;l)h^{2l}\Big),
\end{equation}
where $b(r;1):=1-\frac{(2r-1)r}{3}, \,b(r;l)=(2r-1)\Big(\frac{(2r)_{2l-3}}{(2l-1)!}-\frac{(2r)_{2l-1}}{(2l+1)!}\Big)$ for $l\ge 2$. Now check $b(1/2;1)=1$,  $b(1/2;l)=0$ for $l\ge 2$, and $b(r;l)>0$ for $1/2<r<3/2$ and $l\ge 1$. (Moreover, $b(3/2;l)=0$ for $\l\ge 1$, and $b(r;l)<0$ for $r>3/2$ and $\l\ge 1$.) Recall $h=h_1(\sigma)h_1(\rho)$ for $h_1(\rho)=\rho/(1+\epsilon(\rho))$, which is injective.  Hence, again,  for $\rho_1\ne\rho_2$  and  $(c_1,c_2)=(1,-1)$ one easily checks 
 $\sum_{i,j=1}^2c_ic_jk^{(r)}_0(\rho_i,\rho_j)<0$,  thus ending the proof.\qed\\
 
We like to point out that an easy proof of the fact that $\breve{g}_r$, $0<r<\frac{1}{2}$ and $K_r$, $0<r<\frac{3}{2}$ are not positive definite is given in (\ref{ANC}).

\section{Lorentz invariant kernels}\label{GTC}

The kernels $\breve{g}$ on $\R^3$ given by  $\breve{g}(k,p):=g\big(\epsilon(k)\epsilon(p)-kp\big)$  for some $g:[1,\infty[\to \C$ are characterized as the functions $G:\R^3\times\R^3\to \C$, which are Lorentz invariant, i.e., which satisfy $G(L\cdot k,L\cdot p)=G(k,p)$ for $L\in O(1,3)_0$, the set of proper orthochronous Lorentz transformations. Here $L\cdot p$ denotes the spatial vector part of $L\mathfrak{p}$ for $\mathfrak{p}:=\big(\epsilon(p),p\big)$.\\
\hspace*{6mm}
 Indeed, for the less trivial implication assume  $G$ to  be Lorentz invariant. For $k,p\in \R^3$ there is $L\in O(1,3)$ such that $L\mathfrak{k}=(1,0,0,0)$ and $L\mathfrak{p}=(a,0,0,b)$, $a\ge 0$. By the invariance of the Minkowski one finds $\mathfrak{k}\cdot\mathfrak{p}=L\mathfrak{k}\cdot L\mathfrak{p}=a$ and $1=\mathfrak{p}\cdot\mathfrak{p}=L\mathfrak{p}\cdot L\mathfrak{p}=a^2-b^2$: This shows $G(k,p)=G\big(0, (0,0,\sqrt{ (\mathfrak{k}\cdot\mathfrak{p})^2-1})\big)$ for all $k,p$, whence $G$ is symmetric and  the claim holds by 
 \begin{equation}\label{MEG}
 \breve{g}=G \text{ \;for  } g(t):=G\big(0, (0,0,\sqrt{ t^2-1})\big)
 \end{equation} 
 
 \begin{Def}\label{LIPGK} Let $\mathcal{G}$ denote the set of measurable Lorentz invariant positive definite kernels $G$ on $\R^3$ with $G(0,0)=1$.
 \end{Def}\\
 \hspace*{6mm}  
(a)   $G\in\mathcal{G}$ is automatically continuous. Indeed,   continuity of $G$  follows from (\ref{MEG}) and (\ref{CRIK}).\\
\hspace*{6mm}  
(b) $G\in\mathcal{G}$ is automatically real-valued since it is symmetric and positive definite.\\
\hspace*{6mm}
(c) $G\in\mathcal{G}$ is normalized, i.e., $G(p,p)=G(0,0)=1$ for all $p\in \R^3$, whence $G$ is bounded by $1$ (\ref{CSK}). \\
 \hspace*{6mm}  
(d)    If $G$, $G' \in\mathcal{G}$, then also $GG'\in\mathcal{G}$. Moreover, if  $G$ is the pointwise limit of  a sequence  in $\mathcal{G}$, then      $G\in\mathcal{G}$.\\
 \hspace*{6mm}
 (e) If $g$ determines $K\in\mathcal{K}$, then $\breve{g}\in\mathcal{G}$. Indeed, recall (\ref{KIPDG}).\\
 \hspace*{6mm}
(f) If   $K\in\mathcal{K}$ and $G\in\mathcal{G}$, then  $KG\in\mathcal{K}$. Indeed, recall (\ref{GMWRTK}).

  The crucial facts, why we are interested  in $\mathcal{G}$, are (e) and (f). Up to now the only elements of  $\mathcal{G}$ we know are the maximal element $\textbf{1}$ and  $\breve{g}_r$, $r\ge1/2$ and pointwise limits of convex combinations of these.

 \subsection{Positive definite Lorentz invariant kernels, reps of $SL(2,\C)$ with $SU(2)$-invariant vectors} \label{PLKIV}
 Let $G$ be a positive definite kernel on $\R^3$.  We consider  the RKHS $\mathcal{H}_G$ associated to the kernel $G$ (see e.g.\,\cite{F08}).  $\mathcal{H}_G$  consists of  functions  $f:\R^3\to \C$. Put $G_p:=G(\cdot, p)$. $\mathcal{H}_G$ is the completion of the span of $\{G_p:p\in\R^3\}$, on which  the inner product is determined by 
 $\langle G_k,G_p\rangle=G(k,p)$. For $f\in \mathcal{H}_G$
and $p\in\R^3$ one has  $\langle G_p,f\rangle=f(p)$.\\
\hspace*{6mm}
Lorentz invariance of $G$ gives rise to a  rep of $SL(2,\C)$. For $A\in SL(2,\C)$ and $p\in\R^3$ let  $A\cdot p$ denote the spatial vector part of $A\cdot \mathfrak{p}$ for $\mathfrak{p}:=\big(\epsilon(p),p\big)$. Moreover, let  $A_\kappa$ denote the boost  $\e^{\kappa \sigma_3/2}$ in direction $(0,0,1)$ with rapidity $\kappa\in\R$. One has $A_\kappa \cdot p=(p_1,p_2,\epsilon(p)\sinh \kappa+p_3\cosh\kappa)$. Put $\cosh^{-1}(t) =\ln\big(t+\sqrt{t^2-1}\,\big)), \,t\ge 1$.

\begin{Lem} Let $G\in\mathcal{G}$.
 Then $\big(U_G(A)f\big)(k):=f(A^{-1}\cdot k)$ for $A\in SL(2,\C)$, $f\in \mathcal{H}_G$ defines a rep $U_G$ of  $SL(2,\C)$ on $\mathcal{H}_G$, and $\mathcal{H}_G$ is separable.  Moreover, $G=\breve{g}$ for
   \end{Lem} 
\begin{equation}\label{RMEG}
 g(t):=\langle G_0, U_G(A_\kappa)G_0\rangle \text{\; for } \kappa =\cosh^{-1}t
\end{equation}

  {\it Proof.} Define $U(A)G_p:=G_{A\cdot p}$ for all $p$. Since  $G_{A\cdot p}(k) =G(k,A\cdot p)=G(A^{-1}\cdot k,p)=G_p(A^{-1}\cdot k)$, one has $\big(U(A)G_p\big)(k)=G_p(A^{-1}\cdot k) $. $U(A)$ is norm preserving since  $\langle U(A)G_k,U(A)G_p\rangle=\langle G_{A\cdot k}, G_{A\cdot p}\rangle=G( A\cdot k, A\cdot p)=G(k,p)=\langle G_k,G_p\rangle$. One easily checks $U(A)U(A')=U(AA')$.  By linear  extension $U$ becomes a unitary homomorphism on  $V:=\operatorname{span}\{G_p:p\in\R^3\}$. Since the latter is dense in $\mathcal{H}_G$, $U$ extends further to 
 a unitary homomorphism on  $\mathcal{H}_G$. It follows $U=U_G$. Indeed, let $f_n$ in $V$ converge to $f\in\mathcal{H}_G$. Then $U(A)f_n\to U(A)f$ and also $\big(U(A)f_n\big)(p)=f_n(A^{-1}\cdot p)=
 \langle G(A^{-1}\cdot p),f_n\rangle\to  \langle G(A^{-1}\cdot p),f\rangle=f(A^{-1}\cdot p)$.\\
\hspace*{6mm} 
 Now, the matrix elements $(k,p,A)\mapsto \langle G_k,U(A)G_p\rangle=G(k,A\cdot p)$ are continuous, whence $\mathcal{H}_G$ is separable and $U$ is continuous. \\
\hspace*{6mm}
 This is quite obvious. Indeed, recall that $V$ is dense in $\mathcal{H}_G$. Then even $\operatorname{span}\{G_p:p\in\mathbb{Q}^3\}$ is dense in $\mathcal{H}_G$, since $||G_{p_n}-G_p||^2=G(p_n,p_n)-2G(p_n,p)+G(p,p)\to 0$ if $p_n\to p$. Hence  $\mathcal{H}_G$ is separable. --- As known (see \cite{K15} for more information) it suffices to show the weak continuity of $U$. First note that obviously $A\mapsto \langle f,U(A)g\rangle$ is continuous for $f,g$ in $V$. Let $f_0\in \mathcal{H}_G$, There is $f_n\in V$, $f_n\to f_0$. Let $A_k\to A_0$ in $SL(2,\C)$. Then $|\langle f_0, U(A_k)g\rangle -\langle f_0, U(A)g\rangle|\le |\langle f_0-f_n, U(A_k)g-U(A_0)g\rangle|+|\langle f_n,U(A_k)g-U(A_0)g\rangle|\le 2||g||\,||f_0-f_n||+|\langle f_n,U(A_k)g\rangle-\langle f_n,U(A_0)g\rangle|$. Now choose $n_0$ such that $2||g||\,||f_0-f_{n_0}||\le \epsilon/2$. Note $|\langle f_{n_0},U(A_k)g\rangle-\langle f_{n_0},U(A_0)g\rangle|\to 0$ for $k\to\infty$. Thus the continuity of $A\mapsto \langle  f_0,U(A)g\rangle$ for all $f_0\in\mathcal{H}_G$ and $g\in V$ follows. Similarly one replaces $g$ by $g_0\in\mathcal{H}_G$ showing the assertion.\\
\hspace*{6mm}
In view of (\ref{MEG}) note that $A_\kappa\cdot 0=(0,0,\sinh \kappa)$, 
 whence (\ref{RMEG}).
\qed\\

One reminds that the matrix element in (\ref{RMEG}) is invariant  under  unitary equivalence transformations and that $U_G(B)G_0=G_0$ holds for all $B\in SU(2)$, i.e., $G_0$ is $SU(2)$-invariant.  

\begin{Lem}\label{IKFR}
Suppose that $U$ is a rep of $SL(2,\C)$ with an $SU(2)$-invariant normalized vector $\Gamma_0$. Put 
\begin{equation}\label{ARMEG}
g(t):=\langle \Gamma_0, U(A_\kappa)\Gamma_0\rangle \text{\; for } \kappa =\cosh^{-1}t
\end{equation}
Then $\breve{g}(0,0)=1$ and  $\breve{g}$ is real continuous positive definite.
\end{Lem}

 {\it Proof.}  Note $g(1)=1$ and that $g$ is continuous. We show at once that $g$ is real. The matrix $\tau:=\left( \begin{array}{cc} 0 & 1\\ -1 & 0 \end{array}\right)\in SU(2)$ satisfies $U(\tau)\Gamma_0=\Gamma_0$ and $\tau A_\kappa \tau^{-1}=A_{-\kappa}$. 
 So $\overline{\langle \Gamma_0, U(A_\kappa)\Gamma_0\rangle}=\langle U(A_\kappa)\Gamma_0, \Gamma_0\rangle=\langle \Gamma_0, U(A_{-\kappa})\Gamma_0\rangle=\langle \Gamma_0, U(\tau^{-1})U(A_{-\kappa})U(\tau)\Gamma_0\rangle=\langle \Gamma_0, U(A_\kappa)\Gamma_0\rangle$.
  --- It remains to show the positive definiteness of $\breve{g}$.  Let $k\in\R^3$. Then there is $C_k\in SL(2,\C)$ with $C_k\cdot 0=k$. Set $\Gamma_k:=U(C_k)\Gamma_0$. Here $\Gamma_k$ is well-defined, since $A\cdot 0=k$ implies $A^{-1}C_k\cdot 0=0$, whence $A^{-1}C_k\in SU(2)$ and hence $U(A)\Gamma_0=U(A)U(A^{-1}C_k)\Gamma_0=U(C_k)\Gamma_0$. Put $G(k,p):=\langle \Gamma_k,\Gamma_p\rangle$. Clearly $G$ is positive definite. $G$ is also Lorentz invariant, since $G(A\cdot k,A\cdot p)=G(AC_k\cdot 0,AC_p\cdot 0)=
 \langle U(AC_k)\Gamma_0,U(AC_p)\Gamma_0\rangle= \langle U(C_k)\Gamma_0,U(C_p)\Gamma_0\rangle=G(k,p)$. Now note $G\big(0,(0,0,\sqrt{t^2-1})\big)= G(0,A_\kappa\cdot 0)=\langle \Gamma_0, \Gamma_{A_\kappa\cdot 0}\rangle=\langle \Gamma_0,U(A_\kappa)\Gamma_0\rangle=g(t)$ for $ \kappa =\cosh^{-1}t$, whence the claim by (\ref{MEG}).\qed\\
 
In (\ref{IKFR})  the carrier space $\mathcal{H}_U$ of $U$ is not assumed to be separable.  One concludes

\begin{Cor}\label{CERPDK} $\mathcal{G}=\{\breve{g}: \,g \text{ in } \emph{(\ref{ARMEG})}, \,\mathcal{H}_U \text{ separable}\}$
\end{Cor}

The matrix elements $g$ in (\ref{ARMEG}) are closely related to the zonal (or elementary) spherical functions on $SL(2,\C)$, i.e., the $SU(2)$-bi-invariant functions
$\phi$ on $SL(2,\C)$ associated to  irreps $U$ on $SL(2,\C)$ given by $\phi(A):=\langle \Gamma_0,U(A)\Gamma_0\rangle$, where up to normalization $\Gamma_0$ is the unique $SU(2)$-invariant vector (see below). Recall that for every $A\in SL(2,\C)$ there are $B,B'\in SU(2)$ with $A=B'A_\kappa B$. The rapidity $\kappa\ge 0$ is uniquely determined by $A$, since $A^*A=B^*A^2_\kappa B$, whence $\e^\kappa$, $\e^{-\kappa}$ are the eigenvalues of $A^*A$, and since   $\tau A_\kappa \tau^{-1}=A_{-\kappa}$ (see the proof of (\ref{IKFR})).

\subsection{$\mathcal{G}$ is the set of   convex combinations of its irreducible elements} The theory of reps of $SL(2,\C)$ is treated exhaustively  in \cite{N63}, to which we will refer in the sequel. The irreps of  $SL(2,\C)$ of the principal series are characterized by a pair $(m,\lambda)\in \mathbb{Z}\times\R$ of invariant numbers. For $m\ne 0$ they do not contain $SU(2)$-invariant vectors $\ne 0$, and for $(0,\lambda)$ there is up to a phase just one $SU(2)$-invariant normalized vector. $(0,\lambda)$ and  $(0,-\lambda)$ determine equivalent reps. The matrix element (\ref{ARMEG}) reads 
\begin{equation}\label{LIKIR}
g^{\textsc{p}}_{\lambda}(t):=\frac{\sin \lambda\kappa}{\lambda \sinh \kappa}  \text{\; for } \kappa =\cosh^{-1}t \text{\; and } \lambda\in [0,\infty[
\end{equation}
 See \cite[III\ \S 11.5\,  formula after Eq.\,$(10)$]{N63}  for   $\langle \Gamma_0, U(A_{2\kappa})\Gamma_0\rangle$ with $\rho=2\lambda$. For $\lambda\ne 0$ it is  not positive and hence $\breve{g}_\lambda^{\textsc{p}}$ is not positive.\\
\hspace*{6mm}
The irreps  of $SL(2,\C)$ of the supplementary series are characterized by one invariant number $\lambda\in]0,1[$. They all contain just one one-dimensional  $SU(2)$-invariant subspace.
The matrix element (\ref{ARMEG}) reads
\begin{equation}\label{LIKIRP}
g^\textsc{s}_{\lambda}(t):=\frac{\sinh \lambda\kappa}{\lambda \sinh \kappa} \text{\; for } \kappa =\cosh^{-1}t  \text{\; and } \lambda\in ]0,1[
\end{equation}
See \cite[III\ \S 12,7 (3)]{N63} for $\rho=2\lambda$. It is positive.  Hence, by (\ref{CERPDK}), $\breve{g}_\lambda^{\textsc{s}}$ is positive for $0\le \lambda\le 1$. The cases $\lambda= 0,1$ hold by continuity. Actually 
$$g^\textsc{s}_0(t)=g^\textsc{p}_0(t)=\frac{\kappa}{\sinh \kappa}, \quad \kappa =\cosh^{-1}t$$
 for the case $\lambda=0$, and  the trivial rep yields the case $\lambda= 1$. Note the ordering
 \begin{equation}\label{OK}
|g^\textsc{p}_\lambda|\le  g^\textsc{p}_0= g^\textsc{s}_0\le g^\textsc{s}_{\lambda_1}\le g^\textsc{s}_{\lambda_2}\quad \forall\;\lambda\ge 0, \;\;0\le\lambda_1\le\lambda_2\le 1
\end{equation}

We call $g^\textsc{p}_\lambda$ and $g^\textsc{s}_\lambda$ irreducible and 
$\breve{g}^{{\textsc{p}}}_{\lambda}$, $\breve{g}^{{\textsc{s}}}_{\lambda}$ the \textbf{irreducible elements} of $\mathcal{G}$.

\begin{The}\label{MTGP} Every $G\in\mathcal{G}$ is given by $G=\breve{g}$ for
\begin{equation}\label{GLIPDK}
g(t)=\int_{[0,\infty[}\frac{\sin \lambda\kappa}{\lambda \sinh \kappa}\d \mu^\textsc{p}(\lambda)+\int_{]0,1]} \frac{\sinh \lambda\kappa}{\lambda \sinh \kappa}\d \mu^\textsc{s}(\lambda) 
\end{equation}
with $\kappa =\cosh^{-1}t$, where $\mu^\textsc{p}$ and $\mu^\textsc{s}$ are Borel measures such that $\mu^\textsc{p}([0,\infty[)+\mu^\textsc{s}(]0,1])=1$.
\end{The}\\
{\it Proof.} We apply (\ref{CERPDK}). So let $U$ be a rep of $SL(2,\C)$ in a separable Hilbert space having normalized $SU(2)$-invariant vectors.  Being semisimple connected, $SL(2,\C)$ is a locally compact tame group with countable basis. Therefore by \cite[Second Part \S\,8.4 Theorem 3]{K76} it allows the direct integral (continuous sum) decomposition into its primary components such that

$$U=\int_{\Lambda}W_\lambda\d \mu(\lambda)$$

up to unitary equivalence. Here $\mu$ is a probability measure on the set $\Lambda$ of equivalence classes of irreps of $SL(2,\C)$, $W_\lambda$ is a primary component written in the form $U_\lambda\otimes I_\lambda$. $U_\lambda$ belongs to the class $\lambda\in\Lambda$, and $I_\lambda$ is the trivial rep of dimension $n(\lambda)\in\N\cup\{0,\infty\}$.\\
\hspace*{6mm}
First we describe the $SU(2)$-invariant normalized vectors $\Gamma_0$ of $U$. $\Gamma_0$ decomposes into $SU(2)$-invariant normalized vectors  of the primary components $W_\lambda$. Let $\Lambda_0$ denote  the set of  equivalence classes of reps of the supplementary series and the principal series with $m= 0$. For $\lambda\in\Lambda\setminus \Lambda_0$  there is no such vector  for $W_\lambda$. For $\lambda\in\Lambda_0$ they read
$\gamma_\lambda\otimes v_\lambda$ with $\gamma_\lambda$ the unique (up to a phase) $SU(2)$-invariant normalized vector of $U_\lambda$, and $v_\lambda$ any normalized vector in the carrier space of $I_\lambda$. Hence $\Gamma_0(\lambda)=0$ if $\lambda\not\in\Lambda_0$, and $\Gamma_0(\lambda)=C\,\gamma_\lambda\otimes v_\lambda$ with the normalization constant $C:=\mu(\Lambda_0)^{-1/2}$ otherwise.
\\
\hspace*{6mm}
Determine (\ref{ARMEG}): $\langle \Gamma_0, U(A_\kappa)\Gamma_0\rangle=\int_{\Lambda_0}C^2 \langle \gamma_\lambda\otimes v_\lambda,W_\lambda(A_\kappa)\gamma_\lambda\otimes v_\lambda\rangle\d \mu(\lambda)=\int_{\Lambda_0}C^2 \langle \gamma_\lambda\otimes v_\lambda,\big(U_\lambda(A_\kappa)\gamma_\lambda\big)\otimes v_\lambda\rangle\d \mu(\lambda)=\int_{\Lambda_0}C^2 \langle \gamma_\lambda,U_\lambda(A_\kappa)\gamma_\lambda\rangle\d \mu(\lambda)$. Now, $C^2\mu$ is a probability measure on  $\Lambda_0$, which we write as the sum of two orthogonal measures $\mu^\textsc{p}$ and $\mu^\textsc{s}$ on the disjoint sets $\Lambda^\textsc{p}:=\{[(0,\lambda)]:\lambda\ge 0\}$ and $\Lambda^\textsc{s}:=\{[\lambda]: 0<\lambda\le 1\}$. We identify $\lambda$ with the invariant number determining the irreps of the principal and the supplementary series. The result (\ref{GLIPDK}) follows by the formulae (\ref{LIKIR}), (\ref{LIKIRP}).\qed

Several more or less immediate results  concerning (\ref{MTGP}) and (\ref{GLIPDK}) are gathered in (\ref{RMT}). For the 
 following the conclusions (a) - (d) are indispensable. The remaining  items   illustrate the ordering on $\mathcal{G}$.

\begin{VLem}\label{RMT}  
\begin{itemize}
\item[(a)] {\it Let $G\in\mathcal{G}$ be extreme. Then $g$ is irreducible, i.e., $g=g^\textsc{p}_\lambda$ or $g=g^\textsc{s}_\lambda$ for some $0\le\lambda<\infty$ and $0<\lambda\le 1$, respectively.}
\end{itemize}
{\it Proof.} For $G$ extreme, obviously either $\mu^\textsc{p}=0$ or  $\mu^\textsc{s}=0$. Consider first $\mu^\textsc{s}=0$. Let $B\subset [0,\infty[$ be a Borel set and $B'$ its complement. Using an obvious notation, $g=g_B+g_{B'}$. So one summand must vanish, say $g_{B'}=0$. This implies $0=g_{B'}(1)=\mu^\textsc{p}(B')$ and hence $\mu^\textsc{p}(B)=1$. Thus  every Borel set has either measure $1$ or $0$. Since $\lim_{N\to \infty}\mu^\textsc{p}([0,N[)=\mu^\textsc{p}([0,\infty[)=1$, there is some $N$ with 
$\mu^\textsc{p}([0,N[)=1$. Then $[0,N[$ contains an interval $I_1$ of length $N/2$ with $\mu^\textsc{p}(I_1)=1$, and $I_1$ contains an interval $I_2$ of length $N/4$ with $\mu^\textsc{p}(I_2)=1$, and so on. The intersection of these intervals consists of one point $\lambda_0$. Hence $\mu^\textsc{p}=\delta_{\lambda_0}$. --- The case $\mu^\textsc{p}=0$ follows analogously.\qed

\begin{itemize}
\item[(b)] {\it Let $\mu^\text{p}(]0,\infty[)=0$.   Then obviously $G$  is positive. Moreover, $g\ge g^\textsc{s}_0=   \frac{\kappa}{\sinh \kappa}$, whence  $\lim_{t \to \infty} t\, g(t)\to\infty$.}
\item[(c)] {\it If $\mu^\textsc{s}= 0$, then $|g(t)|\le \frac{\kappa}{\sinh \kappa}$. If $\mu^\textsc{s}\ne 0$, then $\lim_{t \to \infty} t^r\, g(t)\to\infty$ for some $r<1$. Finally, $\mu^\textsc{s}= 0$ if and only if  $|g(t)|\le \frac{\kappa}{\sinh \kappa}$.}
\end{itemize}
{\it Proof}.   Write $g=g^\textsc{p}+g^\textsc{s}$ in the obvious way. Keep the ordering (\ref{OK}) in mind. 
Then $|g^\textsc{p}(t)|\le \frac{\kappa}{\sinh \kappa}$, which implies the first part of the claim.  Now let  $\mu^\textsc{s}\ne 0$. Then  there is $0<\delta<1$ with $\mu^\textsc{s}([\delta,1])>0$. Hence $g^\textsc{s}(t)\ge \frac{\sinh\delta\kappa}{\delta\sinh\kappa}\mu^\textsc{s}([\delta,1])$. So $t^r\,g(t)\ge \frac{\cosh^r\kappa}{\sinh\kappa}(\alpha  \sinh\delta\kappa-\kappa)$ for some $\alpha>0$, whence the result for $r=1-\delta/2$. The last part of the claim is an easy consequence of the prior results.\qed

\begin{itemize}
\item[(d)]  {\it Note $g_{1^/2}=g^\textsc{s}_{1/2}$. In view of  \emph{(\ref{NCKC})} and \emph{(\ref{NCCK})} we show  that  $\breve{g}\in\mathcal{G}$ satisfies}  $$g\le g_{1/2}\;\Rightarrow\, \mu^\textsc{s}(]1/2,1])=0\;\Rightarrow\, |g|\le g_{1/2}$$
\end{itemize}
{\it Proof}.   Write $g=g^\textsc{p}+g^\textsc{s}$ in the obvious way. Keep the ordering (\ref{OK}) in mind.\\
\hspace*{6mm}
($\alpha$) Let $g\le g_{1/2}$. Assume $\mu^\textsc{s}(]1/2,1])>0$. Then $|g^\textsc{p}(t)|\le \frac{\kappa}{\sinh \kappa}\mu^\textsc{p}([0,\infty[)$, and there is $\lambda_0> 1/2$ with 
$\mu^\textsc{s}([\lambda_0,1])>0$, whence $g^\textsc{s}(t)\ge \frac{\sinh \lambda_0\kappa}{\lambda_0\sinh \kappa}\mu^\textsc{s}([\lambda_0,1])>0$. Hence $\alpha  \frac{\sinh \lambda_0\kappa}{\sinh \kappa}\le \beta \frac{\kappa}{\sinh \kappa} +\frac{1}{\cosh \frac{\kappa}{2}}$ for some $\alpha>0$ and $\beta\ge 0$. However this inequality obviously does not hold for large $\kappa$. One concludes $\mu^\textsc{s}(]1/2,1])=0$.\\
\hspace*{6mm}
($\beta$) Let $\mu^\textsc{s}(]1/2,1])=0$. Then  $|g^\textsc{p}(t)|\le \frac{\kappa}{\sinh \kappa}\mu^\textsc{p}([0,\infty[)=g^\textsc{s}_0(t)\,\mu^\textsc{p}([0,\infty[)$, whence $|g^\textsc{p}|\le g^\textsc{s}_{1/2}\,\mu^\textsc{p}([0,\infty[)$. Further  $g^\textsc{s}\le g^\textsc{s}_{1/2}\,\mu^\textsc{s}([0,1/2])$. It follows $|g|\le g^\textsc{s}_{1/2}$. Recall $g^\textsc{s}_{1/2}= g_{1/2}$.\qed

\begin{itemize}
\item[(e)] {\it For every $G\in\mathcal{G}\setminus\{1\}$ there is $G_1\ne 1$ with $\mu^\textsc{p}(]0,\infty[)=0$ such that $G\le G_1$.}
\end{itemize}
{\it Proof.} The claim holds by the ordering (\ref{OK}).\qed
\begin{itemize}
\item[(f)] {\it For every $G\in\mathcal{G}\setminus \{1\}$  
there is $\lambda_0\in]0,1[$ such that}
$\breve{g}^{\textsc{s}}_{\lambda_0}\not\le G$.
\end{itemize}
{\it Proof.} By (e) let without restriction $\mu^\textsc{p}(]0,\infty[)=0$. Fix $\kappa>0$. Recall $\operatorname{sinch}(y)=\frac{\sinh y}{y}$ for $y\ge 0$  is strictly increasing. It suffices to show $\int_{]0,1]}\frac{\operatorname{sinch}x \kappa}
{\operatorname{sinch}\lambda_0 \kappa}\d\mu^\textsc{s}(x)<1$ for some $0<\lambda_0<1$. Let $0< \lambda_n<1$, $\lambda_n\to 1$. Since $\frac{\operatorname{sinch}x \kappa}
{\operatorname{sinch}\lambda_n \kappa}\le \operatorname{sinch} \kappa$ $\forall$ $n,x$, dominated convergence yields $\lim_{n\to\infty}\int_{]0,1]}\frac{\operatorname{sinch}x \kappa}
{\operatorname{sinch}\lambda_n \kappa}\d\mu^\textsc{s}(x)=\int_{]0,1]}\frac{\operatorname{sinch}x \kappa}
{\operatorname{sinch} \kappa}\d\mu^\textsc{s}(x)=:A$. The integrand is $< 1$ for $x<1$ and $=1$ for $x=1$. Hence $A< 1$ since otherwise $\mu^\textsc{s}(\{1\})=1$, whence the contradiction $G=1$. The result follows. \qed

\begin{itemize}
\item[(g)] {\it On the other hand there are $G\in\mathcal{G}_+$ satisfying}\,
$G\,\not\le \,\breve{g}^{\textsc{s}}_\lambda \quad \forall \;\;0< \lambda<1$.
\end{itemize}
For an example take $\d \mu^\textsc{s}(\lambda):=2\lambda\, 1_{]0,1]}(\lambda)\d \lambda$, $\mu^\textsc{p}=0$  yielding $g(t)=\frac{\tanh \kappa/2}{\kappa/2}$. Then for $\lambda<1$ and $\kappa\to \infty$ one has $\frac{\kappa}{2}\frac{\sinh \lambda\kappa}{\lambda \sinh \kappa}=\frac{\kappa}{2\lambda}\e^{-(1-\lambda)\kappa }\frac{1-\exp(-2\lambda\kappa)}{1-\exp(-2\kappa)}\to 0$, whereas $\tanh \frac{\kappa}{2}\to 1$.\qed
\end{VLem}

\section{The Lorentz invariant kernels due to the principal series }

It is easy to see that there are  positive $G$ with $\mu^\text{s}=0$.   As an example take $\d \mu^\textsc{p}(\lambda):=1_{[0,1]}(\lambda)\d \lambda$. Then $g(t)=\frac{\operatorname{Si}(\kappa)}{\sinh \kappa}>0$ with $\operatorname{Si}$  the sine integral. Actually one has the following characterization of $$\mathcal{G}^\textsc{p}:=\{\breve{g}\in\mathcal{G}: \mu^\textsc{s}=0 \text{ in } (\ref{GLIPDK}) \}$$

\begin{Pro}\label{IFSFK}  Let $g$ denote a  function on $[1,\infty[$ and put  $\kappa=\cosh^{-1} t$. The following statements \emph{$(\alpha),(\beta), (\gamma)$} are equivalent.
\begin{itemize}

\item[\emph{$(\alpha)$}]  $\breve{g}\in\mathcal{G}$ with $|g(t)|\le \frac{\kappa}{\sinh \kappa}$ for all $t\ge 1$.

\item[\emph{$(\beta)$}]  $\breve{g}\in \mathcal{G}^\textsc{p}$, i.e., $g(t)=\int_{[0,\infty[}\frac{\sin \lambda\kappa}{\lambda \sinh \kappa}\d \mu(\lambda)$   with $\mu$  a  probability Borel measure on $[0,\infty[$. 
\item[\emph{$(\gamma)$}]  $g(t)= \frac{\Psi(\kappa)}{\sinh (\kappa)}$ for $\Psi(\kappa):=\int_0^\kappa \psi(x)\d x$ with  $\psi$ a  real continuous  function  on $\R$ of positive type normalized by $\psi(0)=1$. 
 \end{itemize}
By  \emph{\cite[Theorem IX.9 (Bochner's theorem)]{RS75}} there is a unique even  probability measure $\mu_{even}$ on $\R$ such that 
\begin{equation}\label{MUEV}
\psi(x)=\int_\R \e^{\i \lambda\,x}\d \mu_{even}(\lambda)
\end{equation}
 Then  $g$, $\psi$, $\mu$, and $\mu_{even}$ determine uniquely each other. In particular one has
\begin{itemize}
\item[\emph{$(\delta)$}] $\mu(B)=\mu_{even}(B\cap\{0\})+2\,\mu_{even}(B\setminus\{0\})$ for all Borel sets $B\subset [0,\infty[$.
\end{itemize}
\end{Pro}
{\it Proof.} $(\alpha)$ $\Leftrightarrow$ $(\beta)$ holds by (\ref{MTGP}) and (\ref{RMT})(c).\\
\hspace*{6mm} As to  $(\beta)$ $\Rightarrow$ $(\gamma)$,\, $g(t)\sinh\kappa=\int_{[0,\infty[}\frac{\sin \lambda\kappa}{\lambda}\d \mu(\lambda)=\int_{[0,\infty[}\frac{1}{2}\int_{-\kappa}^\kappa\e^{\i \lambda x}\d x \d \mu(\lambda)=\int_0^\kappa\int_\R\e^{\i \lambda x}\d \mu_{even}(\lambda)\d x$,  where  $\mu_{even}$ is  the even probability measure on $\R$ given by $\mu_{even}(B):=\frac{1}{2}\mu(B)$ for a Borel set $B\subset ]0,\infty[$, $\mu_{even}(B):=\frac{1}{2}\mu(-B)$ for a Borel set $B\subset ]-\infty,0[$, and $\mu_{even}(\{0\}):=\mu(\{0\})$. Now put $\psi(x):=\int_\R\e^{\i \lambda x}\d \mu_{even}(\lambda)$, $x\in\R$. Clearly $\psi$ is real, $\psi(0)=1$, continuous and, by Bochner's theorem, of positive type. \\
\hspace*{6mm} We turn to $(\beta)$ $\Leftarrow$ $(\gamma)$ and ($\delta$). Recall (\ref{MUEV}).
Then going backward the prior proof, $\frac{\Psi(\kappa)}{\sinh (\kappa)}$ equals $g(t)$ in $(\beta)$ for $\mu$ given in $(\delta)$.\\
\hspace*{6mm} Finally, obviously $g$ and $\psi$ determine each other. Hence so do  $g$, $\psi$, $\mu_{even}$ and, by ($\delta)$, also $\mu$.\qed

Referring to (\ref{IFSFK}), the characterization  of $g$ in $(\gamma)$ gives rise to an inversion formula of the formula in $(\beta)$.

\begin{Cor}\label{CIF} Let $\breve{g}\in\mathcal{G}$ with $g$  a  function on $[1,\infty[$. Suppose   $|g(t)|\le \frac{\kappa}{\sinh \kappa}$  and $\psi\in L^1(\R)$ for $\psi(x):=\frac{\d}{\d x}\big(\sinh x \;g(\cosh x)\big)$, $x\in\R$. Put 
\begin{equation}\label{MFIF}
w(\lambda):= \frac{1}{\pi}\int_{-\infty}^\infty \psi(x)\e^{-\i \lambda x}\d x
\end{equation}
Then 
\begin{equation}\label{AMFIF}
w\ge 0, \;1=\int_0^\infty w(\lambda)\d \lambda, \text{ and \;} g(t)=\int_0^\infty\frac{\sin \lambda\kappa}{\lambda \sinh \kappa}w(\lambda)\d\lambda 
\end{equation}
\end{Cor}

{\it Proof.} By (\ref{IFSFK})($\gamma$) $\psi$ is real continuous bounded by $1=\psi(0)$ of positive type. So by Bochner's theorem $\psi(x)=\int_\R\e^{\i \lambda x}\d \mu_{even}(\lambda)$, $x\in\R$  for some even probability measure  $ \mu_{even}$. As by assumption $\psi$ is integrable,
(\ref{PDDOO3})(d) applies. Accordingly $w_{even}:=(2\pi)^{-1/2}\mathcal{F}\psi$ is integrable nonnegative. It satisfies $\psi=(2\pi)^{1/2}\mathcal{F}^{-1} w_{even}$. Thus the Fourier transforms of the finite measures $\mu_{even}$ and $w_{even}m$ with $m$ the Lebesgue measure on $\R$ coincide. Thus the measures itself coincide. \\
\hspace*{6mm}
Now recall (\ref{IFSFK})($\delta$). Put $w:=2\, w_{even}|_{[0,\infty[}$. The claim holds by  (\ref{IFSFK})$(\beta)$.\qed

\begin{Exa}\label{WFGOH} Let $g=g_1$, i.e.,   $g(t)=\frac{2}{1+t}$.  \\
\hspace*{6mm} We apply (\ref{CIF}) to compute the weight function $w$ satisfying (\ref{AMFIF}). First check that $\psi(x)=2(1+\cosh x)^{-1}$.  The integral (\ref{MFIF}) can be routinely evaluated by the theorem of residues.  The poles of the integrand $\psi$ are at $z_k:=\i(2k+1)\pi$, $k\in\mathbb{Z}$. They are all of second order and  give   rise to the residues $2\i \lambda \e^{-\i\lambda z_k}$. One finds 
$$w(\lambda)=\frac{4\lambda}{\sinh \pi \lambda}$$
(One easily confirms the result by the formulae from Gradshteyn \cite[3.521(1)]{GR07}    and  \cite[17.34(28))]{GR07} for $a:=\i\kappa$, $b:=\pi$, $x:=\lambda$, $\xi:=0$.)\\
\hspace*{6mm} 
Similarly one finds 
$$w(\lambda)=\frac{8\lambda^3}{\sinh \pi \lambda}$$ for $g=g_2$, i.e., $g(t)=4(1+t)^{-2}$. (Confirm the normalization of $w$ by  \cite[3.523(6)]{GR07}.)
\end{Exa}\\
In view of (\ref{KKKR}), (\ref{MTKK}) the next example is particularly interesting. 

\begin{Exa} Let $g=g_{3/2}$, i.e.,   $g(t)=\big(\frac{2}{1+t}\big)^{3/2}$.  \\
\hspace*{6mm} We proceed as in (\ref{WFGOH}). One finds $\psi(x)=\big(2-\cosh^2\frac{\kappa}{2}\big)\cosh^{-3}\frac{\kappa}{2}$. Note that $\psi\not\ge 0$\,(!) despite the fact that $g_{3/2}>0$.\\
\hspace*{6mm}
The poles of $\psi$ are at $z_k:=\i(k+\frac{1}{2})\pi$, $k\in\mathbb{Z}$. They are all of third  order and  give   rise to the residues $-4\i \lambda^2 (-1)^k\e^{-2\i \lambda z_k}$. One finds
$$ w(\lambda)=\frac{8\lambda^2}{\cosh\pi\lambda}$$
(Confirm the normalization of $w$ by  \cite[3.523(5)]{GR07}.)
\end{Exa}\\
A further result using  decisively (\ref{IFSFK}) is 

\begin{Cor} \label{EIP}
Let $G\in \mathcal{G}^\text{p}$.  Then
\begin{center}
$G$ is extreme  in $\mathcal{G}^\text{p}$ $\Leftrightarrow$ $G$ is extreme in  $\mathcal{G}$  $\Leftrightarrow$ $G$ is irreducible
\end{center}
\end{Cor}

{\it Proof.} As to the first $\Rightarrow$, let $G=\alpha G_a+\beta G_b$ with $\alpha, \beta > 0$, $\alpha+\beta=1$, and $G_i\in\mathcal{G}$, $i=a,b$. Then by (\ref{RMT})(c) it follows immediately $G_i\in \mathcal{G}^\text{p}$. So by assumption $G=G_a=G_b$. --- The first $\Leftarrow$ is trivial.\\
\hspace*{6mm}
The second $\Rightarrow$ holds by (\ref{RMT})(a). We turn to the second $\Leftarrow$. Let $G= \breve{g}^{\text{p}}_{\lambda_0}$ for  $\lambda_0\in[0,\infty[$ and let  $g^\text{p}_{\lambda_0}=\alpha g_a+\beta  g_b$,  with $\alpha, \beta> 0$,  $\alpha+\beta=1$, and $\breve{g}_a, \breve{g}_b \in\mathcal{G}$. Then again, by (\ref{RMT})(c) it follows $\breve{g}_i\in \mathcal{G}^\text{p}$, $i=a,b$. Hence due to the uniqueness result in  (\ref{IFSFK}) one has $\delta_{\lambda_0}=\alpha \mu^\textsc{p}_a+\beta  \mu^\textsc{p}_b$. This implies 
$\mu^\textsc{p}_a=  \mu^\textsc{p}_b=\delta_{\lambda_0}$.\qed

\section{The maximal causal kernel}\label{TMCK}

It turns out that the positive definite Lorentz invariant kernels  $\breve{g}\in\mathcal{G}$, which determine causal kernels, are due to the principal series only, i.e., belong to  $\mathcal{G}^\textsc{p}$. 
The result is obtained by checking the necessary condition NC below. It discards also  all irreducible  elements of $\mathcal{G}$.  (\ref{ANC}) reveals NC to be rather sharp. It is the main tool in proving the important result that $K_{3/2}$ is the greatest element of $\mathcal{K}$ (\ref{MTKK}).

 We deal with the causal kernels on $\R^3$  (\ref{KKK}).     
  Referring to the expansion of $K$  (\ref{PDEC}), we exploit the fact that by (\ref{CKPOLMB}) $$k_0(\sigma,\rho)=\frac{\epsilon(\sigma)+\epsilon(\rho)}{2\sqrt{\epsilon(\sigma)\epsilon(\rho)}}\int_{-1}^1g\big(\epsilon(\sigma)\epsilon(\rho)-\sigma\rho\, x\big)\frac{\d x}{2}$$
 is positive definite on $[0,\infty[$. Hence it satisfies the inequality $k_0(\sigma,\rho)^2\le k_0(\sigma,\sigma)k_0(\rho,\rho)$ (\ref{CSK}). For $\sigma=0$ this reads
 \begin{equation*}
 \frac{\big(1+\epsilon(\rho)\big)^2}{4\,\epsilon(\rho)}\,g\big(\epsilon(\rho)\big)^2\le \frac{1}{2\rho^2}\int_1^{1+2\rho^2}g(t)\d t\tag{NC}
 \end{equation*}
 
So,  if $g$ determines a causal kernel, then $g$ satisfies NC.

 \begin{ANC}\label{ANC} The necessary condition for positive definiteness  of rotational invariant normalized kernels  introduced above
\begin{equation*}
k_0(0,\rho)^2\le k_0(0,0)\,k_0(\rho,\rho)
\end{equation*} is shown to be not satisfied by $\breve{g}_r$ for $0<r<\frac{1}{2}$ and by  $K_r$ for  $0<r<\frac{3}{2}$ thus furnishing  straightforward  alternative proofs to (\ref{PDGSOH}) and (\ref{NPDKR}),  by which these kernels are not positive definite. Moreover, what is important, the necessary condition holds by equality for the kernels $\breve{g}_{1/2}$ and $K_{3/2}$. 
\begin{itemize}
\item[(a)]  {\it $\breve{g}_r$ for $0<r<\frac{1}{2}$ is not positive definite}
\end{itemize}
{\it Proof.} The left hand side  LS of NC reads $2^{2r}\big(1+\epsilon(\rho)\big)^{-2r}$ (there is no prefactor here), the right hand side  RS of NC  is $\frac{1}{1-r}\frac{1}{\rho^2}\left(\epsilon(\rho)^{2-2r}-1\right)$, $0\le r\ne 1$. Hence  NC imposes 
$$f(r)\le R(\rho)$$
with $f(r):=2^{2r}(1-r)$ and $R(\rho):=\frac{\epsilon(\rho)^2}{\rho^2}\left(\frac{1+\epsilon(\rho)}{\epsilon(\rho)}\right)^{2r}-\frac{\big(1+\epsilon(\rho)\big)^{2r}}{\rho^2}\to 1$ for $\rho\to \infty$. Hence $f(r)\le1$. However this does not hold, since $f(0)=f(\frac{1}{2})=1$, $f(\frac{1}{4})=\frac{3}{4}\sqrt{2}>1$, and $f'(r)=0$ has exactly one solution ($r=1-\frac{1}{\ln 4}$).\qed
\begin{itemize}
\item[(b)]  {\it   $g_{1/2}\big(\epsilon(\rho)\big)^2 = \frac{1}{2\rho^2}\int_1^{1+2\rho^2}g_{1/2}(t)\d t$, which is \emph{NC} by equality for $\breve{g}_{1/2}$ }
\end{itemize}
{\it Proof.} The claim is easily checked.\qed

\begin{itemize}
\item[(c)]  {\it $K_r$ for $0<r<\frac{3}{2}$ is not positive definite}
\end{itemize}
{\it Proof.} Here LS has  the additional prefactor $\frac{\big(1+\epsilon(\rho)\big)^2}{4\epsilon(\rho)}$. Hence for $r\ne 1$
NC imposes
$$L(\rho)\le R(\rho)$$
with $L(\rho):=2^{2r-2}\rho^2\,\epsilon(\rho)^{-1}\big(1+\epsilon(\rho)\big)^{2-2r}\to \infty$,  $R(\rho):=\frac{1}{1-r}\left(\epsilon(\rho)^{2-2r}-1\right)\to\frac{1}{r-1}$ for $\rho\to \infty$ if $r>1$, and $L(\rho):=2^{2r-2}\big(1+\epsilon(\rho)\big)^{2-2r}\epsilon(\rho)^{2r-2}\to 2^{2r-2}$, $R(\rho):=\frac{1}{1-r}\rho^{-2}\epsilon(\rho)\left(1-\epsilon(\rho)^{2r-2}\right)\to 0$  for $\rho\to \infty$ if $r<1$. Hence NC is not satisfied for $r\ne 1$. \\
\hspace*{6mm}
For $r=1$, LS equals $\epsilon(\rho)^{-1}$ and RS reads $2\rho^{-2}\,\ln \epsilon(\rho)$. Hence NC imposes $\rho\,\epsilon(\rho)^{-1}\le 2\rho^{-1} \ln\epsilon(\rho)$, which does not hold.\qed

\begin{itemize}
\item[(d)] {\it $ \frac{\big(1+\epsilon(\rho)\big)^2}{4\,\epsilon(\rho)}\,g_{3/2}\big(\epsilon(\rho)\big)^2= \frac{1}{2\rho^2}\int_1^{1+2\rho^2}g_{3/2}(t)\d t$, which is \emph{NC} by equality for $K_{3/2}$ }
\end{itemize}
{\it Proof.} The claim is easily checked. \qed

\end{ANC}

 \begin{Lem}\label{DIEG} Let $g$ be irreducible, i.e., equal to $g^\textsc{p}_\lambda$ or $g^\textsc{s}_\lambda$ for some $\lambda$. Then
 \begin{equation*}
 g\big(\epsilon(\rho)\big)^2\;= \frac{1}{2\rho^2}\int_1^{1+2\rho^2}g(t)\d t \tag{1}
 \end{equation*}
 Explicitly \emph{(1)} equals
 \begin{equation*}\left(\frac{\sin \lambda\, l(\rho)}{\lambda\rho}\right)^2 \text{ for \;} g^\textsc{p}_\lambda \text{ and \;} \left(\frac{\sinh \lambda\, l(\rho)}{\lambda\rho}\right)^2 \text{ for \;} g^\textsc{s}_\lambda\tag{2}
 \end{equation*}
 with $l(\rho):=\ln\big(\epsilon(\rho)+\rho\big)$. The limiting case $\lambda=0$ reads $\big(\frac{l(\rho)}{\rho}\big)^2$. Finally, $g$ does not satisfy \emph{NC}. So  an irreducible $g$  does not determine a causal kernel.
 \end{Lem}\\
{\it Proof.} As to  $g\big(\epsilon(\rho)\big)^2$, note $\kappa =\cosh^{-1}\epsilon(\rho)=\ln\big(\epsilon(\rho)+\sqrt{\epsilon(\rho)^2-1}\big)=l(\rho)$ and  $2\sinh l(\rho)=\epsilon(\rho)+\rho-(\epsilon(\rho)+\rho)^{-1}=\epsilon(\rho)+\rho-(\epsilon(\rho)-\rho)=2\rho$, whence $(2)$.\\
\hspace*{6mm}
We turn to the integral  $\frac{1}{2\rho^2}\int_1^{1+2\rho^2}g(t)\d t$, which for $g=g^\textsc{p}_\lambda$, $\lambda>0$,  becomes  $\frac{1}{2\rho^2\lambda}\int_0^{2\,l(\rho)}\sin\lambda x \d x$ using the substitution $x=\cosh^{-1}t$, whence $(2)$. Analogously one obtains $(2)$ for  $\lambda=0$ and for $g=g^\textsc{s}_\lambda$.\\
\hspace*{6mm}
Finally, $g$ does not satisfy NC because of  $(1)$,$(2)$ and the fact $\frac{1}{\rho}\, \frac{\big(1+\epsilon(\rho)\big)^2}{4\,\epsilon(\rho)}\to \frac{1}{4}>0$ for $\rho\to\infty$,  \qed

\begin{Cor}\label{CDIEG} Put  $l(\rho):=\ln\big(\epsilon(\rho)+\rho\big)$. For $g$ in \emph{(\ref{GLIPDK})} one has 
$$g(\epsilon(\rho))=\int_{[0,\infty[}\frac{\sin \lambda\,l(\rho)}{\lambda\,\rho}\d\mu^\textsc{p}(\lambda)+\int_{]0,1]}\frac{\sinh \lambda\,l(\rho)}{\lambda\,\rho}\d\mu^\textsc{s}(\lambda) 
$$
and
$$ \frac{1}{2\rho^2}\int_1^{1+2\rho^2}g(t)\d t =\int_{[0,\infty[}\left(\frac{\sin \lambda\,l(\rho)}{\lambda\,\rho}\right)^2\d\mu^\textsc{p}(\lambda)+\int_{]0,1]}\left(\frac{\sinh \lambda\,l(\rho)}{\lambda\,\rho}\right)^2\d\mu^\textsc{s}(\lambda)$$
\end{Cor}

\begin{Pro}\label{CKK} Let $g$ determine a causal kernel. Then $\breve{g}\in\mathcal{G}^\textsc{P}$, but $\breve{g}$ is not extreme. Recall \emph{(\ref{IFSFK})}.
\end{Pro}\\
{\it Proof.}
 There is the representation (\ref{GLIPDK}). Put  $g=g^\textsc{p}+g^\textsc{s}$ in the obvious way. Assume $g^\textsc{s}\ne 0$. Keep in mind  (\ref{CDIEG}).\\
\hspace*{6mm}
 As to the left hand side LS of NC, $g^\textsc{p}(\epsilon(\rho))=\int_{[0,\infty[}\frac{\sin \lambda\,l(\rho)}{\lambda\,\rho}\d\mu^\textsc{p}(\lambda)$, whence $|g^\textsc{p}(\epsilon(\rho))|\le \alpha\frac{l(\rho)}{\rho}$ with $\alpha:=\mu^\textsc{p}([0,\infty])\ge 0$. Further $g^\textsc{s}(\epsilon(\rho))=\int_{]0,1]}\frac{\sinh \lambda\,l(\rho)}{\lambda\,\rho}\d\mu^\textsc{s}(\lambda)$. 
By the mean value theorem of integrals there is $\lambda_{l,\rho}\in]0,1]$ satisfying $\beta\,\frac{\sinh \lambda_{l,\rho}\,l(\rho)}{\lambda}=\int_{]0,1]}\frac{\sinh \lambda\,l(\rho)}{\lambda}\d\mu^\textsc{s}(\lambda)$ with $\beta:=\mu^\textsc{s}(]0,1])>0$. Since $\frac{\sinh \lambda\,l(\rho)}{\lambda}$ is strictly increasing regarding $\lambda$ and $\rho$, the position 
$ \lambda_{l,\rho}$ is uniquely determined and is strictly increasing with respect to $\rho$. 
Thus the estimation 
$g^\textsc{p}(\epsilon(\rho))^2\ge \big(-\alpha\frac{l(\rho)}{\rho}+ \beta\,\frac{\sinh \lambda_{l,\rho}\,l(\rho)}{\lambda\,\rho}\big)^2$ holds. Put $\lambda_l:=\lim_{\rho\to \infty} \lambda_{l,\rho}$. \\
\hspace*{6mm}
The right hand side RS $=\int_{[0,\infty[}\left(\frac{\sin \lambda\,l(\rho)}{\lambda\,\rho}\right)^2\d\mu^\textsc{p}(\lambda)+\int_{]0,1]}\left(\frac{\sinh \lambda\,l(\rho)}{\lambda\,\rho}\right)^2\d\mu^\textsc{s}(\lambda)$ of NC is treated similarly getting RS $\le \alpha\big(\frac{l(\rho)}{\rho}\big)^2+\beta\left(\frac{\sinh \lambda_{r,\rho}\,l(\rho)}{\lambda\,\rho}\right)^2$. Put  $\lambda_r=\lim_{\rho\to \infty} \lambda_{r,\rho}$. \\
\hspace*{6mm}
Note $\lambda_{l,\rho}\le \lambda_{r,\rho}$ and $0<\lambda_l\le\lambda_r$. Indeed, this holds because $\left(\beta\,\frac{\sinh \lambda_{l,\rho}\,l(\rho)}{\lambda}\right)^2=\left(\int_{]0,1]}\frac{\sinh \lambda\,l(\rho)}{\lambda}\d\mu^\textsc{s}(\lambda)\right)^2\le \beta \int_{]0,1]}\left(\frac{\sinh \lambda\,l(\rho)}{\lambda\,\rho}\right)^2\d\mu^\textsc{s}(\lambda)= \beta^2\left(\frac{\sinh \lambda_{r,\rho}\,l(\rho)}{\lambda\,\rho}\right)^2 $ by the Cauchy-Schwarz inequality.\\
\hspace*{6mm}
Due to NC one infers
\begin{equation*}
 \frac{\big(1+\epsilon(\rho)\big)^2}{4\,\epsilon(\rho)}\,\left(-\alpha\frac{l(\rho)}{\rho}+ \beta\,\frac{\sinh \lambda_{l,\rho}\,l(\rho)}{\lambda\,\rho}\right)^2\le \alpha\left(\frac{l(\rho)}{\rho}\right)^2+\beta\left(\frac{\sinh \lambda_{r,\rho}\,l(\rho)}{\lambda\,\rho}\right)^2
\end{equation*}
Hence one easily verifies 
\begin{equation*}
\rho^{1-2 (\lambda_{r,\rho}- \lambda_{l,\rho})}L(\rho)\le R(\rho)\tag{1}
\end{equation*}
with $L(\rho):=\rho^{-1}\; \frac{\big(1+\epsilon(\rho)\big)^2}{4\,\epsilon(\rho)}\;\rho^{-2 \lambda_{l,\rho}}\,\left(-\alpha \,l(\rho)+ \beta\,\frac{\sinh \lambda_{l,\rho}\,l(\rho)}{\lambda}\right)^2\to \frac{1}{4}\left(\beta\frac{2^{\lambda_l}}{2\lambda_l}\right)^2>0$ for $\rho\to\infty$, and similarly $R(\rho)\to \beta \left(\frac{2^{\lambda_r}}{2\lambda_r}\right)^2$ for $\rho\to\infty$.\\
\hspace*{6mm}
Now one remembers  (\ref{NCCK}),\,(\ref{RMT})(d), according to which $\mu^\textsc{s}(]1/2,1])=0$. This implies $\lambda_r\le\frac{1}{2}$. It follows  $2 (\lambda_{r,\rho}- \lambda_{l,\rho})\le 1-\delta$ for some $\delta>0$ and all $\rho\ge \rho_0$ for some $\rho_0$. Thus $(1)$ requires $\beta=0$  contradicting the assumption  $g^\textsc{s}\ne 0$.\\
\hspace*{6mm}
Finally recall (\ref{EIP}) and (\ref{DIEG}).
\qed

\begin{The}\label{MTKK} $|K|\le K_{3/2}$ pointwisely for all  causal kernels $K$.
\end{The}\\
{\it Proof.} Let $g$ determine a causal kernel (\ref{KKK}). For $x\ge 1$ put $\alpha(x):=\frac{(1+x)^2}{4x}$.\\
\hspace*{6mm}
(a) {\it $g$ is integrable.} Indeed, $g$ satisfies NC and, by (\ref{CKK}), $|g|\le g_0^\textsc{p}$. Hence
$\alpha\big(\epsilon(\rho)\big)\,g\big(\epsilon(\rho)\big)^2\le \frac{1}{2\rho^2}\int_1^{1+2\rho^2}g(t)\d t\le \frac{1}{2\rho^2}\int_1^{1+2\rho^2}g_0^\textsc{p}(t)\d t=
g_0^\textsc{p}\big(\epsilon(\rho)\big)^2$ by (\ref{DIEG})(1), whence $|g\big(\epsilon(\rho)\big)|\le \alpha\big(\epsilon(\rho)\big)^{-1/2}\,g_0^\textsc{p}\big(\epsilon(\rho)\big)$. The latter means $|g(t)|\le \frac{2\sqrt{\cosh \kappa}}{1+\cosh\kappa}\; \frac{\kappa}{\sinh\kappa}$ with $\kappa=\cosh^{-1}t$.  It implies 
\begin{equation*}
A:=\int_1^\infty |g(t)|\d t<\infty\tag{1}
\end{equation*}
since $\int_1^\infty |g(t)|\d t=\int_0^\infty |g(\cosh x)|\sinh x \d x\le \int_0^\infty \frac{2x\sqrt{\cosh x}}{1+\cosh x} \d x\le \int_0^\infty \frac{x\e^{x/2}}{\cosh^2x/2} \d x\le \int_0^\infty 4x\e^{-x/2}\d x=8$.\\
\hspace*{6mm}
(b) {\it $g$ is dominated by a multiple of $g_{3/2}$.} Indeed, the right hand side of NC is bounded by $1$, since $|g|\le 1$. This implies $|g\big(\epsilon(\rho)\big)|^2\le \alpha\big(\epsilon(\rho)\big)^{-1}=g_{3/2}\big(\epsilon(\rho)\big)^2\;\frac{1}{2} \epsilon(\rho)\big(1+\epsilon(\rho)\big)\le 2\,g_{3/2}\big(\epsilon(\rho)\big)^2$ for $\rho\le 1$. By (1)
the right hand side of NC is bounded by $\rho^{-2}\,A$, whence $|g\big(\epsilon(\rho)\big)|^2\le g_{3/2}\big(\epsilon(\rho)\big)^2\;A\,\frac{1}{2} \frac{\epsilon(\rho)}{\rho}\big(1+\frac{\epsilon(\rho)}{\rho}\big) \le 2A\, g_{3/2}\big(\epsilon(\rho)\big)^2$ for $\rho\ge 1$. Thus
\begin{equation*}
|g|\le C\;g_{3/2}\tag{2}
\end{equation*}
for some constant $1\le C<\infty$.\\
\hspace*{6mm}
(c) By NC and (2) one has  due to (\ref{ANC})(d): $\alpha\big(\epsilon(\rho)\big)\,g\big(\epsilon(\rho)\big)^2\le \frac{1}{2\rho^2}\int_1^{1+2\rho^2}g(t)\d t\le C\; \frac{1}{2\rho^2}\int_1^{1+2\rho^2}g_{3/2}(t)\d t=C\;\alpha\big(\epsilon(\rho)\big)\,g_{3/2}\big(\epsilon(\rho)\big)^2$ . Therefore $|g|\le C^{1/2}\;g_{3/2}$. Iterating this step the result follows.\qed

The result in (\ref{MTKK}) is improved by 

\begin{Cor}\label{IMTKK} $|K(k,p)|<K_{3/2}(k,p)$ for all $k\ne p$ and  all  $K\in\mathcal{K}\setminus\{K_{3/2}\}$.
\end{Cor}\\
{\it Proof.} Consider $h:[1,\infty[\to\R$ with $\breve{h}=K/K_{3/2}$.  $h$ is continuous and by (\ref{MTKK})  $h(1)=1$, $|h|\le 1$ holds. The claim is $|h(t)|<1$ for  $t>1$.\\
\hspace*{6mm}
 Since $g_{3/2}h$ determines $K$, it satisfies NC. Therefore
 \begin{equation*}
 h\big(\epsilon(\rho)\big)^2\le h(t_\rho)\tag{1}
 \end{equation*}
 where  $t_\rho\in[1,1+2\rho^2]$ satisfies 
 \begin{equation*}
\int_1^{1+2\rho^2}h(t)\,g_{3/2}(t)\d t= h(t_\rho)\int_1^{1+2\rho^2}g_{3/2}(t)\d t\tag{2}
\end{equation*}
 Indeed,  (1) follows from NC for $g_{3/2}\, h$ with (2) according to  the mean value theorem of integrals.
Recall (\ref{ANC})(d).\\
\hspace*{6mm}
Now assume $h(t^*)=1$ for some $t^*>1$. Then for $\rho^*>0$ satisfying $\epsilon(\rho^*)=t^*$ one has $ h(t_{\rho^*})=1$ by (1), whence 
 $h(t)=1$ for $1\le t\le 1+2\rho^{*2}$ due to (2). Note that $t^*<1+2\rho^{*2}$. Therefore this result implies that $\{t\ge 1: h(t)=1\}$ is connected and unbounded and hence equal to $[0,\infty[$ contradicting  $h\ne \textbf{1}$. --- Now assume  $h(t^*)=-1$ for some $t^*>1$. Then still $ h(t_{\rho^*})=1$ by (3). Apply the foregoing result.\qed

\section{Localization in bounded regions}\label{LBR}

As known a POL  $T$ of the massive scalar boson with CT, i.e.,  for which  time evolution  is causal, does not localize the boson in any bounded region $\Delta\subset \R^3$, which means that $T(\Delta)$ has no eigenstate with eigenvalue $1$. One has the general result

\begin{Theo} \emph{\cite[(8) Theorem]{C17}\textbf{.}}\label{EDLR}
Let $T$ be a POL with CT. Let the relativistic relation $H\ge |P|$ hold. Suppose that 
there is a state localized in the region $\Delta$. Then  $\Delta$ is essentially dense, i.e., $\overline{\Delta \setminus N}=\R^3$ for every  Lebesgue null set $N$.
\end{Theo}

 However, the property $\norm{T(\Delta)}=1$ may  be regarded as physically equivalent to the presence of the eigenvalue $1$.
 Indeed, as
\begin{equation*} \label{CID}
||T(\Delta)||= \sup \left\{\langle \phi, T(\Delta)\varphi\rangle:\, ||\phi||=1\right\}
\end{equation*}
 norm $1$  means that  the system can be localized within that region  $\Delta$ by a suitable preparation, not strictly but as accurately as desired. So for obvious physical reasons one is interested in POL with $||T(B)||=1$ for every however small  open ball $B\ne \emptyset$.  
 We call them \textbf{separated}. In this case for every $b\in\R^3$ 
there is a sequence $(\phi_n)$ of states satisfying 
\begin{equation}\label{SSL}
\big\langle \phi_n,T(B )\, \phi_n\big\rangle \to 1,  \quad n\to \infty
\end{equation}
for every  open ball $B$ around $b$. This means that by a suitable preparation the system can be localized around $b$ as good as desired, thus distinguishing $b$ from any other point. According to \cite[sec.\,G]{CL15},  any $(\phi_n)$ satisfying  (\ref{SSL}) is called a \textbf{sequence of states localized  at} $b$.

The main result of this section is the following criterion on the separateness of a POL.

\begin{The}\label{SSL0} Let $K$ be the kernel on $\R^3\setminus\{0\}$ of a POL $T$  for a massive scalar boson. Suppose that
\begin{itemize}
\item[\emph{(i)}] $\lim_{\lambda\to\infty} K(k, \lambda p)$\; exists for every $k$ and almost all $p$
\item[\emph{(ii)}]  $\lim_{\lambda,\lambda'\to\infty} K(\lambda p,\lambda' p)=1$  for almost all  $p$
\item[\emph{(iii)}]  there is  $k_0$  such that, for any $\lambda_n\to\infty$, $K(k_0,\lambda_n p)$  does not vanish for almost all $p$.
\end{itemize}
Then $T$ is separated.
\end{The}\\
{\it Proof.} We use the representation (\ref{AKPOLMB}) of $T$. Let $\mathcal{H}_J$  be the RKHS associated to the kernel $K$ (see e.g.\,\cite{F08}) and put $J_p:=K(\cdot, p)$. The orthogonal projection $P$  on  $L^2(\R^3,\mathcal{H}_J)$, $(P\varphi)(p):=\langle J_p,\varphi(p)\rangle J_p$ maps onto the subspace 
$j\big(L^2(\R^3)\big)$. There is the representation $D$ of the dilation group acting on $L^2(\R^3,\mathcal{H}_J)$ by $(D_\lambda\varphi)(p)=\lambda^{3/2}\varphi(\lambda p)$. It  satisfies $D_\lambda E(\Delta)D_\lambda^{-1}=E(\lambda\Delta)$ for $E:= \mathcal{F}E^{can}\mathcal{F}^{-1}$. Put $P_\lambda :=D_\lambda P D_\lambda^{-1}$. The result holds by \cite[Theorem 7]{CL15} if the strong limit $Q:=\lim_{\lambda\to\infty}P_\lambda$ exists with $Q\ne 0$.\\
\hspace*{6mm}
Check $(P_\lambda\varphi)(p)=\langle J_{\lambda p},\varphi(p)\rangle J_{\lambda p}$. For the Cauchy criterion compute 
$\norm{P_\lambda\varphi-P_{\lambda'}\varphi}^2=\int\d^3p\, | \langle J_{\lambda p},\varphi(p)\rangle - \langle J_{\lambda' p},\varphi(p)\rangle|^2 + 2 \operatorname{Re} \big[(1-K(\lambda p,\lambda' p))    \langle J_{\lambda' p},\varphi(p)\rangle     \overline{ \langle J_{\lambda p},\varphi(p)\rangle}\,\big]  $.\\
\hspace*{6mm}
We evaluate this formula for $\varphi(p)=1_\Delta(p) J_k$ with $\Delta$ a Borel set of finite measure. The integrand $1_\Delta(p)\,\Big(|K(\lambda p,k)-K(\lambda' p,k)|^2+2\operatorname{Re}\big[\big(1-K(\lambda p,\lambda' p)\big)     K(\lambda' p,k)    \overline{K(\lambda p,k)}\,\big]\Big)$ vanishes for almost  every $p$ when $\lambda,\lambda'\to\infty$ due to (i), (ii) and since $|K|\le 1$. So the integral vanishes by dominated convergence. We conclude $\norm{P_\lambda\varphi-P_{\lambda'}\varphi}^2\to 0$ for $\lambda,\lambda'\to \infty$.\\
\hspace*{6mm}
Actually this result follows for all $\varphi\in L^2(\R^3,\mathcal{H}_J)$ since the set   of all $\varphi$ of the above kind is total in  $L^2(\R^3,\mathcal{H}_J)$  as $\{J_k:k\in\R^3\setminus\{0\}\}$ is total in $\mathcal{H}_J$. So $P_\lambda$ converges strongly  to some projection $Q$.\\
\hspace*{6mm}
Finally show $Q\ne 0$. Let $\varphi_0(p):=\e^{-|p|^2}J_{k_0}$ for $k_0$ in (iii). Then $\norm{P_\lambda \varphi_0}^2=\int \e^{-|p|^2} |K(\lambda p, k_0)|^2\d^3p$ converges to 
$\norm{Q\varphi_0}^2$ for $\lambda\to\infty$. Hence $Q\varphi_0 \ne 0$ due to (iii).\qed

\begin{Cor}\label{PLSS} Let $K$ be from \emph{(\ref{SSL0})}.  Let $k_0$ satisfy \emph{(\ref{SSL0})(iii)}  and let $b\in\R^3$. Then 
$$\phi_n(p):=c_n \e^{-\i bp}\e^{-|p|^2/n^2}K(k_0,p)$$
 with $c_n>0$  the normalizing  constant  constitutes  a sequence of states localized at $b$ with respect to $T$. 
\end{Cor}

{\it Proof.} Consider at once $b=0$ \cite[(18)]{C17}.
Then for $\varphi_0=\e^{-|\cdot|^2}J_{k_0}$in the proof of   (\ref{SSL0}) one has $PD_\lambda^{-1}\varphi_0(p)=\lambda^{-3/2}\e^{-|p|^2/\lambda^2}K(p,k_0)J_p$, whence the claim by \cite[Theorem 7]{CL15}.
\qed

The foregoing criterion on the separateness of POL  is applied to the POL $T^{tct}$, $T^{TM}$, and POL with causal kernel.

\begin{Cor}\label{SPOL} The POL  with CT   $T^{TM}$   \emph{(sec.\,\ref{POLTM})} and $T^{tct}$   \emph{(sec.\,\ref{TCT})},
are separated. Every $k_0\in\R^3\setminus\{0\}$ yields a point localized sequence of states by \emph{(\ref{PLSS})}.
\end{Cor}

{\it Proof.} (a) Check easily  (\ref{SSL0})(i)-(iii) for $\textsc{t}^{TM}$  (\ref{KTM}). In particular  $\lim_{\lambda\to\infty}\textsc{t}^{TM}(k,\lambda p)=
\frac{1}{2}\big(1+\frac{kp}{\epsilon(k)|p|}\big)\ge \frac{1}{2}\big(1-\frac{|k|}{\epsilon(k)}\big)>0$ for all $k,p$.\\
(b) Similarly, $\textsc{t}^{tct}$  (\ref{KPOLCT}) satisfies   (\ref{SSL0})(i)-(iii). In particular $\lim_{\lambda\to\infty}\textsc{t}^{tct}(k,\lambda p)=\frac{1}{2}\big(1+m/\epsilon(k)\big)^{1/2}+\frac{1}{2|p|}\big(\epsilon(k)(m+\epsilon(k))\big)^{-1/2}kp>0$ for all $k,p$.
\qed\\

As to causal kernels recall (\ref{KKK}) and the remarks following it. One has

\begin{Lem} 
\begin{itemize}
\item[\emph{(a)}] There is no causal kernel $K$, which satisfies the condition \emph{(\ref{SSL0})(ii)}.
\item[\emph{(b)}] Let $K$ be a map  \emph{(\ref{KKK})}, where $g:[m^2,\infty[\to\C$ is continuous. Then $K$ satisfies  \emph{(\ref{SSL0})(ii)} if and only if $g=g_{1/2}$. In view of  \emph{(\ref{SSL0})(i),(iii)} check $\lim_{\lambda\to\infty}K_{1/2}(k, \lambda p)=m\big(2 \epsilon(k)(\epsilon(k)-\frac{1}{|p|}kp)\big)^{-1/2}$.
\end{itemize}
\end{Lem}

{\it Proof.}
 Let  $m=1$.  
(b) Check (\ref{SSL0})(i)-(iii) for $K_{1/2}$. Now, conversely let  $p\ne 0$, $\alpha\ge 1$. Check  $\lim_{\lambda\to \infty} (\epsilon(\lambda p) \epsilon(\alpha\lambda  p)-\alpha \lambda^2 |p|^2 )=\frac{1+\alpha^2}{2\alpha}$ and $\lim_{\lambda\to \infty} \frac{\epsilon(\lambda p)+\epsilon(\alpha\lambda p)}{2\sqrt{\epsilon(\lambda p)\epsilon(\alpha\lambda p)}}=\frac{1+\alpha}{2\sqrt{\alpha}}$. Fix $t\in [1,\infty]$. Then $t=\frac{1+\alpha^2}{2\alpha}$ for $\alpha=t+\sqrt{t^2-1}$, and for this $\alpha$ one has $\frac{1+\alpha}{2\sqrt{\alpha}}=
 \sqrt{(1+t)/2}$. Hence continuity of $g$ and condition (\ref{SSL0})(ii) imply $ \sqrt{(1+t)/2} \,g(t)=1$, as asserted.\\
\hspace*{6mm} (a) follows from (b), since $K_{1/2}$ is not a positive definite kernel. \qed\\

Hence a POL $T$ of the massive scalar boson  with causal kernel  (\ref{KKK}) is a candidate for a non-separated POL, i.e., $\norm{T(B)}<1$ might hold for some small  ball $B\ne \emptyset$.

\section{Discussion}\label{D}
A POL can be regarded as  a  physically accessible data  set, namely the set of the probabilities of localization of the particle in every state  in every region. Therefore among the infinitely many POL for a massive scalar boson (\ref{APOL}) only a particular one describes truly the position of the boson. The question is how to identify this POL.\\
\hspace*{6mm}
Many, as e.g.  NWL, are ruled out by the requirement of causality CC (sec.\,\ref{CKD}). The remaining causal POL are still infinitely many. We studied thoroughly  the causal POL  related to a conserved density current analyzing the set $\mathcal{K}$ of causal kernels (\ref{KKK}).\\
\hspace*{6mm}
In search of the best POL we recall the observation that $KG\in\mathcal{K}$ if $K\in\mathcal{K}$ and $G\in\mathcal{G}$ (\ref{LIPGK}). Probably multiplying $K$ by  $G$ deteriorates the localization features of the corresponding  POL. More generally, if $K,K'\in\mathcal{K}$ with $K'\le K$, then $K$ is better than $K'$? 
Take as evidence the fact $K'\le K\le \textbf{1}$ with $\textbf{1}$ the kernel of NWL distinguished by its  abundance of  boundedly localized states. If it were true, the POL    with kernel $K_{3/2}$, being the greatest element (\ref{MTKK}),  would be the only valid POL.\\
\hspace*{6mm}
For the localization features of the POL $T_{3/2}$ with kernel $K_{3/2}$ it is decisive to know whether  $||T_{3/2}(B)||=1$ for all balls $B\ne \emptyset$. Assume this. Then  $T_{3/2}$ is separated (sec.\,\ref{LBR}) and  for every point $b\in \R^3$ there is a sequence of  states localized at $b$ (\ref{SSL}), which is  very satisfactory from a physical point of view.\\
\hspace*{6mm}
 However, also the case  $||T_{3/2}(B_*)||=\delta <1$  for some ball $B_*\ne\emptyset$ were rather interesting. Assume this. By translation covariance $B_*$ may be centered at $0$.
 The probability for the boson of being localized in $B_*$ is at most  $\delta$ in every state. Note $||T_{3/2}(mB_*)||\to 1$ for $m\to \infty$ since $mB_*\uparrow_m\R^3$. \\
\hspace*{6mm} 
 Now let $T^m$ denote the POL for the boson with mass $m>0$ with kernel  (\ref{KKK}). $T$ means $T^1$. It is easy to check
 $$ T(m\Delta)=D_mT^m(\Delta)D_m^{-1}$$
for all regions $ \Delta$ with $(D_m\phi)(p):=m^{3/2} \phi(mp)$ unitary on $L^2(\R^3)$. Hence $||  T(m\Delta)||=||T^m(\Delta)||$, whence in particular $||T^m_{3/2}(B_*)||\to 1$ for $m\to \infty$.\\
\hspace*{6mm}
 The  physically reasonable result would be that a more massive scalar boson is better localizable in bounded regions than a less massive one.\\

\textbf{Acknowledgements} 

 I am indebted to R.F. Werner having drawn my attention to this topic and agreeing to quote  his personal communications in this regard, and I am very grateful to V. Moretti  for  valuable discussions and suggestions, and above all because of  his work \cite{M23} which enabled me to accomplish this work.

 \appendix

\section{Criterion for CT} \label{AppB}

This section is concerned with a criterion for  CT.
 It is based on the existence of a nonspacelike  conserved current for the probability density of localization. Originally this is an approach due to Petzold et al. \cite{GGP67} when treating  their  covariant kernels of the density and proving CT for the spatial probability \cite{GGPR68}. Moretti  \cite{M23} improves the technique  considerably showing rigorously CT for the POL  proposed by Terno  \cite{T14}. From Moretti's proof  we distill  the  general criterion for CT   (\ref{GCCT}).\\ 
\hspace*{6mm}
The following lemma, which is an extension of a result in   \cite{M23}, essentially says that CT and  CC, respectively, 
already holds, if it holds for   regions  $\Delta$ being the union of finitely many disjoint closed balls.

\begin{Lem} \label{MLCC}
Let $\varepsilon=\{0\}\times \R^3\equiv \R^3$ be the standard spacelike hyperplane and  let $\sigma\subset \R^4$ be any other spacelike hyperplane  of Minkowski space. Let $T$ and $T'$ be POM on $\varepsilon$ and $\sigma$, respectively, vanishing at  Lebesgue null sets. Let $V$ be a fixed   open subset of $\R^3$. Suppose that 
\begin{equation}
T(\Delta)\le T'(\Delta_\sigma) \tag{1}
\end{equation}
holds for every $\Delta\subset V$ being a union of finitely many disjoint closed balls. Then \emph{(1)} holds for all  measurable $\Delta\subset V$.
 \end{Lem}

 {\it Proof.} Provide  a total set of unit vectors $\{\phi_j:j\in\N\}$. Then $\mu(\Delta):=\sum_j 2^{-j}\langle \phi_j,T(\Delta)\phi_j\rangle$ defines a probability measure controlling $T$.
The latter means that $$\mu(\Delta^{(n)})\to 0 \Leftrightarrow T(\Delta^{(n)})\to 0 \text{ strongly}$$ for every sequence $(\Delta^{(n)})$ in $\mathcal{L}^d$. Plainly, $\Leftarrow$ holds by dominated convergence. As to $\Rightarrow$, note $||T(\Delta)\phi||^2\le ||\sqrt{T(\Delta)}\sqrt{T(\Delta)}\phi||^2\le ||\sqrt{T(\Delta)}\phi||^2=\langle\phi,T(\Delta)\phi\rangle$.  
One recalls that $\mu$ is a tight Radon measure (see e.g. \cite[Chapter 2 (67)]{CR08}). --- $\mu'$ with respect to $T'$ is defined quite analogously.\\
\hspace*{6mm} Henceforth all sets are  subsets of $V$.

\hspace*{6mm}
(a) Let (1) hold for $\Delta^{(n)}$ with $\Delta^{(n)} \uparrow\Delta$.  Then (1) holds for $\Delta$. Indeed, note first  $\Delta^{(n)}_\sigma \uparrow\Delta_\sigma$. Since
$\mu(\Delta\setminus\Delta^{(n)})\to 0$, one has $T(\Delta^{(n)})=T(\Delta)-T(\Delta\setminus\Delta^{(n)})\to T(\Delta)$. Similarly  $T'(\Delta_\sigma^{(n)})\to T'(\Delta_\sigma)$ follows.\\
\hspace*{6mm}
(b)  Let (1) hold for $\Delta^{(n)}$ with $\Delta^{(n)} \downarrow\Delta$ and suppose that $\Delta_\sigma^{(n)} \downarrow\Delta_\sigma$. Then, arguing as in (a) analogously, (1) holds for $\Delta$.\\
\hspace*{6mm}
 (c) Now  consider an open bounded set $U$. Then according to Vitali's covering theorem \cite[2.8 Theorem]{M95} there  are countably many disjoint closed balls $B^{(i)}\subset U$ with $\mu(U\setminus C)=0$ for $C:=\bigcup_iB^{(i)}$. By assumption (1) holds for $C^{(n)}:=\bigcup_{i=1}^nB^{(i)}$, whence (1) holds for $C$ by (a). Moreover, since $\mu(U\setminus C^{(n)})\to 0$, one has $T(U\setminus C^{(n)})\downarrow 0$. Therefore, $T(U)=T(C)+T(U\setminus C) \le T'(C_\sigma)+T(U\setminus C)\le T'(U_\sigma)+T(U\setminus C^{(n)})\to  T'(U_\sigma)$. Hence (1) holds for $U$.\\
\hspace*{6mm}
(d) Next let $K$ be compact. Choose bounded open sets $U^{(n)}$ satisfying $\overline{U^{(n+1})}\subset U^{(n)}$, $U^{(n)}\downarrow K$. Then $U^{(n)}_\sigma\downarrow K_\sigma$.
 Hence (1) holds for $K$ by (b).\\
\hspace*{6mm}
(e) Finally let $\Delta$ be any measurable set. Since $\mu$ is tight, there are compact $K^{(i)}\subset \Delta$ satisfying $\mu(\Delta\setminus C)=0$ for $C:=\bigcup_i K^{(i)}$. By (d), (1) holds for $C^{(n)}:=\bigcup_{i=1}^nK^{(i)}$. Now proceed as in (c). Hence (1) holds for $\Delta$.\qed\\

 The applications  of (\ref{MLCC}) in (\ref{GCCT}) and (\ref{GCCPOL}) regard the case $T'=T$ as they  concern  a POL $T$ of the massive scalar boson, which due to the Poincar\'e covariance  (\ref{PCEPOL}) is defined on all spacelike hyperplanes. Moreover the particular case  in (\ref{MLCC})  of parallel hyperplanes $\varepsilon$ and $\sigma=\{t\}\times \R^3$ for $t\in\R$  and for  $V=\R^3$ is used for the CT criterion below.

\begin{Def}\label{DCPOLK} Let $\textsc{t}$ be the kernel of POL $T$ of a massive scalar boson.  
Suppose that $\textsc{t}$ is continuous. 
$\textsc{t}$ is said to be (i) conserved, (ii)  timelike definite,
  if there are continuous Hermitian  kernels $\textsc{j}_i:\R^3\setminus\{0\}\times\R^3\setminus\{0\}\to \C$, $i=1,2,3$ satisfying respectively
\begin{itemize}
\item [(i)] $\big(\epsilon(k)-\epsilon(p)\big)\textsc{t}(k,p)=\sum_{i=1}^3(k_i-p_i)\textsc{j}_i(k,p)$ \, for all $k,p\ne 0$ 
\item [(ii)] $\big(\sum_{a,b=1}^n\overline{c}_ac_b\textsc{t}(p_a,p_b)\big)^2\ge \sum_{i=1}^3\big(\sum_{a,b=1}^n\overline{c}_ac_b\textsc{j}_i(p_a,p_b)\big)^2$ 
 for all $(c_1,\dots,c_n)\in\C^n$, $(p_1,\dots,p_n)\in(\R^3\setminus\{0\})^n$, $n\in\N$
\end{itemize}
\end{Def} 

From (\ref{DCPOLK})(ii) it follows that $|\textsc{j}_i(p,p)|\le 1$, $ \textsc{t}-\textsc{j}_i$ is positive definite, and in particular boundedness $|\textsc{j}_i(k,p)|\le 3$ for  $k,p\ne 0$, $i=1,2,3$.

\begin{The}\label{GCCT} Let $T$ be  a POL of a massive scalar boson.  Suppose that its kernel $\textsc{t}$ is continuous and  conserved timelike definite. Then  time evolution is causal.
\end{The}

{\it Proof.} We treat the particle case $\eta=+$ (the antiparticle case is analogous) and use  the shell rep in sec.\,\ref{CTSR}.  Let $\d{o}(\mathfrak{p})$ denote the Lorentz invariant measure on the mass shell. Put $\textsc{j}_0:=\textsc{t}$. Define for $i=0,1,2,3$
\begin{equation}\label{FVCC}
 J_i(\Phi,\mathfrak{x}):=(2\pi)^{-3}\int\int 
  \,   \sqrt{\epsilon(k)\epsilon(p)}    \textsc{j}_i(k,p)\,       \e^{\i(\mathfrak{k}-\mathfrak{p})\cdot\mathfrak{x}}\overline{\Phi(\mathfrak{k})}\Phi(\mathfrak{p})\,\d o(\mathfrak{k}) \d o(\mathfrak{\mathfrak{p}})
  \end{equation}
  for $\mathfrak{x}\in\R^4$, and $\Phi$ square integrable and $\sqrt{\epsilon}\, \Phi$ integrable. Recall that $\textsc{j}_i$ is bounded. Put $J:=(J_1,J_2,J_3)$ and $\mathfrak{J}:=(J_0,J)$.
  \begin{itemize}
\item[(a)] {\it Let $\Delta\subset\R^3\equiv \{0\}\times\R^3$ be measurable, $t\in\R$. Then $\langle W(t)^{-1}\Phi,T(\Delta)W(t)^{-1}\Phi\rangle=
\langle \Phi, W(t)T(\Delta)W(t)^{-1}\Phi\rangle=\langle \Phi,T(\{t\}\times\Delta)\Phi\rangle$. The latter  equals the surface integral $\int_{\{t\}\times\Delta}J_0(\Phi,\mathfrak{x}) \d S(\mathfrak{x})$.}
\end{itemize} 
Indeed, by (\ref{POLSR}) $\langle W(t)^{-1}\Phi,T(\Delta)W(t)^{-1}\Phi\rangle=\int_\Delta J_0(W(t)^{-1}\Phi,(0,x))\d^3x$. Since $\big(W(t)^{-1}\Phi\big)(\mathfrak{p})=\e^{-\i (t,0)\cdot \mathfrak{p}}\Phi(\mathfrak{p})$, the former equals $\int_\Delta J_0(\Phi,(t,x))\d^3x$, whence the claim.
\begin{itemize}
\item[(b)] $\operatorname{div}\, \mathfrak{J}(\Phi,\cdot)=0$
\end{itemize}
This follows immediately from (\ref{DCPOLK})(i).
\begin{itemize}
\item[(c)] $J_0(\Phi,\mathfrak{x})\ge |J(\Phi,\mathfrak{x})|$ 
\end{itemize}
Put $\phi(p):=\Phi\big(\epsilon(p),p\big)$. Clearly it suffices to prove the claim for $\Phi$, for which $\phi$ is continuous with compact support in $\R^3\setminus\{0\}$. This  follows from (\ref{DCPOLK})(ii) just approximating the integrals in (\ref{FVCC})  by appropriate Riemann sums. 

\hspace*{2mm} 
(d) Now following \cite{M23} the main step toward the result is an application of the divergence theorem, i.e., Gauss's or Ostrogradsky's theorem. Assume $t>0$. The case $t<0$ is analogous. Let $B\subset \R^3$ be a closed ball of radius $R$ and with center at the origin $0$. Let $M$ be the compact subset of $\R^4$, the boundary of which is the union of the base and top  $\{0\}\times B_t$, $\{t\}\times B$, and the lateral surface $L:=\{(s,x)\in \R^4:|R+t-s|=|x|,0\le s\le t\}$, which is the piece of the past  light cone with apex  $(R+t,0)$ between the base and the top. Note that $\mathfrak{n}(\mathfrak{x}):=\frac{1}{\sqrt{2}}\big(1,x/|x|\big)$ is  outward-pointing orthonormal at $\mathfrak{x}=(s,x)\in L$.
\\
\hspace*{6mm}
By (b) the  volume integral $\int_M \operatorname{div}\mathfrak{J}(\Phi,\mathfrak{x}) \,\d  V(\mathfrak{x})$ vanishes, whence using (a) one has 
\begin{equation*}
\langle \Phi, W(t)T(B)W(t)^{-1}\Phi\rangle- \langle \Phi, T(B_t)\Phi\rangle=-  \int_L \mathfrak{J}(\Phi,\mathfrak{x})\,\mathfrak{n}(\mathfrak{x}) \d S(\mathfrak{x})\le 0\tag{1}
\end{equation*}
since $\sqrt{2}\, \mathfrak{J}(\Phi,\mathfrak{x})\,\mathfrak{n}(\mathfrak{x})=J_0(\Phi,\mathfrak{x})+J(\Phi,\mathfrak{x})\,\frac{x}{|x|}\ge 0$ by (c).
So the causality condition  (\ref{CTEPOL}) holds for $B$ and hence due to translation covariance for any ball. \\
\hspace*{6mm}
We conclude the proof still following \cite{M23}. In view of (\ref{MLCC})
it suffices to verify (\ref{CTEPOL}) for $\Delta$ being the union of finitely many disjoint closed balls $B_i$. The boundary of the compact set $M:=\bigcup M_i$ is the union 
of the base and top  $\{0\}\times \Delta_t$, $\{t\}\times \Delta$, and a piecewise smooth lateral surface $L$. Note $L\subset\bigcup L_i$. So, as $L$ is composed by finitely many pieces of light cones, (1) holds true also for $\Delta$ in place of $B$.\qed\\

\section{POL with conserved covariant kernel satisfies CC}\label{AppC}

Extending Moretti's  method we show the result  on causality  (\ref{GCCPOL})  based on the existence of a  conserved covariant current for the probability density of localization. The corollary (\ref{CCKKK}) on conserved covariant POL kernels for a massive scalar boson  essentially is due to Petzold et al. \cite{GGP67}.

Put  $m=1$ without restriction, and let     $\mathfrak{p}=\big(\epsilon(p),p\big)$ with $\epsilon(p)=\sqrt{1+|p|^2}$ for $p\in\R^3$.

\begin{Def} \label{LCFV} $\mathfrak{v}: \R^3\setminus \{0\}\times R^3\setminus \{0\} \to \C^4$ is called  (i) conserved, (ii) (Lorentz) covariant, if 
\begin{itemize}
\item[(i)]  $(\mathfrak{k}-\mathfrak{p}) \cdot \mathfrak{v}(k,p)=0$
\item[(ii)]  $L\,\mathfrak{v}(k,p)=\mathfrak{v}(L\cdot k,L\cdot p)$ for all $p,k\ne 0$, $L\in O(1,3)_0$
\end{itemize}
Here $L\cdot p$ denotes the spatial vector part of $L\mathfrak{p}$.
\end{Def}

\begin{Lem}\label{PLCFV} Let $\mathfrak{v}: \R^3\setminus\{0\}\times \R^3\setminus\{0\} \to \C^4$ be covariant. Then
\begin{itemize}
\item[\emph{(a)}]  $v_0$ determines uniquely  $\mathfrak{v}$.
\item[\emph{(b)}] there are unique  $\alpha,\beta: [1,\infty]\to  \C$ such that 
$\mathfrak{v}(k,p)=\alpha(\mathfrak{k}\cdot \mathfrak{p}) \,\mathfrak{k} + \beta(\mathfrak{k}\cdot \mathfrak{p})\, \mathfrak{p}$.
\item[\emph{(c)}] $\mathfrak{v}$ is conserved, iff  \,   $\mathfrak{v}(k,p)=\gamma(\mathfrak{k}\cdot \mathfrak{p}) \,(\mathfrak{k} + \mathfrak{p})$ for some unique $\gamma:[1,\infty[\to \C$.
\item[\emph{(d)}] $v_0$ is Hermitian, iff \,$\mathfrak{v}(k,p)= g(\mathfrak{k}\cdot \mathfrak{p})(\mathfrak{k} + \mathfrak{p}) +\i f(\mathfrak{k}\cdot \mathfrak{p})(\mathfrak{k} - \mathfrak{p})$ with unique  $g,f:[1,\infty[\to \R$.
\item[\emph{(e)}] $\mathfrak{v}$ is conserved if  $v_0$ is real symmetric.
\item[\emph{(f)}] if $v_0$ is positive definite, then $v_0$ is even  timelike definite (see \emph{(\ref{DCPOLK})(ii)}). Cf. \emph{\cite[sec.\,3]{GGPR68}}.
\end{itemize}
\end{Lem}

{\it Proof.} (a) Verify $v_i(k,p)=\sqrt{2}\,v_0(p,k) - v_0\big(L_i^{-1}\cdot k, L_i^{-1} \cdot p \big)$ with 
$L_i$  the boost along the $\i$th     direction with velocity $1/\sqrt{2}$ evaluating the equation $v_0(k,p)=L(1,0,0,0)\,\cdot \mathfrak{v}(L\cdot k,L\cdot p)$ for $L=L_i$.\\
\hspace*{6mm} (b)  It is no restriction to assume that $\mathfrak{v}$ is real, since covariance does not mix the real and the imaginary part. ---  Consider first the case $k=p$. 
 By covariance the fixed point subgroup of $\mathfrak{p}$ leaves $\mathfrak{v}(p,p)$ invariant. Therefore $\mathfrak{v}(p,p)=\gamma(p)\mathfrak{p}$ for some scalar $\gamma(p)\in\C$. Then covariance implies that $\gamma$ is constant. ---  Now let $k\ne p$.
By a Lorentz transformation map $\mathfrak {k}$ to $(1,0,0,0)$ and $\mathfrak{p}$ to $(a',0,0,b')$ with $a',b'\ge 0$ and subsequently by the boost in the third direction, $(1,0,0,0)$ to $(\sqrt{2},0,0,1)$ and  $(a',0,0,b')$ to $(a,0,0,b)$. So there is  $L\in O(1,3)_0$  with  $L\mathfrak{k}=(\sqrt{2},0,0,1)$,  $L\mathfrak{p}=(a,0,0,b)$. One has 
 $b=b(\mathfrak{k}\cdot \mathfrak{p})$ for 
$b(t):=t+\sqrt{2}\sqrt{t^2-1}$ and similarly for $a=a(\mathfrak{k}\cdot \mathfrak{p})$.
Hence $L\mathfrak{v}(k,p)=
\mathfrak{s}(\mathfrak{k}\cdot \mathfrak{p})$ with $\mathfrak{s}(t):=\mathfrak{v} \big((0,0,1),(0,0,b(t))\big)$. Note $b\ne 1$ since $k\ne p$ and, by covariance, $R\,\mathfrak{s}(t)=\mathfrak{s}(t)$
 for all rotations $R$ around the third axis. Therefore $\mathfrak{s}(t)=\alpha(t)(\sqrt{2},0,0,1)+\beta(t)(a(t),0,0,b(t))$ with unique $\alpha(t),\beta(t)$, whence the result.\\
\hspace*{6mm}
(c), (d), and (e) follow immediately from (b).\\
\hspace*{6mm} (f) Since $v_0$ is positive definite,  the zero component of the four vector 
$$\mathfrak{z}:=\sum_{a,b}\overline{c}_ac_b\,\mathfrak{v} (p_a, p_b)$$
is nonnegative for every choice of $(c_1,\dots,c_n)$ and $(p_1,\dots,p_n)$. Consider any Lorentz transformation  $L\in O(1,3)_0$. Put $\tilde{\mathfrak{z}}:=L\mathfrak{z}$, $\tilde{\mathfrak{p}}_a:=L\mathfrak{p}_a$. Then $\tilde{\mathfrak{z}}=\sum_{a,b}\overline{c}_ac_b\, \mathfrak{v}(\tilde{p}_a,\tilde{p}_b)$ by Lorentz covariance,
whence the zero component of $L\mathfrak{z}$ is nonnegative.  \qed

\begin{Cor}\label{CCKKK}  Let $\textsc{t}$ be the kernel of a POL of a massive scalar boson. Let $\textsc{t}$  be conserved covariant, which means that there is a  conserved covariant   $\mathfrak{v}$ with $v_0(k,p)=\textsc{t}^{shell} (\mathfrak{k},\mathfrak{p}) =\sqrt{\epsilon(k)\epsilon(p)}\textsc{t}(k,p)$.
Then there  is a continuous $g:[1,\infty] \to \R$ with $g(1)=1$ such that 
\begin{equation}\label{CKKK}
\textsc{t}(k,p) =\frac{\epsilon(k)+\epsilon(p)}{2\sqrt{\epsilon(k)\epsilon(p)}}\,\,g(\mathfrak{k}\cdot \mathfrak{p})
\end{equation}
The right hand side is continuous on $\R^3\times\R^3$ and by continuity it stays positive definite on $\R^3$. The extension is still denoted by $\textsc{t}$. \\
\hspace*{6mm}
Conversely let $g:[1,\infty] \to \R$  be measurable with $g(1)=1$ such that the right hand side of \emph{(\ref{CKKK})} is positive definite. Then the latter is a conserved covariant POL  kernel and $g$ is continuous.
\end{Cor}\\
{\it Proof.} $g$ is continuous by   (\ref{CRIK}). For the second part of the assertion recall   (\ref{CKPOLMB}). \qed

\begin{The}\label{GCCPOL} Let $T$ be  a POL of a massive scalar boson.  Suppose that its kernel $\textsc{t}$ is   conserved covariant \emph{(\ref{CKKK})}. Then  $T$ is causal.
\end{The}

{\it Proof.} By (\ref{CCKKK}),  (\ref{PLCFV})(f)  $\textsc{t}$  is continuous timelike definite. So  (\ref{GCCT}) applies with $v_i(k,p)= \sqrt{\epsilon(k)\epsilon(p)}    \textsc{j}_i(k,p)$. Recall  the beginning of the proof of (\ref{GCCT}), in particular (\ref{FVCC}).
\begin{itemize}
\item[(a)] {\it Let $\Gamma\in\mathfrak{S}$. Then 
\begin{equation}\label{SFI}
\langle \Phi,T(\Gamma)\Phi\rangle =\int_\Gamma \mathfrak{J}(\Phi, \mathfrak{x})\,\mathfrak{n}_\Gamma\,\d S(\mathfrak{x})
\end{equation}
Regarding the surface integral there is the (Euclidean)  scalar product of $\mathfrak{J}$ with    the (Euclidean) future-pointed orthonormal $\mathfrak{n}_\Gamma$ on $\Gamma$.}
\end{itemize}
Indeed, there is $g=(\mathfrak{a},A)\in\tilde{\mathcal{P}}$ and $\Delta\subset \R^3\equiv \{0\}\times\R^3$ with $\Gamma=g\cdot \Delta$. Then $\langle \Phi,T(\Gamma)\Phi\rangle =\langle W(g)^{-1}\Phi,T(\Delta) W(g)^{-1}\Phi\rangle =\int_\Delta J_0(W(g)^{-1}\Phi,(0,x))\d^3x$. Inserting  $\big(W(g)^{-1}\Phi\big)(\mathfrak{p})=\e^{-\i L^{-1}\mathfrak{a}\,\cdot\mathfrak{p}}\,\Phi(L\mathfrak{p})$, $L:=\Lambda(A)$, in (\ref{FVCC}) for $i=0$  one obtains $J_0(W(g)^{-1}\Phi,(0,x))=(2\pi)^{-3}\int\int \textsc{j}_0^{shell}(L^{-1}\mathfrak{k},L^{-1}\mathfrak{p}) \e^{\i (\mathfrak{k}-\mathfrak{p})\cdot(g\cdot (0,x)) } \overline{\Phi(\mathfrak{k)}} \Phi(\mathfrak{p)} \d o(\mathfrak{k}) \d o(\mathfrak{\mathfrak{p}})$
by Lorentz invariant integration. Now, since $\textsc{t}$ is  covariant, the latter becomes  $\big(L^{-1}\mathfrak{J}(\Phi,g\cdot (0,x))\big)_0$, and further using the normal $\mathfrak{e}=(1,0,0,0)$, $=\big(L^{-1}\mathfrak{J}(\Phi,g\cdot (0,x))\big)\cdot \mathfrak{e}=\mathfrak{J}(\Phi,g\cdot (0,x))\cdot \mathfrak{e}_L$, $ \mathfrak{e}_L:= L\mathfrak{e}$. Hence one has  $\langle \Phi,T(\Gamma)\Phi\rangle =\int_\Delta \mathfrak{J}(\Phi,g\cdot (0,x))\cdot  \mathfrak{e}_L \,\d^3x $. The Minkowski product by  $\mathfrak{e}_L$ becomes the scalar product by $\mathfrak{e}^-_L$ 
with the negative spatial part of $\mathfrak{e}_L$. Note that $\mathfrak{e}^-_L$ is future-oriented orthogonal to $\Gamma$. Therefore (\ref{SFI}) holds.

\hspace{6mm}
(b) Let $\Delta\in\mathfrak{S}$ and let $\sigma$ be a spacelike hyperplane. We are going to show  causality  
\begin{equation}
T(\Delta)\le T(\Delta_\sigma)\tag{1}
\end{equation}
Due to the Poincar\'e covariance of $T$ it suffices to treat the case  $\Delta\subset \varepsilon=  \{0\}\times\R^3 \equiv \R^3$ and $\sigma=\{\mathfrak{x}\in\R^4: x_0=\alpha x_3\}$ for $0<\alpha<1$. This excludes only the case of parallel hyperplanes $\varepsilon$ and  $\sigma$ already treated in (\ref{GCCT}).\\
\hspace*{6mm}
Put $H:=\{x\in\R^3:x_3=0\}$,    $H^\pm:=\{x\in\R^3:x_3\gtrless0\}$ , $\Delta^{\pm}:=\{x\in\Delta:x_3\gtrless 0\}$. Note that $H$ is a null set and that  $\Delta^\pm_\sigma\subset H^\pm$.
So it suffices to have (1) for $\Delta^+$ and $\Delta^-$, since, assuming this, $T(\Delta)=T(\Delta^-\cup \Delta^+)=T(\Delta^+)+T(\Delta^-)\le T(\Delta^+_\sigma)+T(\Delta^-_\sigma)=T(\Delta^+_\sigma \cup \Delta^-_\sigma)\le T(\Delta_\sigma)$.  Therefore by the reduction achieved so far one may assume in (1) that $\Delta\subset H^+$ or $\Delta\subset H^-$. We do the case $\Delta\subset H^+$ applying (\ref{MLCC}) for $ V:=H^+$. The case $\Delta\subset H^-$ is  analogous.\\
\hspace*{6mm}
First let $\Delta$ be a closed ball $B$ with center $(0,0,\beta)$ and  radius $R>0$ so that $0<R<\beta$. Let $M$ be the compact subset of $\R^4$, the boundary of which is the union of the base and top $B$, $B_\sigma$ and the lateral surface $L$, which is the piece of the future light cone with apex  $(-R,0,0,\beta)$ between the base and the top. Exploit the  volume integral $\int_M \operatorname{div}\mathfrak{J}(\Phi,\mathfrak{x}) \,\d  V(\mathfrak{x})$ as in part (d) of the proof of (\ref{GCCT}) applying the formula (\ref{SFI}). The relation (1) for $\Delta=B$  follows. Now achieve the result still following the proof of (\ref{GCCT}).\qed

\section{On functions of positive type} \label{EQOD}

Here some results on functions of positive type are provided. 

 \begin{Lem}\label{PDDOO3} 
\emph{(a)} Let $g$ be from \emph{(\ref{KKOSD})}, i.e., $g:[1,\infty[\to\R$ measurable with $g(1)=1$. The three properties
 $(\alpha)$ $\tilde{g}$ is positive definite\footnote{Recall $\tilde{g}(k,p)=g\big(\epsilon(k)\epsilon(p)-kp\big)$  after (\ref{KKOSD}).}, $(\beta)$ $(x,y)\to g\big(\cosh(x-y)\big)$ is  positive definite, i.e., $g\circ\cosh$ is a function of positive type, and $(\gamma)$ $g\circ \cosh$ equals a.e. the Fourier transform of an even finite measure on $\R$, are equivalent.\footnote{ Regarding functions of positive type which are not continuous see Problem 18. in \cite[Chapter IX]{RS75}.}\\
\hspace*{6mm}
\emph{(b)}  Let $k\in L^1(\R)$ with its Fourier transform $\hat{k}\in L^1(\R)$, $\hat{k}\ge 0$; then $k$ is even and $\tilde{g}$ for $g:=k\circ \cosh^{-1}$  is positive definite.\footnote{$\cosh^{-1}(t) =\ln\big(t+\sqrt{t^2-1}\,\big)), \,t\ge 1$}\\
 \hspace*{6mm}
\emph{(c)}   Consider $ k\in L^1(\R)\cap L^2(\R)$. Then $k$ is equal a.e. to a continuous function of  positive type if and only if its Fourier transform is real nonnegative.\\
 \hspace*{6mm}
\emph{(d)}  Let  $ k\in L^1(\R)$ be  a continuous function of  positive type. Then  its Fourier transform $\hat{k}$  is integrable and nonnegative. 
\end{Lem} 

{\it Proof.} (a) $(\alpha)$ $\Leftrightarrow$ $(\beta)$ is obvious.  $(\beta)$ $\Leftrightarrow$ $(\gamma)$  holds by \cite[Proposition after Theorem IX.10 and Theorem IX.9 (Bochner's theorem)]{RS75}. 

(b)  holds also by Bochner's theorem.

(c) Suppose $k \in L^1\cap L^2$.
For $u,v$ in the Schwartz space $\mathcal{S}$, consider  $\phi:=\mathcal{F}u$, $\psi:=\mathcal{F}v$ in  $\mathcal{S}$. By \cite[21.32]{HS69},  $k\ast \psi\in L^2$ and, by  \cite[21.54(a)]{HS69}, $k\ast \psi=\mathcal{F}\big((\mathcal{F}^{-1}k)(\mathcal{F}^{-1}\psi)\big)$. 
Hence $\langle\phi,k\ast \psi\rangle=\langle\mathcal{F}^{-1}\phi,\mathcal{F}^{-1}(k\ast \psi)\rangle=\langle u, (\mathcal{F}^{-1}k)v\rangle$. For $v=u$ this implies $(2\pi)^{-1/2}\int\int k(x-y)\overline{\phi(y)}\phi(x)\d x\d y=\int |u(x)|^2\,(\mathcal{F}^{-1}k)(x)\d x$.  The right hand side is nonnegative for all $u$ if and only if $\mathcal{F}^{-1}k\ge 0$ a.e. Therefore, looking at the left hand side, $k$ is of weak positive type if and only if $\mathcal{F}^{-1}k\ge 0$ a.e. or equivalently $\mathcal{F}k\ge 0$ a.e.  Recall \cite[Proposition after Theorem IX.10]{RS75}.

(d) Since $k$ is bounded by $k(0)\ge 0$,  $k$ is also square integrable so that (c) applies. So $\hat{k}\ge 0$.  Note $k=\mathcal{F}^{-1}\hat{k}$. We introduce the measure $\nu:=\hat{k}\,m$, where $m$ denotes the Lebesgue measure on $\R$. Since $\nu([-D,D])\le (2D)^{1/2}\norm{\hat{k}}_2$, the measure $\nu$ is polynomially bounded \cite[V: Example 4]{RS75} so that $S(\varphi):=\int \varphi \d m$ defines a tempered distribution. Its Fourier transform is given by $\hat{S}(\varphi)=S(\hat{\varphi})=\int \hat{\varphi}\d \nu=\int \hat{\varphi}\,\hat{k}\d m=\int \varphi \,(\mathcal{F}^{-1}\hat{k})\,\d m=\int \varphi\,k\,\d m$.\\
\hspace*{6mm}
On the other hand by \cite[Theorem IX.9 (Bochner's theorem)]{RS75}, there is a finite measure $\mu$ on $\R$ such that its Fourier transform $\hat{\mu}$ equals $k$. This means that the Fourier transform of the tempered distribution given by $T(\varphi):=\int \varphi\,\d \mu$ reads $\hat{T}(\varphi)=\int \hat{\mu}\,\varphi \,\d m=\int \varphi\,k\,\d m$ \cite[after Theorem IX.8]{RS75}. \\
\hspace*{6mm}
Thus $\hat{S}=\hat{T}$, whence $S=T$ \cite[Theorem IX.2]{RS75}. This obviously means $\nu=\mu$ implying $\hat{k}\in L^1(\R)$.
\qed\\

\section{Proofs concerning $T^{tct}$}\label{AppA} 

{\it Proof of} (\ref{KPOLCT}),\,(\ref{POLTCT}). First recall (\ref{FKDNWL}). --- The second part of the assertion  follows from \cite[sec.\,E, Cor. 5,6,7]{CL15}  (for a concise summary see \cite[sec.\,28.1]{C17}) and   \cite[sec.\,F, Cor. 9]{CL15}. Accordingly, $(W^{m,0,+},T^{tct})$ is obtained as the trace of the 2.nd system with CT $(W_2,\mathcal{F}E^{can}\mathcal{F}^{-1})$ on the carrier space of the subrep   $W^{m,0,+}$ of  the
 rep $W_2$ of $\tilde{\mathcal{P}}$ on  $L^2(\R^3,\C^4\otimes \C^2)$.  For $(t,b,A) \in \tilde{\mathcal{P}}$  
 the latter is given by
\begin{equation}\label{IMCS}
\big(W_2(t,b,A)\varphi\big)(p) :=\operatorname{e}^{-\operatorname{i}bp}
\sum_{\eta =\pm 1}\left( \operatorname{e}^{\operatorname{i}t\eta\epsilon(p)}  \pi^\eta(p)s(A)^{*-1} \otimes R(\mathfrak{p}^\eta,A)  \right)    \varphi(q^\eta)
\end{equation}
for $p\ne 0$, where $s(A):=\operatorname{diag}(A,A^{* -1})$, $\pi^\eta(p):=\frac{1}{2}\big(I_2+\frac{\eta}{\epsilon(p)}h(p)\big)$, and $ h(p):=\sum_{k=1}^3\alpha_k p_k + \beta m$ with $\beta,\alpha_k$  the Dirac matrices  in the Weyl representation (see also \cite[sec.\,13 and sec.\,28.1]{C17}). \\
\hspace*{6mm}
We are going to compute $T^{tct}$ explicitly. Change to the energy rep $Y_2^{-1}W_2Y_2$ by means of the unitary transformation $Y_2:=Y\otimes I_2$
\begin{displaymath} 
(Y\varphi)(p)=Y(p)\varphi(p), \;Y(p):=\left(\frac{m}{2\epsilon(p)}\right)^{1/2}
\left(\begin{array}{cc} Q(\mathfrak{p}^+)& Q(\mathfrak{p}^+)^{-1}\\ Q(\mathfrak{p}^+)^{-1} & -Q(\mathfrak{p}^+)\end{array}\right)
 \end{displaymath}
 One obtains $\oplus_\eta W^{\eta}_2$ with $\big(W^{\eta}_2(t,b,A)\varphi\big)(p):=\sqrt{\epsilon(q^\eta)/\epsilon(p)}\,\operatorname{e}^{\operatorname{i}(\eta t \epsilon(p)-bp)}
 R(\mathfrak{p}^\eta,A)\otimes R(\mathfrak{p}^\eta,A) \;\varphi(q^\eta)$, where $\varphi\in L^2(\R^3,\C^2\otimes\C^2)$,
  cf. \cite[sec.\,30.4]{C17}. By $D^{(1/2)}\otimes D^{(1/2)}\simeq D^{(0)}\oplus D^{(1)}$ one has $W^+_2\simeq W^{m,0,+}\oplus W^{m,1,+}$.
 So the carrier space of the subrep of $W_2$ equivalent to $W^{m,0,+}$ follows to be $Y_2 L^2(\R^3,\C\d)$ with $\d:=\frac{1}{\sqrt{2}}\big((1,0,0,0)^T\otimes(0,1)^T-(0,1,0,0)^T\otimes(1,0)^T\big)$. Let $j_{\d}:L^2(\R^3)\to L^2(\R^3,\C^4\otimes \C^2)$ be  the injection $j_{\d}\phi:=\phi\d$. The result is
 \begin{equation}\label{POLCTMB}
  T^{tct}=j_{\d}^*\,Y_2^{-1}\mathcal{F}E^{can}\mathcal{F}^{-1}Y_2\,j_{\d}
\end{equation}
From this one gets  $\textsc{t}^{tct}(k,p):= \langle Y_2(k)\d,Y_2(p)\d\rangle$ for the kernel of $T^{tct}$.  Then a straight forward  computation, using the explicit expression for $Q(\mathfrak{k})$ \cite[Eq.\,(3.5)]{C17}, yields (\ref{KPOLCT}).\qed\\

{\it Proof of} (\ref{ETCT}).  Regarding the  spinor space it is appropriate   to switch  from the tensor basis to the standard basis. So let $l:\C^4\otimes\C^2\to\C^8$ be an isomorphism such that $D:= D^{(0)}\oplus D^{(1)}\oplus D^{(0)}\oplus D^{(1)}=l\,(D^{(1/2)}\oplus D^{(1/2)})\otimes D^{(1/2)}l^{-1}$. In particular let $l\d=e_1$, $l\d_+=e_3$ with $\d_+:=\frac{1}{\sqrt{2}}\big((1,0,0,0)^T\otimes(0,1)^T+(0,1,0,0)^T\otimes(1,0)^T\big)$, and similarly $l^{-1}e_5$, $l^{-1}e_7$.
Accordingly one has the Hilbert space isomorphism  $L:L^2(\R^3,\C^4\otimes\C^2)\to L^2(\R^3,\C^8)$, $(L\varphi)(p):=l\varphi(p)$. It satisfies $D_{\tilde{\mathcal{E}}}=L \,W_2|_{\tilde{\mathcal{E}}}L^{-1}$. Further put $P_0:L^2(\R^3,\C^8)\to L^2(\R)$, $P_0\varphi:=\varphi_1$. Note $P_0^*\phi=\phi\,e_1$ and $P_0D_{\tilde{\mathcal{E}}}P_0^*=U^{(0)}$. Finally recall that $Y_2$ and
 $W_2|_{\tilde{\mathcal{E}}}$ commute. \\
\hspace*{6mm}
 Now, the extension of $W^{m,0,\eta}|_{\tilde{\mathcal{E}}}=U^{(0)}$ considered above for the construction of $T^{tct}$  is $U':=L\,Y_2\,W_2|_{\tilde{\mathcal{E}}}\,Y_2^{-1}L^{-1}$. Of course it equals $D_{\tilde{\mathcal{E}}}$ and hence satisfies $P_0 U' P_0^*=U^{(0)}$.  However $U'=\iota\,D_{\tilde{\mathcal{E}}}\iota^{-1}$ for $\iota:L^2(\R^3,\C^8)\to L^2(\R^3,\C^8)$,
$\iota:=L\,Y_2^{-1}L^{-1}$ so that $U'$ is endowed with the covariant PM $E'=\iota \mathcal{F}E^{can}\mathcal{F}^{-1}\iota^{-1}$. It gives rise to the POL $(W^{m,0,\eta}|_{\tilde{\mathcal{E}}}, T_{\e})$ with $T_{\e}=P_0E'P_0^*$ for some spinor choice $\e$ as shown in sec.\,\ref{CPOLMSB}.\\One easily verifies that $T_{\e}= P_0L\,Y_2^{-1}L^{-1}\mathcal{F}E^{can}\mathcal{F}^{-1}L\,Y_2L^{-1}P_0^*$ coincides with $T^{tct}$ in (\ref{POLCTMB}). It remains to compute $\e$ 
determined by $j_{\e}=XL\,Y_2L^{-1}P_0^*$. Put $X':=L^{-1}X\,L$, $(X'\varphi)(p)=\big((B(p)^{-1}\oplus B(p)^{-1}) \otimes B(p)^{-1}\big)\varphi(p)$, and  $Y'_2:=Y'\otimes I_2$ with 
\hspace*{6mm}

$(Y'\varphi)(p)=Y'(p)\varphi(p)$ and 
\begin{displaymath} 
 Y'(p):=\big(4\epsilon(p)(m+\epsilon(p)) \big)^{-1/2}
\left(\begin{array}{cc} m+\epsilon(p)+|p|\sigma_3& m+\epsilon(p)-|p|\sigma_3\\ m+\epsilon(p)-|p|\sigma_3 & -m-\epsilon(p)-|p|\sigma_3\end{array}\right)
 \end{displaymath}
Then $X'Y_2=Y'_2X'$ due to the relation Eq.\,\cite[(27.9)]{C17} concerning the helicity cross section. Hence $j_{\e}\phi=L\,Y'_2L^{-1}XP_0^*\phi=L\,Y'_2L^{-1}X\phi\,e_1=L\,Y'_2L^{-1}\phi\,e_1=LY'_2\phi\,\d$, whence $\e(|p|)=lY'_2(p)\d $, and (\ref{ETCT}) follows.\qed

\end{document}